\documentclass{article}
\usepackage[utf8]{inputenc}
\usepackage{amsmath}
\usepackage{amsfonts}
\usepackage{graphicx}
\graphicspath{ {./figures/} }
\usepackage{booktabs}
\usepackage{natbib}

\usepackage{subcaption}
\usepackage{diagbox}
\usepackage{geometry}
\usepackage{algorithm}
\usepackage{algorithmic}
\usepackage{bbm}
\usepackage{color}
\usepackage{hyperref}
\usepackage{mathtools}
\usepackage{authblk}

\newcommand{\kth}{{(k)}}

\newcommand{\mN}{\mathcal{N}}
\newcommand{\mO}{\mathcal{O}}
\newcommand{\mbP}{\mathbb{P}} 
\newcommand{\mQ}{\mathcal Q}
\newcommand{\mR}{\mathbb{R}}
\newcommand{\mth}{\text{(m)}}

\newcommand{\argmax}{\operatornamewithlimits{argmax}}
\newcommand{\argmin}{\operatornamewithlimits{argmin}}
\newcommand{\diag}{\mathop{\mathrm{diag}}}
\newcommand{\indep}{\perp\!\!\!\perp}
\newcommand{\ld}{\ln \det}
\newcommand{\txvec}{\mathrm{vec}}
\newcommand{\cov}{\mathrm{cov}}

\newcommand{\var}{\mathrm{var}}
\newcommand{\se}{\mathrm{se}}
\newcommand{\tr}{\mathrm{tr}}

\newcommand{\bb}{\bar\beta}
\newcommand{\be}{\bar\epsilon}
\newcommand{\bmu}{\bar\mu}
\newcommand{\bSig}{\bar\Sigma}
\newcommand{\bUp}{\bar \Upsilon}
\newcommand{\bX}{\bar X}
\newcommand{\bY}{\bar Y}
\newcommand{\bzt}{\bar \zeta}
\newcommand{\e}{\epsilon}
\newcommand{\Gb}{\Gamma_\beta}
\newcommand{\Ge}{\Gamma_\epsilon}

\newcommand{\hT}{\hat\Theta}
\newcommand{\lone}{\lambda_{1}}
\newcommand{\ltwo}{\lambda_{2}}
\newcommand{\rij}{\rho_{ij}}
\newcommand{\rk}{\rho^\kth}

\newcommand{\SnN}{\Sigma_{n=1}^N}

\newcommand{\tij}{\theta_{ij}}
\newcommand{\Tinv}{\Theta^{-1}}

\newcommand{\tz}{\tilde z}
\newcommand{\Up}{\Upsilon}
\newcommand{\up}{\upsilon}
\newcommand{\vY}{\mathrm{vec}Y}
\newcommand{\Xb}{X\beta}

\title{Flexible and Accurate Methods for Estimation and Inference of Gaussian Graphical Models with Applications}
\author{Yueqi Qian, Xianghong Hu, and Can Yang}
\affil{Department of Mathematics, The Hong Kong University of Science and Technology}
\affil{\href{mailto:yqianai@connect.ust.hk}{yqianai@connect.ust.hk}, \href{mailto:maxhu@ust.hk}{maxhu@ust.hk}, \href{mailto:macyang@ust.hk}{macyang@ust.hk}}

\begin{document}

\maketitle


\begin{abstract}

The Gaussian graphical model (GGM) incorporates an undirected graph to represent the conditional dependence between variables, with the precision matrix encoding partial correlation between pair of variables given the others.
To achieve flexible and accurate estimation and inference of GGM, we propose the novel method FLAG, which utilizes the random effects model for pairwise conditional regression to estimate the precision matrix and applies statistical tests to recover the graph.
Compared with existing methods, FLAG has several unique advantages: 
(i) it provides accurate estimation without sparsity assumptions on the precision matrix, (ii) it allows for element-wise inference of the precision matrix, (iii) it achieves computational efficiency by developing an efficient PX-EM algorithm and a MM algorithm accelerated with low-rank updates, and (iv) it enables joint estimation of multiple graphs using FLAG-Meta or FLAG-CA.
The proposed methods are evaluated using various simulation settings and real data applications, including gene expression in the human brain, term association in university websites, and stock prices in the U.S. financial market. The results demonstrate that FLAG and its extensions provide accurate precision estimation and graph recovery.

\end{abstract}


\section{Introduction}

Quantifying the relationships among components in a complex system based on observations is a fascinating yet challenging problem. Graphical models utilize probability models to represent relationships as a graph, where the nodes are random variables and the edges denote their dependencies.
Graphical models have wide real-world applications in various research fields, including genetics \cite{feng2019high, zhao2019cancer, yi2022information}, economics \cite{anufriev2015connecting, bernardini2022new}, psychology \cite{epskamp2018gaussian, williams2021bayesian}, and environmental science \cite{engelke2020graphical}.


To model the components in the system, we consider a $p$-dimensional random vector distributed in a multivariate normal distribution with zero mean, without loss of generality, as $z \sim \mN(0,\Sigma)$, with $\Theta \coloneqq \Sigma^{-1}$.
Then, we have $p(z_i|z_{-i}) \propto \exp(-\frac12\Theta_{ii}z_i^2-\Sigma_{j\neq i}z_i\Theta_{ij}z_j)$ which implies that the probability of the $i$-th component only depends on the components with nonzero entries in the $i$-th column of the precision matrix $\Theta$. Also, 
$p(z_i,z_j|z_{-ij})\propto \exp(-\frac12
\begin{bmatrix}z_i & z_j\end{bmatrix}
\begin{bmatrix}\Theta_{ii} & \Theta_{ij}\\ \Theta_{ji} & \Theta_{jj}\end{bmatrix} \\
\begin{bmatrix}z_i\\ z_j\end{bmatrix} )$
as a bivariate normal distribution form, which implies that the conditional dependence between $i$ and $j$-th components, given the remaining $p-2$ variables, can be quantified by the conditional correlation
$\rho_{ij}=-\frac{\tij}{\sqrt{\theta_{ii}\theta_{jj}}}$.
On the contrary, a zero value of $\theta_{ij}$ indicates the conditional independence between the $i$-th and $j$-th random variables, i.e., $z_i \indep z_j \mid z_{-ij} \iff \theta_{ij}=0$ \cite{rue2005gaussian}.
Let $G=(V,E)$ be a graph representation of the conditional dependence between the random variables in $z$, where the vertex set is $V=\{z_1, \dots, z_p\}$, and the edge set is $E=\{(z_i,z_j) \mid \Theta_{ij} \neq 0\}$. The structure of this undirected graph depends on the nonzero elements in the precision matrix $\Theta$.
The main task of the Gaussian graphical model is to estimate the precision matrix and infer the graph structure indicating conditional dependence.


Many existing methods for estimating the precision matrix rely on structural assumptions, including the sparsity of the entire precision matrix or each column, and graph structures such as hubs, clusters, and bands. The penalized likelihood-based methods \cite{friedman2008sparse, rothman2008sparse, fan2009network, cai2011constrained, tan2014learning, jankova2015confidence} impose penalty based on assumptions to the likelihood function, while conditional regression-based methods \cite{meinshausen2006high, ren2015asymptotic, chen2016asymptotically} transfer precision matrix estimation to regression problems using conditional Gaussian property.
Bayesian methods \cite{atay2005monte, dobra2011bayesian, williams2021bayesian} assume prior probabilities for the precision matrix and use posterior probabilities for inference and sampling process is required. 

However, theses assumptions may not reflect the actual structure of the precision matrix, and the penalties can induce bias in the estimates.
These methods often require hyperparameter tuning and their results depend on the selection of hyperparameters.
As for the Bayesian Gaussian graphical models (BGGM), empirical intervals of the magnitude of precision entries are required to be specified manually based on domain knowledge. In conclusion, the usage of these methods is limited in real-world applications, as they may lead to unstable results when the data is at various scales.

In terms of the joint estimation of multiple graphs, one common limitation of existing methods, including penalized likelihood-based joint estimation \cite{danaher2014joint, shan2018joint}, hierarchical-based methods \cite{guo2011joint, lee2015joint}, prior information incorporated estimation \cite{ma2016joint, ma2016joint, bilgrau2020targeted, saegusa2016joint, hao2018simultaneous, price2021estimating}, is that they are at the precision level, while the diagonal of precision matrices might vary, due to the unbalanced sample sizes in different groups, even when they indeed share some common structures.
Furthermore, when the underlying common pattern is encoded in the same partial correlation between pairs across groups, for example, $\rho_{ij}^\kth = \rho_{ij}^{(k')}$ where $k \neq k'$ denote different groups, the precision element $\tij=-\rho_{ij}\sqrt{\theta_{ii}\theta_{jj}}$ is influenced by the diagonal elements in the precision matrix, which might not remain the same across groups.
Therefore, the joint estimation methods that rely solely on precision matrices might not be sufficient to capture the full picture of the underlying structures from the data.

In summary, the Gaussian graphical models present several challenges, particularly in high-dimensional settings where the number of dimensions $p$ is large. These challenges include method feasibility when the sample size $n$ is smaller than $p$, high computational cost when $p$ and $n$ are large, and quantifying the uncertainty of conditional dependence of each pair.
Furthermore, leveraging the partially shared structure from different groups within the same domain to achieve joint estimation and inference has not been studied extensively.

The remaining content is organized as follows: In Section 2, we propose a flexible and accurate method for the estimation and inference of Gaussian graphical model.
Section 3 introduces the accelerated algorithms for multiple pairs and the extended model for joint estimation of multiple graphs.
We present numerical results from various simulation studies and real data analysis in Section 4. Finally, we conclude this manuscript with a brief discussion in the Section 5.


\section{Methods}


The proposed method does not depend on any explicit structural assumptions on the precision matrix, and it does not introduce bias into the estimates. Instead, it utilizes the conditional Gaussian property and rewrites the estimation of each entry in the precision matrix as the estimation of the covariance of residuals obtained by regressing two variables on the remaining $p-2$ variables.
Unlike the asymptotic normal thresholding (ANT) method from \cite{ren2015asymptotic}, we neither impose sparsity assumptions on the regression coefficients nor assume column-wise sparsity in the precision matrix, as the shrinkage on parameters may introduce bias to the residuals of regressions and the precision entries.
In addition to estimation, the proposed method enables inference and quantification of the uncertainty of each entry in the precision matrix and the corresponding edge in the graph.


\subsection{Model Setting}
Utilizing the conditional Gaussian property, the proposed method estimates a two-by-two submatrix of the precision matrix each time by taking the inverse of the residual oebatined from a two-versus-rest bivariate regression. To achieve an unbiased estimation of covariance of residuals and the precision entries, each regression is solved by random effect model.

Consider a pair of random variables $a=\{i,j\}$ versus the other $p-2$ variables each time. Take the $i$-th and $j$-th elements from the $p$-dimensional random vector $z$ as responses $y=[z_i, z_j]$, while the remaining $p-2$-dimensional random vector $x=[z_i,z_j]^c$ indicates the explanatory variables. The conditional probability $y|x \sim \mN(x^T \beta, \Theta_{aa}^{-1})$ can be expressed as $y=x^T \beta + \epsilon$, where $\epsilon \sim \mN(0, \Ge)$ and $\Theta_{aa} = \Ge^{-1}$.

Let $Z\in\mathbb R^{n\times p}$ denote a collection of $n$ realizations of the random vector $z$, the $i$-th and $j$-th columns from observation $Z$ as responses $Y=[Z_{\cdot i},Z_{\cdot j}]\in\mR^{n\times2}$, while the remaining columns $X=[Z_{\cdot i},Z_{\cdot j}]^c\in\mR^{n\times(p-2)}$ indicate the explanatory variables.
Subsequently, a bivariate regression model is constructed based on the conditional Gaussian property, as
$Y = X \boldsymbol{\beta} + \boldsymbol{\epsilon}$, 
where the coefficient matrix $\boldsymbol{\beta} \in \mR^{(p-2)\times2}$, and the covariance of each row $\epsilon_{k\cdot}$ in the $\epsilon \in \mR^{n\times2}$ is a 2-by-2 positive definite matrix $\Ge$, satisfying $\cov(\epsilon_{k\cdot}^T)=\Ge=\Theta_{aa}^{-1}$.

To solve this bivariate regression, we consider a random effects model 
\begin{equation}
\begin{aligned}
& Y = X \boldsymbol{\beta} + \boldsymbol{\epsilon}, \\
& \beta_{k\cdot}^T \sim \mN(0,\Gb), \epsilon_{k\cdot}^T \sim \mN(0,\Ge),
\label{REM}
\end{aligned}
\end{equation}
where $\beta$ is treated as random effects, and the $k-$th row $\beta_{k\cdot}$ is assumed to be distributed in a normal distribution with zero mean and covariance as $\Gb$.

After vectorizing $Y$ and $X \boldsymbol{\beta}$, we obtain $\vY|X \sim \mN(\txvec(X \boldsymbol{\beta}), \Ge \otimes I_n)$.
By integrating over $\beta$, the random effects model can be expressed as
\begin{equation}
\begin{aligned}
& \vY\sim N(0, \Omega^{-1}), \\
& \Omega=\Gb\otimes XX^T+\Ge\otimes I_n.
\label{model}
\end{aligned}
\end{equation}
The parameters in this model are denoted as
$\Gb = \begin{bmatrix}\sigma_1^2 & \tau\\ \tau & \sigma_2^2\end{bmatrix},
\Ge = \begin{bmatrix}\sigma_3^2 & \eta\\ \eta & \sigma_4^2\end{bmatrix},$
where $\Gb$ and $\Ge$ are symmetric and positive semi-definite matrices.

Firstly, the variance components $\Gb$ and $\Ge$ are estimated for the pair $(i,j)$, using efficient algorithms designed in the random effects model.
Based on the conditional Gaussian property, the submatrix of the precision matrix with respect to this pair can be estimated by $\Theta_{aa}=\Ge^{-1}$.
Furthermore, to quantify the uncertainty of each entry in the precision matrix, inference can be performed based on the proposed model. In addition, edges in the graph are detected through hypothesis testing, rather than relying solely on point estimates.

\subsection{Algorithms}

To estimate the variance component $\Ge$, two approaches based on maximum likelihood estimation (MLE) are provided: the minorize-maximization (MM, \cite{hunter2000quantile}) algorithm and the parameter-expanded expectation-maximization (PX-EM, \cite{liu1998parameter}) algorithm.

According to the random effects model as shown in the formula \ref{model}, the log-likelihood function with respect to the random components $\Gamma = \{\Gb, \Ge\}$ in a two-versus-rest conditional regression for each pair is
\begin{equation}
\begin{aligned}
\ell(\Gamma)=\ln\mbP(Y|X;\Gb,\Ge)
 =-\frac12\ld\Omega-\frac12\vY^T\Omega^{-1}\vY+c,
\label{incomplete_ll}
\end{aligned}
\end{equation}
where $c$ is a trivial constant. Two MLE-based algorithms have been developed for estimating the variance components in order to achieve unbiased estimation and statistical inference.

\subsubsection{MM Algorithm}
Direct maximum likelihood estimation of variance components models is numerically challenging. The minorize-maximization (MM) algorithm first finds a surrogate function $g$ that minorizes the log-likelihood function \ref{incomplete_ll}, such that $g(\Gamma|\Gamma^{(m)}) \leq \mathcal{L}(\Gamma)$. Then, the optimization variable is updated according to the current surrogate function, i.e., $\Gamma^{(m+1)}=\argmax_{\Gamma} g(\Gamma|\Gamma^{(m)})$.

The surrogate function for the log-likelihood function with respect to variance components is constructed using two minorizations based on two inequalities \cite{zhou2019mm}. The convexity of the negative log determinant function implies
$-\ld\Omega \geq -\ld \Omega^{(m)}-\tr[\Omega^{-(m)} (\Omega-\Omega^{(m)})].$
Since the variance components $\Gb$ and $\Ge$ are positive definite matrices, $\Omega$ is also positive definite, then we have
$\Omega^{-1} \preceq \Omega^{-(m)} [(\Gb^{(m)}\Gb^{-1}\Gb^{(m)})\otimes XX^T + (\Ge^{(m)}\Ge^{-1}\Ge^{(m)})\otimes I_n] \Omega^{-(m)}.$
The surrogate function for the MM algorithm is then given by
\begin{equation}
\begin{aligned}
g(\Gamma|\Gamma^{(m)}):=
&-\tr[\Omega^{-(m)}(\Gb\otimes XX^T)] 
 - \tr[\Gb^{(m)}R^{(m)T}XX^TR^{(m)}\Gb^{(m)}\Gb^{-1}] \\
&-\tr[\Omega^{-(m)}(\Ge\otimes I_n)]
 -\tr[\Ge^{(m)}R^{(m)T}R^{(m)}\Ge^{(m)}\Ge^{-1}] + c^{(m)},
\label{surrogate}
\end{aligned}
\end{equation}
where $c^{(m)}$ is a constant in the $m$-th iteration, and the matrix $R\in\mR^{n\times2}$ satisfies $\txvec(R^{(m)})=\Omega^{-(m)}\vY$ in all iterations.

In each iteration, the parameters in $\Gamma$ are updated by setting the derivative of $g(\Gamma|\Gamma^{(m)})$ to be zero, as $\Gb$ is updated by $\nabla_{\Gb} g(\Gamma|\Gamma^{(m)})=0$
and $\Ge$ is updated by $\nabla_{\Ge} g(\Gamma|\Gamma^{(m)})=0$. The log-likelihood is then calculated after update. Once the change in log-likelihood becomes arbitrarily small, the MM algorithm is considered to have converged.



Due to the high computational cost of inverting the large matrix $\Omega\in\mR^{(2n)\times(2n)}$ in each iteration, eigen-decomposition is used to reduce such consumption on frequent matrix inverting.
Let the eigen-decomposition of $XX^T$ be $U^TXX^TU=D=\diag(d)$, where $D$ is a diagonal matrix with its diagonal elements denoted by the vector $d \in \mR^n$.
The simultaneous congruence decomposition of $(\Gb,\Ge)$ is $(\Lambda, \Phi)$, such that 
$\Phi^T\Gb\Phi=\Lambda, \Phi^T\Ge\Phi=I_2$.
Then, $\Gb=\Phi^{-T}\Lambda\Phi^{-1}, \Ge=\Phi^{-T}I_2\Phi^{-1}$. The inverse of $\Omega$ can be efficiently calculated in each iteration according to the following equations
\begin{equation}
\begin{aligned}
\Omega^{(m)} = (\Phi^{-(m)}\otimes U^{-1})^T (\Lambda^{(m)}\otimes D+I_2\otimes I_n) (\Phi^{-(m)}\otimes U^{-1}),\\
\Omega^{-(m)} = (\Phi^{(m)}\otimes U) (\Lambda^{(m)}\otimes D+I_2\otimes I_n)^{-1} (\Phi^{(m)}\otimes U)^T.
\end{aligned}
\end{equation}
Additionally, the determinant of $\Omega$ can be calculated accordingly as $|\Omega^{(m)}|=| \Lambda^{(m)}\otimes D + I_n\otimes I_n | |\Ge^{(m)}|^n$.

In each iteration, $\Gb$ is updated by setting the derivative of $g(\Gamma|\Gamma^{(m)})$ with respect to $\Gb$ to zero, and $\Ge$ is updated similarly.
The former trace term in \ref{surrogate} is linear to $\Gamma$, with the coefficients collected in the $(2n)\times(2n)$ matrices $M_\beta$ and $M_\epsilon$ as
\begin{equation}
\begin{aligned}
M_\beta = \Phi^{(m)} \diag\{ \tr[ D(\lambda_l^{(m)}D+I_n)^{-1} ] \} \Phi^{(m)T}, \\
M_\epsilon = \Phi^{(m)} \diag\{ \tr[ (\lambda_l^{(m)}D+I_n)^{-1} ] \} \Phi^{(m)T}.
\label{mm_mbe}
\end{aligned}
\end{equation}

The latter trace term, which involves the inverse of $\Gamma$ can be rewritten in the general form $-\tr[A\Gamma^{-1}]$. Its derivative with respect to $\Gamma$ is $\Gamma^{-1}A\Gamma^{-1}$. For positive definite matrices $A$ and $M$, the unique positive definite solution for $\Gamma$ with respect to the Riccati equation $M=\Gamma^{-1}A\Gamma^{-1}$ is given by $L^{-T}(L^TAL)^\frac12L^{-1}$, where $L$ is the Cholesky factor of $M$.

After bypassing the computational cost of matrix inversion and solving the updated $\Gamma$ in the Riccati equation, we further reduce the computational cost by simplying the coefficients of the latter trace term in the surrogate function that were generalized as $A$ before. Specifically, $A_\beta=\Gb^{(m)}R^{(m)T}XX^TR^{(m)}\Gb^{(m)}$ and $A_\epsilon=\Ge^{(m)}R^{(m)T}R^{(m)}\Ge^{(m)}$ are $2\times2$ symmetric matrices, but the inner calculation involves matrix multiplication of a large matrix with dimension $n$ which is repeated in each iteration. To moderate this computational cost, the coefficients of inverse terms are denoted by matrices $N_\beta^TN_\beta$ and $N_\epsilon^TN_\epsilon$ as
\begin{equation}
\begin{aligned}
\Gb^{(m)}R^{(m)T}XX^TR^{(m)}\Gb^{(m)} = N_\beta^TN_\beta, \\
\Ge^{(m)}R^{(m)T}R^{(m)}\Ge^{(m)} = N_\epsilon^TN_\epsilon.
\end{aligned}
\end{equation}

Taking all the aforementioned techniques into consideration to solve the optimization problem of maximizing the surrogate function \ref{surrogate} and further speed it up, the MM algorithm can be summarized as follows,

\begin{algorithm}
\caption{MM Algorithm with Eigen-decomposition}
\label{MM_eigen}
\hspace*{\algorithmicindent} \textbf{Input}: X, Y \\
\hspace*{\algorithmicindent} \textbf{Output}: $\hat\Gamma_\beta, \hat\Gamma_\epsilon$
\begin{algorithmic}[1]
\STATE Eigen decomposition: $U^TXX^TU=D=\diag(d)$.
\STATE Transform data: $\tilde Y\leftarrow U^TY$.
\STATE Initialization: $\Gb^{(0)}=\Ge^{(0)}=\frac12\cov(Y)$. \\
\REPEAT
\STATE Simultaneous congruence decomposition: $(\Lambda^{(m)}, \Phi^{(m)}) \leftarrow (\Gb^{(m)},\Ge^{(m)}), $
\STATE $ \Omega^{(m)} \leftarrow (\Phi^{-(m)}\otimes U^{-1})^T (\Lambda^{(m)}\otimes D +I_2\otimes I_n) (\Phi^{-(m)}\otimes U^{-1}), $
\STATE $ \text{Cholesky }L_\beta^{(m)}L_\beta^{(m)T}\leftarrow\Phi^{(m)}\diag( \tr(D(\lambda_l^{(m)}D+I_n)^{-1}), l=1,2 )\Phi^{(m)T}, $
\STATE $ \text{Cholesky }L_\epsilon^{(m)}L_\epsilon^{(m)T}\leftarrow\Phi^{(m)}\diag( \tr((\lambda_l^{(m)}D+I_n)^{-1}), l=1,2 )\Phi^{(m)T}, $
\STATE $ N_\beta^{(m)}\leftarrow D^{\frac12} [(\tilde Y \Phi^\mth) \oslash (d\lambda^{(m)T}+\mathbf{1}_n\mathbf{1}_2^T)] \Lambda^{(m)}\Phi^{-(m)}, $
\STATE $ N_\epsilon^{(m)}\leftarrow [(\tilde Y \Phi^\mth) \oslash (d\lambda^{(m)T}+\mathbf{1}_n\mathbf{1}_2^T)] \Phi^{-(m)}, $
\STATE $ \Gb^{(m+1)} \leftarrow L_\beta^{-(m)T} (L_\beta^{(m)T} N_\beta^{(m)T} N_\beta^{(m)} L_\beta^{(m)})^{\frac12} L_\beta^{-(m)}, $
\STATE $ \Ge^{(m+1)} \leftarrow L_\epsilon^{-(m)T} (L_\epsilon^{(m)T} N_\epsilon^{(m)T} N_\epsilon^{(m)} L_\epsilon^{(m)})^{\frac12} L_\epsilon^{-(m)}. $
\UNTIL{the log-likelihood $\mathcal{L}(\Gamma)$ stop increasing or maximum iteration reached}
\end{algorithmic}
\end{algorithm}

Note that $\oslash$ denotes the Hadamard quotient. The MM algorithm estimates two variance component matrices, $\Gb$ and $\Ge$, and the estimate of the corresponding $2\times2$ submatrix of the precision matrix can be estimated using the inverse of $\hat\Ge$.

\subsubsection{PX-EM Algorithm}

The parameter-expanded expectation-maximization (PX-EM) algorithm \cite{liu1998parameter} is an accelerated version of the EM algorithm that is fast and stable in estimating variance-covariance components in linear mixture models \cite{foulley2000px}.

The linear model \ref{REM} is reconstructed for a parameter expanded version as 
\begin{equation}
\begin{aligned}
& Y=\delta X \boldsymbol{\beta} + \boldsymbol{\epsilon}, \\
& \beta_{k\cdot}^T \sim \mN(0,\Gb), \epsilon_{k\cdot}^T \sim \mN(0,\Ge),
\end{aligned}
\end{equation}
where $\delta\in \mathbb{R}^1$ is the expanded parameter.
The data and parameters are vectorized as follows,
$\bX=I_2\otimes X\in\mathbb{R}^{2n\times2(p-2)},
\bb=\txvec \beta \in\mathbb{R}^{2(p-2)}$ with $
\bb \sim \mN(0,\Gb\otimes I_{p-2}),
\be=\txvec \epsilon \in\mathbb{R}^{2n}$ with $
\be\sim \mN(0,\Ge\otimes I_{n}), $ and $
\bY=\vY=\delta\bX\bb+\be \in\mathbb{R}^{2n}.$

The complete data log-likelihood is
\begin{equation}
\begin{aligned}
\ell(\Gamma) = & \text{logPr} (\bY, \bb|\Gb,\Ge;\bX) \\
=& -\frac n2 \text{log}|\Ge|
    -\frac12 (\bY-\delta\bX\bb)^T
        (\Ge^{-1}\otimes I_{n})
        (\bY-\delta\bX\bb) \\
 & -\frac{p-2}2 \text{log}|\Gb|
    -\frac12 \bb^T
        (\Gb^{-1}\otimes I_{p-2})
        \bb.
\label{complete_ll}
\end{aligned}
\end{equation}

The terms involving $\bb$ are in a quadratic form given by
$\bb^T (-\frac{\delta^2}2 \Ge^{-1} \otimes X^TX -\frac12 \Gb^{-1} \otimes I_{p-2}) \bb
+\delta\bY^T (\Ge \otimes X) \bb$.
The posterior distribution of $\bb$ is $N(\bb|\mu_{\bb}, \Sigma_{\bb})$, where
$$ \Sigma_{\bb}^{-1}= \delta^2 \Ge^{-1} \otimes X^TX +\Gb^{-1} \otimes I_{p-2}, $$
$$ \mu_{\bb}= (\delta^2 \Ge^{-1} \otimes X^TX +\Gb^{-1} \otimes I_{p-2})^{-1}
\delta(\Ge^{-1} \otimes X^T)\bY. $$

During the E-step of the PX-EM algorithm, the $\mQ$-function is evaluated by taking the expectation of the complete data log-likelihood with respect to the posterior $N(\bb|\mu_{\bb}, \Sigma_{\bb})$.
The quadratic terms involving $\bb$ are taken as expectation values:
\begin{equation}
\begin{aligned}
& \mathbb{E}[(\bY-\delta\bX\bb)^T
        (\Ge^{-1}\otimes I_{n})
        (\bY-\delta\bX\bb)] \\
=& (\bY-\delta\bX \mu_{\bb})^T (\Ge^{-1}\otimes I_{n}) 
    (\bY-\delta\bX \mu_{\bb})
    +\delta^2 \text{tr}[ (\Ge^{-1} \otimes X^TX) \Sigma_{\bb} ]. \\
\end{aligned}
\end{equation}
$$
\mathbb{E}[\bb^T (\Gb^{-1}\otimes I_{p-2}) \bb] \\
= \mu_{\bb}^T (\Gb^{-1}\otimes I_{p-2}) \mu_{\bb}
+ \text{tr}[ (\Gb^{-1}\otimes I_{p-2}) \Sigma_{\bb} ]
$$

The $\mQ$-function given the estimated parameter in the previous iteration as $\theta_{old}$, is expressed as follows,
\begin{equation}
\begin{aligned}
\mQ(\theta|\theta_{old})
=& -\frac n2 \text{log}|\Ge| -\frac{p-2}2 \text{log}|\Gb| \\
 & -\frac12 \text{tr}\Big[ (\Ge^{-1}\otimes I_{n})
    [ (\bY-\delta\bX \mu_{\bb})(\bY-\delta\bX \mu_{\bb})^T
     +\delta^2 \bX \Sigma_{\bb} \bX^T] \Big] \\
 & -\frac12 \text{tr}\Big[ (\Gb^{-1}\otimes I_{p-2})
    ( \mu_{\bb} \mu_{\bb}^T + \Sigma_{\bb} ) \Big] \\
=& -\frac n2 \text{log}|\Ge| -\frac{p-2}2 \text{log}|\Gb| \\
 & -\frac12 \text{tr}\Big[ \Ge^{-1} 
    \begin{pmatrix} \text{tr}[S_{11}] & \text{tr}[S_{12}] \\ \text{tr}[S_{21}] & \text{tr}[S_{22}] \end{pmatrix} \Big]
   -\frac12 \text{tr}\Big[ \Gb^{-1} 
    \begin{pmatrix} \text{tr}[W_{11}] & \text{tr}[W_{12}] \\ \text{tr}[W_{21}] & \text{tr}[W_{22}] \end{pmatrix} \Big],
\end{aligned}
\end{equation}
where $S = \begin{pmatrix} S_{11} & S_{12} \\ S_{21} & S_{22} \end{pmatrix} = (\bY-\delta\bX \mu_{\bb})(\bY-\delta\bX \mu_{\bb})^T
     +\delta^2 \bX \Sigma_{\bb} \bX^T$, 
$W = \begin{pmatrix} W_{11} & W_{12} \\ W_{21} & W_{22} \end{pmatrix} = \mu_{\bb} \mu_{\bb}^T + \Sigma_{\bb}$.
Denote $\bY = \begin{pmatrix} \bY_1 \\ \bY_2 \end{pmatrix}$,
$\mu_{\bb} = \begin{pmatrix} \bmu_1 \\ \bmu_2 \end{pmatrix}$,
$\Sigma_{\bb} = \begin{pmatrix} \bSig_{11} & \bSig_{12} \\ \bSig_{21} & \bSig_{22} \end{pmatrix}$. Then, for $i=1,2; j=1,2$,
\begin{equation}
\begin{aligned}
&\text{tr}[S_{ij}]= \text{tr}[ (\bY_i-\delta X\bmu_i)(\bY_j-\delta X\bmu_j)^T
+\delta^2 X \bSig_{ij} X^T], \\
&\text{tr}[W_{ij}]= \text{tr}[ \bmu_i \bmu_j^T + \bSig_{ij} ].
\end{aligned}
\end{equation}

In the subsequent M-step, the new estimates of the parameters are obtained by setting the derivative of the $\mQ$-function to be zero.
From the detailed calculations in the supplementary materials \ref{supp_pxem}, the updated parameters are $\Ge = \frac1n \begin{pmatrix} \text{tr}[S_{11}] & \text{tr}[S_{12}] \\ \text{tr}[S_{21}] & \text{tr}[S_{22}] \end{pmatrix}$, $\Gb = \frac1{p-2} \begin{pmatrix} \text{tr}[W_{11}] & \text{tr}[W_{12}] \\ \text{tr}[W_{21}] & \text{tr}[W_{22}] \end{pmatrix}$, $\delta=\frac {\bY^T(\Ge^{-1}\otimes X) \mu_{\bb}}
                {\text{tr}[(\Ge^{-1} \otimes X^TX)
                (\mu_{\bb}\mu_{\bb}^T + \Sigma_{\bb})]}$.



To avoid frequent inversion the $2(p-2)\times2(p-2)$ matrix $\Sigma_{\bb}^{-1}$ in the iterations, an eigen-decomposition $X^TX=VQV^T$ is performed, where $Q\in\mathbb{R}^{(p-2)\times(p-2)}$ is a diagonal matrix with diagonal elements given by the vector $q$ of eigenvalues.
Hence, matrix $\Sigma_{\bb}^{-1}$ can be written as
\begin{equation}
\begin{aligned}
&\Sigma_{\bb}^{-1}
= \begin{pmatrix} 
\delta^2(\Ge^{-1})_{11}X^TX + (\Gb^{-1})_{11}I_{p-2} & \delta^2(\Ge^{-1})_{12}X^TX + (\Gb^{-1})_{12}I_{p-2} \\ \delta^2(\Ge^{-1})_{21}X^TX + (\Gb^{-1})_{21}I_{p-2} & \delta^2(\Ge^{-1})_{22}X^TX + (\Gb^{-1})_{22}I_{p-2} \end{pmatrix} \\
&\hspace{-0.1cm}=\begin{pmatrix} V & 0 \\ 0 & V \end{pmatrix}
\underbrace{
\small
\begin{pmatrix}
\delta^2(\Ge^{-1})_{11}Q + (\Gb^{-1})_{11}I_{p-2} & \delta^2(\Ge^{-1})_{12}Q + (\Gb^{-1})_{12}I_{p-2} \\ \delta^2(\Ge^{-1})_{21}Q + (\Gb^{-1})_{21}I_{p-2} & \delta^2(\Ge^{-1})_{22}Q + (\Gb^{-1})_{22}I_{p-2} \end{pmatrix}
}_{
\begin{pmatrix} A & B \\ C & H \end{pmatrix}
= \begin{pmatrix} \diag(a) & \diag(b) \\ \diag(c) & \diag(h) \end{pmatrix}
} 
\begin{pmatrix} V & 0 \\ 0 & V \end{pmatrix}^T.
\end{aligned}
\end{equation}
Since $X^TX$ is a real symmetric matrix, $V$ is an orthogonal matrix, as is the block matrix $\begin{pmatrix} V & 0 \\ 0 & V \end{pmatrix}$, whose inverse is trivial to obtain.
The matrix $\begin{pmatrix} A & B \\ C & H \end{pmatrix}$ in the middle consists of blocks with diagonal matrices, which makes it easier to calculate the inverse.
Specifically, the inverse of the middle matrix is
$ \begin{pmatrix} A & B \\ C & H \end{pmatrix} ^{-1}
= \begin{pmatrix} \diag( h \oslash (a\odot h - c\odot b) ) &
    \diag( -b \oslash (a\odot h - c\odot b) ) \\
    \diag( -c \oslash (a\odot h - c\odot b) ) &
    \diag( a \oslash (a\odot h - c\odot b) ) \end{pmatrix}, $
and then $\Sigma_{\bb}=
\begin{pmatrix} V & 0 \\ 0 & V \end{pmatrix}
\begin{pmatrix} A & B \\ C & H \end{pmatrix} ^{-1}
\begin{pmatrix} V^T & 0 \\ 0 & V^T \end{pmatrix}
$.
The PX-EM algorithm with the eigen-decomposition of $X^TX$ is summarized as follows,

\begin{algorithm}
\caption{PX-EM algorithm with eigen-decomposition}
\label{EM_eigen}
\begin{algorithmic}[1]
\STATE Initialization: $\Gb=\Ge=\frac{\cov(Y)}{2}$.
\STATE Eigen-decomposition: $X^TX=VQV^T$.
\REPEAT
\STATE E-step: set $\delta^{(m)}=1$,
\begin{equation}
\begin{aligned}
\Sigma_{\bb}=
& \begin{pmatrix} V^T & 0 \\ 0 & V^T \end{pmatrix} \\
&\hspace{-0.3cm} \begin{pmatrix} \diag( \delta^2(\Ge^{-1})_{11}q +                                         (\Gb^{-1})_{11}\mathbbm{1}_{p-2} ) &
        \diag( \delta^2(\Ge^{-1})_{12}q +                                       (\Gb^{-1})_{12}\mathbbm{1}_{p-2} ) \\
        \diag( \delta^2(\Ge^{-1})_{21}q +                                       (\Gb^{-1})_{21}\mathbbm{1}_{p-2} ) &
        \diag( \delta^2(\Ge^{-1})_{22}q +                                       (\Gb^{-1})_{22}\mathbbm{1}_{p-2} )
    \end{pmatrix} ^{-1} \\
& \begin{pmatrix} V & 0 \\ 0 & V \end{pmatrix},
\end{aligned}
\end{equation}
$$\mu_{\bb}= \Sigma_{\bb}
\delta(\Ge^{-1} \otimes X^T)\bY, $$
$$ELBO^{(m)}=Q(\Omega^{(m)})+\frac12\text{log}|\Sigma_{\bb}|.$$
\STATE M-step: Update the model parameters by
$$\delta^{(t+1)}=\frac {\bY^T(\Ge^{-1}\otimes X) \mu_{\bb}}
        { \mu_{\bb}^T (\Ge^{-1} \otimes X^TX) \mu_{\bb}
        + \tr[ (\Ge^{-1} \otimes X^TX) \Sigma_{\bb} ] }, $$
$$\Ge^{(t+1)} = \frac1n \begin{pmatrix} \tr[S_{11}] & \tr[S_{12}] \\ \tr[S_{21}] & \tr[S_{22}] \end{pmatrix}, $$
$$\Gb^{(t+1)} = \frac1{p-2} \begin{pmatrix} \tr[W_{11}] & \tr[W_{12}] \\ \tr[W_{21}] & \tr[W_{22}] \end{pmatrix}. $$
\STATE Reduction-step: Rescale
$\Gb^{(t+1)} = (\delta^{(t+1)})^2\Gb^{(t+1)}$
and reset $\delta^{(t+1)}=1$.
\UNTIL{the incomplete data log-likelihood $ELBO^{(m)}$ stop increasing}
\end{algorithmic}
\end{algorithm}

\subsubsection{Initialization}
In the previous algorithm design, we simply used the covariance of $Y$ to initialize the parameters, setting $\Gb=\Ge=\frac12\cov(Y)$.
Although The method of moments (MoM) estimators may not be optimal, they are easy to compute and can be used to calculate an initial value of parameters  for MLE-based iterative methods \cite{wasserman2004all} like the MM algorithm and PX-EM algorithm.

The parameters in the variance component set $\Gamma= \{\Gb, \Ge\}$ are denoted by $\gamma = [\sigma_1^2, \sigma_3^2, \sigma_2^2, \\ \sigma_4^2, \tau, \eta]^T$. The MoM estimator is obtained by solving the ordinary least squares (OLS) problem
$$ \argmin_\gamma \left\| \vY \vY^T - (\Gb \otimes XX^T + \Ge \otimes I_n) \right\|_F^2. $$

Denote $Y=[y_1, y_2]$, the MoM estimate of parameter $\gamma$ is
\begin{equation}
    \hat\gamma = \begin{bmatrix}
        \frac12 S_0^{-1} & 0 & 0 \\ 0 & \frac12 S_0^{-1} & 0 \\ 0 & 0 & \frac12 S_0^{-1}
    \end{bmatrix}
    \begin{bmatrix}
        2y_1^TXX^Ty_1 \\ 2y_1^Ty_1 \\
        2y_2^TXX^Ty_2 \\ 2y_2^Ty_2 \\
        4y_2^TXX^Ty_1 \\ 4y_2^Ty_1 \\
    \end{bmatrix},
\end{equation}
where $S_0 = \begin{bmatrix}
    \tr[(XX^T)^2] & \tr[XX^T] \\ \tr[XX^T] & n
\end{bmatrix}$.

\subsection{Inference}
For these maximum likelihood-based methods, as the MM algorithm with respect to the incomplete data log-likelihood function in Formula \ref{incomplete_ll} and the PX-EM algorithm for the complete data log-likelihood function in Formula \ref{complete_ll}, the difference between maximum likelihood estimate and the true parameter converges in distribution to a normal distribution with a mean of zero and a covariance matrix equal to the inverse of the Fisher information matrix as $ \sqrt{n}(\hat\Gamma-\Gamma^*) \xrightarrow{d} \mN(0,I^{-1}). $
The maximum likelihood estimator is $\sqrt{n}$-consistent and asymptotically efficient, with the smallest variance.

In addition to estimating of precision matrix, we further quantify the uncertainty of each entry in the precision matrix, and the existence and weight of the corresponding edge in the graph.

The parameters in the variance component set $\Gamma= \{\Gb, \Ge\}$ are denoted by 
$\gamma = [\gamma_1, \gamma_2, \gamma_3, \gamma_4, \\ \gamma_5, \gamma_6]^T := [\sigma_1^2, \sigma_3^2, \sigma_2^2, \sigma_4^2, \tau, \eta]^T
$.
The covariance matrix of the maximum likelihood estimates can be calculated
unsing the inverse of the Fisher Information Matrix (FIM), where the FIM is
$I(\gamma) = -E[\frac{\partial^2}{\partial\gamma^2}\log \text{Pr}(\vY|X;\Gamma)]$.
Denote
$M_1 = \begin{bmatrix} XX^T & 0 \\ 0 & 0 \end{bmatrix}$,
$M_2 = \begin{bmatrix} I_n & 0 \\ 0 & 0 \end{bmatrix}$,
$M_3 = \begin{bmatrix} 0 & 0 \\ 0 & XX^T \end{bmatrix}$,
$M_4 = \begin{bmatrix} 0 & 0 \\ 0 & I_n \end{bmatrix}$, 
$M_5 = \begin{bmatrix} XX^T & 0 \\ 0 & XX^T \end{bmatrix}$,
$M_6 = \begin{bmatrix} I_n & 0 \\ 0 & I_n \end{bmatrix}$, then we have

\begin{equation}
\begin{aligned}
\frac{\partial^2}{\partial\gamma_i \partial\gamma_j} \ln P(\vY|X;\Gamma)]
=\tr[ (\frac12I_{2n}-\Omega^{-1}\vY \vY^T) (\Omega^{-1}M_i \Omega^{-1}M_j) ].
\end{aligned}
\end{equation}

For MLE-based methods, the covariance matrix of the estimated parameters $\gamma$ is equal to the inverse of the fisher information matrix, denoted as $\cov(\gamma)=I(\gamma)^{-1}$. Using this, the variance of $\eta$ and its standard error can be obtained.

Recall that the non-zero precision entries correspond to edges in the graph and a zero off-diagonal entry $\Theta_{ij}$ are equivalent with a zero $\eta$ in $\Ge$ for the pair $(i,j)$ since zero off-diagonal entries remain after a 2-by-2 matrix inverse operation.

A null hypothesis is set as $H_0: \eta=0$, and the Wald test can be applied with a test statistic given by $W=\frac{(\eta-\eta_0)^2}{\var(\eta)}$, where $\eta_0=0$. The p-value of the test for the existence of an edge between the pair $(i,j)$ is collected.

Alternatively, the likelihood ratio test can be applied alternatively by calculating the difference between the log-likelihoods of the original parameter space $\gamma$ and the restricted parameter space where $\eta$ in $\Ge$ is constrained to zero. The test statistic is given by $-2[\mathcal{L}(\Gamma_0)-\mathcal{L}(\Gamma)]$ where the two parameter sets are optimized separately with respect to the log-likelihood function $\mathcal{L}$ in Formula \ref{incomplete_ll}, and $\Gamma_0$ denotes the parameters when $\eta$ in $\Ge$ is set to zero.

FLAG not only calculates the point estimates of the precision matrix but also computes standard errors and performs hypothesis testing on the precision entries, while many existing methods can only provide point estimates without efficient element-wise inference.
After collecting the p-value of each entry in the precision matrix, large-scale hypothesis testing 
is considered to control the false discovery rate (FDR) based on the Benjamini–Hochberg procedure \cite{benjamini1995controlling}.
Alternatively, the Bonferroni correction can be applied to control the family-wise error rate (FWER), which is relatively conservative \cite{hastie2009elements}.
This inference on the precision matrix can be used to extend the usage of FLAG when utilizing meta-analysis to jointly estimate multiple graphs dealing with data from various groups.


\section{Accelerated Algorithms and Extended Model}

\subsection{Low-rank Update for Multiple Pairs}

The most computationally intensive part of Algorithm \ref{MM_eigen}
designed to estimate the variance components is the eigen-decomposition with a computational complexity of $\mO(n^3)$, which becomes increasingly burdensome as $n$ grows.
Although the eigen-decomposition is performed only once when estimating the precision of each pair of variables, a total of $\frac{p(p-1)}{2}$ eigen-decompositions are required to estimate the entire precision matrix for all pairs.
It is worth noticing that the eigen-decomposition is calculated with respect to $XX^T$, where each $X$ for one pair of variables $(z_i, z_j)$ is the matrix $Z$ with the $i-$th and $j-$th columns removed, denoted as $X=Z_{-\{ij\}}$.
To improve the computational efficiency, the eigen-decomposition of $ZZ^T$ is performed first, followed by the eigen-decomposition of $XX^T=ZZ^T-Z_{-\{ij\}}(Z_{-\{ij\}})^T$ replaced by a low-rank update based on that of $ZZ^T$.

Denote the eigen-decomposition of symmetric matrix $ZZ^T$ is $ZZ^T=UDU^T$, then $XX^T=UDU^T-Z_{\{ij\}}(Z_{\{ij\}})^T$, and the variance-covariance matrix $\Omega=\Gb\otimes XX^T+\Ge\otimes I_n$ in the random effects model \ref{model} can be written as
\begin{equation}
\begin{aligned}
\Omega
&=\Gb\otimes (UDU^T)+\Ge\otimes I_n-\Gb\otimes(Z_{\{ij\}}Z_{\{ij\}}^T) \\
&=(\Phi^{-T}\otimes U) [\Lambda\otimes D+I_2\otimes I_n-\Lambda\otimes(U^TZ_{\{ij\}}(U^TZ_{\{ij\}})^T)](\Phi^{-1}\otimes U^T).
\end{aligned}
\end{equation}

In the MM algorithm, the log-likelihood function \ref{incomplete_ll} involves both the log-determinant and inverse terms with respect to $\Omega$, which need to be revised based on the low-rank update of the eigen-decomposition of $ZZ^T=UDU^T$.

Using the matrix determinant lemma, we have
$ \det \Omega = | \Ge |^n | I_n |^2
\prod_{l=1,2} ( | \lambda_l D + I_n |\,| I_2 - (U^TZ_{\{ij\}})^T (\lambda_l D + I_n)^{-1} U^TZ_{\{ij\}} | ). $
The inverse term is
\begin{equation}
\begin{aligned}
\Omega^{-1}
=& (\Phi\otimes U)
[\Lambda\otimes D+I_2\otimes I_n-\Lambda\otimes(U^TZ_{\{ij\}}(U^TZ_{\{ij\}})^T)]^{-1}(\Phi\otimes U)^T \\
=& \begin{bmatrix} \Phi_{11}U & \Phi_{12}U \\ \Phi_{21}U & \Phi_{22}U \end{bmatrix} \\
& \begin{bmatrix} (\lambda_1D+I_n-\lambda_1U^TZ_{\{ij\}}(U^TZ_{\{ij\}})^T)^{-1} & 0 \\ 0 & (\lambda_2D+I_n-\lambda_2U^TZ_{\{ij\}}(U^TZ_{\{ij\}})^T)^{-1} \end{bmatrix} \\
& \begin{bmatrix} \Phi_{11}U^T & \Phi_{21}U^T \\ \Phi_{12}U^T & \Phi_{22}U^T, \end{bmatrix}
\end{aligned}
\end{equation}
where the block matrix $[\lambda_l D+I_n-\lambda_l (U^TZ_{\{ij\}}(U^TZ_{\{ij\}})^T)]^{-1}, l=1,2$ in the diagonal of the center matrix is the inverse of a diagonal matrix with rank-2 correction.
This inversion can be calculated efficiently using the Woodbury matrix identity, then we have
\begin{equation}
\begin{aligned}
&[\lambda_l D+I_n-\lambda_l (U^TZ_{\{ij\}}(U^TZ_{\{ij\}})^T)]^{-1} \\
=& [(\lambda_l D+I_n)^{-1} + (\lambda_l D+I_n)^{-1} U^TZ_{\{ij\}} 
(\frac1{\lambda_l }I_2 - (U^TZ_{\{ij\}})^T (\lambda_l D+I_n)^{-1} U^TZ_{\{ij\}} )^{-1},
\end{aligned}
\end{equation}
for l=1,2.

Then the log-likelihood function \ref{incomplete_ll} can be rewritten as 
\begin{equation}
\begin{aligned}
&\ell(\Gamma)=-\frac12 \ld \Omega -\frac12\txvec(\tilde Y)^T \\
& \begin{bmatrix} [\lambda_1D+I_n-\lambda_1U^TZ_{\{ij\}}(U^TZ_{\{ij\}})^T]^{-1} & 0 \\ 0 & [\lambda_2D+I_n-\lambda_2U^TZ_{\{ij\}}(U^TZ_{\{ij\}})^T]^{-1} \end{bmatrix}
\txvec(\tilde Y),
\end{aligned}
\end{equation}
where $\txvec(\tilde Y)=(\Phi\otimes U)^T\vY = \txvec(U^T Y \Phi)$ is calculated only once for each pair before the iteration.

The coefficients of the parameters $\Gb$ and $\Ge$ in the gradient of the surrogate function \ref{surrogate}, which are collected in the matrices $M_\beta$ and $M_\epsilon$, are revised accordingly, with the details in Supplementary \ref{supp_low_rank}.


Similarly, the coefficients of the inverse terms $\Gb^{-1}$ and $\Ge^{-1}$ in the gradient of the surrogate function \ref{surrogate}, which are collected in the matrices $N_\beta^TN_\beta$ and $N_\epsilon^TN_\epsilon$, are also revised based on the low-rank update as
$N_\beta^TN_\beta=(R^{(m)}\Gb^{(m)})^T[U(D-U^TZ_{\{ij\}}(U^TZ_{\{ij\}})^T)U^T]R^{(m)}\Gb^{(m)}$,
where the term in the middle can be further simplified as
$EE^T= D-U^TZ_{\{ij\}}(U^TZ_{\{ij\}})^T 
= D^{\frac12}[I_n- D^{-\frac12}U^TZ_{\{ij\}} (D^{-\frac12}U^TZ_{\{ij\}})^T]D^{\frac12}
= (D^{\frac12} F^{\frac12}) (D^{\frac12} F^{\frac12})^T,$
then we have $E=D^{\frac12}F^{\frac12}$.

Let $J=D^{-\frac12}U^TZ_{\{ij\}} \in \mathbb R^{n\times2}$, then
$F^{\frac12}=I_n+J(J^TJ)^{-1}[(I_2-J^TJ)^{\frac12}-I_2]J^T$.
According to the simultaneous congruence decomposition, we have
$\Gb=\Phi^{-(t)T} \Lambda\Phi^{-(t)}, \Ge=\Phi^{-(t)T}\Phi^{-(t)}$. Then the matrices $N_\beta$ and $N_\epsilon$ can be obtained by
$$N_\beta=E^T U^T R^{(t)} \Phi^{-(t)T}\Lambda\Phi^{-(t)}$$
$$N_\epsilon=U^T R^{(t)} \Phi^{-(t)T}\Phi^{-(t)}$$

To further simplify the matrix $N_\beta$, we can vectorize it to obtain
\begin{equation}
\begin{aligned}
\txvec(N_\beta)
= (\Phi^{-T}\Lambda) \otimes E^T \txvec(G)
= \txvec(E^T G \Lambda \Phi^{-1}), \\
   \txvec(N_\epsilon)
= (\Phi^{-T}) \otimes I_n \txvec(G) 
= \txvec(G \Phi^{-1}),
\end{aligned}
\end{equation}
with the details shown in Supplementary \ref{supp_low_rank}.

Hence, the compact equation is $N_\beta=E^T G \Lambda \Phi^{-1}, N_\epsilon=G \Phi^{-1}$, and the expanded form of $N_\beta$ is
\begin{equation}
\begin{aligned}
N_\beta=& E^T G \Lambda \Phi^{-1} 
= F^{\frac12} D^{\frac12} G \Lambda \Phi^{-1} \\
=& \{I_n+J(J^TJ)^{-1}[(I_2-J^TJ)^{\frac12}-I_2]J^T\} D^{\frac12} G \Lambda \Phi^{-1} \\
=& \Big( (D^{\frac12} G) (\Lambda \Phi^{-1}) \Big)
+ \Big( J(J^TJ)^{-1}[(I_2-J^TJ)^{\frac12}-I_2] \Big)
\Big( J^T (D^{\frac12} G \Lambda \Phi^{-1}) \Big),
\end{aligned}
\end{equation}
where the matrix $J$ and the term $(J^TJ)^{-1}[(I_2-J^TJ)^{\frac12}-I_2]$ remain the same in all iterations, and thus they are only calculated once before the iterations.

\subsection{Meta-analysis for Multiple Groups}
A graph can be inferred individually for each group. Nevertheless, the limited samples size, particularly in high-dimensional setting, raises the follow-up research question of how to leverage data from different groups.
For instance, there are university websites for students, faculty, and courses, which may share many common phrases in websites such as "email address" and "home page" with steady relationships between words.
The goal is to leverage the universality across groups to estimate commonly shared pairs more accurately while maintaining the differences in the same pair across different groups, thus preserving the individuality.

\subsubsection{One-to-one Meta-analysis}
Denote
$\Ge
= \begin{bmatrix}\sigma_3^2 & \eta\\ \eta & \sigma_4^2\end{bmatrix}
= \begin{bmatrix}\sigma_3^2 & \rho\sigma_3\sigma_4\\
                 \rho\sigma_3\sigma_4 & \sigma_4^2\end{bmatrix}$,
and the partial correlation is $\rho=\frac{\eta}{\sigma_3\sigma_4}$.
The partial correlations from two groups A and B are denoted as $\rho^{(A)}=\frac{\eta^{(A)}}{\sigma_{3}^{(A)}\sigma_{4}^{(A)}},
\rho^{(B)}=\frac{\eta^{(B)}}{\sigma_{3}^{(B)}\sigma_{4}^{(B)}}$.

The first step is to test whether the partial correlation of a pair of variables across two groups, A and B, is the same or not.
The null hypothesis is $H_0: \rho^{(A)}-\rho^{(B)}=0$, and the test statistic is given by $\frac{\rho^{(A)}-\rho^{(B)}} {\sqrt{\se(\rho^{(A)})^2+\se(\rho^{(B)})^2}}$. The standard error of partial correlation $\rho$ can be obtained using the delta method, as
\begin{equation*}
    \se(\rho)^2=
    \begin{bmatrix}
    -\frac12\sigma_3^{-3}\sigma_4^{-1}\eta &
    -\frac12\sigma_3^{-1}\sigma_4^{-3}\eta &
    \sigma_3^{-1}\sigma_4^{-1}
    \end{bmatrix}
    \Sigma_{\Ge}
    \begin{bmatrix}
    -\frac12\sigma_3^{-3}\sigma_4^{-1}\eta \\
    -\frac12\sigma_3^{-1}\sigma_4^{-3}\eta \\
    \sigma_3^{-1}\sigma_4^{-1}
    \end{bmatrix},
\end{equation*}
where $\Sigma_{\Ge}$ is the covariance matrix of parameters in $\Ge=\begin{bmatrix}\sigma_3^2 & \sigma_4^2 & \eta\end{bmatrix}^T$, which is a submatrix of the inverse of the Fisher information matrix. Specifically, the rows and columns that correspond to these three parameters in $\Ge$ from the inverse of the Fisher information matrix are taken.

If the hypothesis is not rejected in this test, assume that $\rk=\rho+e^\kth$, where $k\in\{A,B\}$ and $e$ is random noise. Then, we use inverse-variance weighting to aggregate $\rho$ from different groups that share similar underlying $\rho$ as $\rho=\frac{\Sigma_k w^\kth \rk}{\Sigma_k w^\kth}$, with $w^\kth=\frac1{\se(\rk)^2}$ as weights. The standard error of the shared underlying $\rho$ is $\se(\rho)=\frac1{\sqrt{\Sigma_k w^\kth}}$. Then, we can adjust the parameter $\eta$ in different groups by $\eta^{(A,meta)}=\rho\sigma_{3}^{(A)}\sigma_{4}^{(A)}, \eta^{(B,meta)}=\rho\sigma_{3}^{(B)}\sigma_{4}^{(B)}$, and the precision will change accordingly.

FLAG-Meta provides a comprehensive analysis of both the similarities and differences between graphs from different groups by adaptively applying hypothesis testing on each edge across groups.
Unlike other methods, such as PNJGL for differences and CNJGL for common parts from \cite{mohan2012structured, mohan2014node}, FLAG-Meta does not require any extra design, in contrast to different penalty functions when target changes.

FLAG-Meta utilizes element-wise by group-wise comparisons to obtain the fine-grained structures across groups, rather than penalizing the same entry across groups equivalently, regardless of the group relations, as in JGL \cite{guo2011joint}, JEMP \cite{lee2015joint}, FGL and GGL \cite{danaher2014joint}, SCAN \cite{hao2018simultaneous}, TFRE \cite{bilgrau2020targeted}, and others.
Furthermore, it is easy for FLAG-Meta to incorporate prior information, such as group relations, group memberships, and relationships of edge subsets within group subsets, if available, into the FLAG-Meta framework.

The majority of existing joint estimation methods are designed at the precision level, typically as $\|\tij^{(k_1)}-\tij^{(k_2)}\|, 1\leq k_1, k_2 \leq K$ \cite{danaher2014joint, price2015ridge, saegusa2016joint, price2021estimating, mohan2012structured}, is penalized to encourage similarity.
In contrast, FLAG-Meta is flexible in testing similarity at the partial correlation, scaled precision level, which is more robust in comparing conditional dependence between the same variables across different groups after adjusting the influence from the varied variance and precision from the diagonal elements in the covariance or precision matrix.

In conclusion, FLAG-Meta incurs only a little extra computational cost of in $\mathcal{O}(K^2p^2)$ based on FLAG.
It is flexible in identifying both similarities and differences with fine-grained structure as element-wise by group-wise, which makes it easier to incorporate with prior information at any granularity, and it is accurate for smaller standard error and larger statistical power.
Moreover, FLAG-Meta only requires summary statistics instead of raw data from different resources, making it more valuable, especially when data from different groups cannot be shared.

\subsubsection{Many-to-one Meta-analysis}
The previous part explains the methodology for aggregating two groups through one-to-one meta-analysis, which can be further extended to more groups.
Suppose that there exist $K$ groups, and the set of all groups are denoted as $G=\{ 1,...,K \}$ with the cardinality $|G|=K$, and we first choose group 1 as the main target for explanation.
For each pair of random variables $(i,j)$ with $i\neq j$, the partial correlation from other groups are compared with $\rij^{(1)}$ separately by testing whether $\rij^{(1)} - \rij^{(k)} = 0$ for $k=2,...,K$.
Then the groups other than group 1 whose tests cannot be rejected are collected in a subset of groups $G$ as $G_1^\text{(meta)} = \{ k \mid k\neq 1, \text{hypothesis } \rij^{(1)} - \rij^{(k)} = 0 \text{ is not rejected}\}$.

For the null set of groups, still with the assumption $\rk = \rho + e^\kth$ for $k\in G_1^\text{(meta)}$, the shared underlying partial correlation is computed by $\rho = \frac{ \Sigma_{k\in G_1^\text{(meta)}} w^\kth \rk }{ \Sigma_i w^\kth }$, where weights $w\kth = \frac{1}{ var(\rk) }$ represent the inverse of the variance of the estimated partial correlation from different groups. The standard error of this shared partial correlation with respect to target group 1 is $\se(\rho^{(1)}) = \frac{1}{ \sqrt{\Sigma_{k\in G_1^\text{(meta)}} w^\kth } }$, and $\eta^\text{(meta)}$ is also adjusted, so as the corresponding entry $\Theta_{ij}^{(1)}$ in the precision matrix for target group 1.
All the pairs of random variables are evaluated through the same approach.
In addition, this whole procedure could be applied to other target groups as well.

Another alternative approach is to use one-to-one meta-analysis for $K-1$ times.
For instance, considering group 1 as the target group as well, we can apply one-to-one meta-analysis between group 1 and group $i$ with $i\in\{2,...,K\}$ for the first time.
Then, the result of partial correlation and precision after meta-analysis with group $i$ is used to apply one-to-one meta-analysis with the result from group $j$ for $j\in G\backslash \{1,i\}$, and so on so force. The strength of this procedure is that the contribution of each additive considered group can be explicitly shown.
The demonstration of this procedure in a real application will be shown in Section \ref{Webpage}, which deals with the university webpage dataset. Specifically, the group with the smallest sample size is considered as the target group and then other groups are used one by one for meta-analysis in the ascending order of sample size.

A special case of the additively applying one-to-one meta-analysis is to follow its original index $1,2,...,K$, where the data in different groups are collected in a time series and the index of the group corresponds to the time steps. One-to-one meta-analysis can be applied sequentially, starting from group 1 and group 2, and then up to the data from group $K$ with the most recent time.

In conclusion, there are various ways to apply meta-analysis in multiple groups, depending on the aims of analysis. FLAG-Meta is flexible because it is based on the most fine-grained granularity across entries and groups.

\subsection{Covariate-adjusted Model for Joint Estimation}

In real-world applications, taking the gene co-expression network from human brain data as an example, sample properties of sample like brain regions and age periods can be considered as covariates.

The conditional Gaussian graphical model (cGGM) is first presented by \cite{yin2011sparse}, which takes covariates into consideration as $z|\up \sim \mN(\zeta \up, \Theta^{-1})$, where $\up\in\mR^q$, and $\zeta\in\mR^{p\times q}$, rather than regarding means of random variables as constants, which is invariant to heterogeneity. The cGGM is estimated by a penalized likelihood-based method, where both $\zeta$ and $\Theta$ are penalized by $\ell_1$ norm based on their sparsity assumptions.

Then, a two-stage method proposed by \cite{cai2013covariate} to solve covariate-adjusted Gaussian graphical model $z=\zeta \up+\tz$ where $\tz$ is a $p\times1$ random vector with mean zero and inverse covariance $\Tinv$, using a constrained $\ell_1$ minimization similar to that of \cite{cai2011constrained}.
The first step is to estimate the regression coefficient matrix $\zeta$ by solving the optimization row by row:
$\hat\zeta = \argmin_{\zeta\in\mR^{p\times q}} |\zeta|_1, \text{s.t. } |S_{\up z} - \zeta S_{\up\up}|\leq\lone$
where $S_{\up z}=\frac1N\SnN(z_i-\bar z)(\up_i-\bar \up)^T$ and $S_{\up\up}=\frac1N\SnN(\up_i-\bar \up)(\up_i-\bar \up)^T$. In the second step, the precision matrix $\Theta$ is estimated when $\hat\zeta$ is fixed from the previous step, by 
$\hT=\argmin_{\Theta\in\mR^{p\times p}} |\Theta|_1, \text{s.t. }|I_p - S_{zz}\Theta|_\infty \leq \ltwo$
where $S_{zz}=\frac1N\SnN(z_i-\bar z)(z_i-\bar z)^T$.

Similarly, a two-step procedure designed by \cite{chen2016asymptotically}, known as asymptotically normal estimation with thresholding after adjusting covariates (ANTAC), to estimate $\zeta$ and $\beta$ separately using scaled lasso.
In the first step, they solve the following optimization problems:
$\hat\zeta_j,\hat\sigma_{jj}=
\argmin_{\zeta_j\in\mR^q, \sigma\in\mR^+} 
\frac{\| Z_j-\up\zeta_j \|_2}{2n\sigma} +
\frac{\sigma_{jj}}{2} +
\lone\Sigma_{k=1}^q\frac{\| \Up_{k} \|}{\sqrt{n}} |\zeta_{jk}|$, for $j=1,...,p$, where the parameter is theoretically specified as $\lone=\sqrt{ \frac{2(1+\frac{\log p}{\log q})}{n} }$.
Next, adjusted data $\tilde Z=Z-\Up\hat\zeta$ is used to estimate the precision matrix, according to the regression residuals after estimating coefficients $\beta$ by solving the optimization as follows,
$\hat\beta_l,\hat\sigma_{ll}=
\argmin_{\beta_l\in\mR^{p-2}, \sigma_{ll}\in\mR^+}
\frac{\| \tilde Z_j-\tilde Z_{A^c}\beta_l \|_2}{2n\sigma_{ll}} +
\frac{\sigma_{ll}}{2} +
\ltwo\Sigma_{k=1}^q\frac{\tilde Z_k \|}{\sqrt{n}} |\beta_{lk}|, l\in A=\{i,j\}$, where the parameter is theoretically specified as $\ltwo=\sqrt{ \frac{2\log p}{n} }$.

One limitation of the methods from \cite{cai2013covariate, chen2016asymptotically} is that the two-stage estimation process induce propagation of errors since the estimation of the precision matrix relies on $\hat\zeta$ from the first step.

When taking covariates into consideration, the random effect model for the Gaussian graphical model as \ref{REM} can be extended to
\begin{equation}
\begin{aligned}
Y =\Up \zeta + X \beta + \epsilon , \beta_i^T \sim \mN(0,\Gb), \epsilon_i \sim \mN(0,\Ge),
\label{REMAC}
\end{aligned}
\end{equation}
where $\Up\in\mR^{n\times q}$ is the covariate matrix and $\zeta\in\mR^{q\times2}$.
The advantage of the flexible and accurate Gaussian graphical model with covariate adjusted (FLAG-CA) is that it evaluates the fixed effect $\zeta$ and the random effect $\beta$ in a single unified model, rather than using two separate steps. 
When adjusting for the effect of covariates, the model can be easily estimated with little extra computational cost in each iteration.

\subsubsection{MM Algorithm for FLAG-CA}
For the revision of MM algorithm, the incomplete-data log-likelihood is
\begin{equation}
\begin{aligned}
\ell(\Gamma)=& \ln\mbP(Y|X;\Gb,\Ge) \\
 =& -\frac12 \ld \Omega -\frac12(\bY-\bUp \bzt)^T\Omega^{-1}(\bY-\bUp \bzt)+c,
\label{mm_ca}
\end{aligned}
\end{equation}
where $\bY=\vY$, $\bUp = I_2\otimes \Upsilon \in \mR^{2n \times 2q}, \bzt=\txvec(\zeta)$, and $c$ is a constant.
The MM algorithm updates the fixed effect $\zeta$ and the variance components $\Gamma$ alternatively, with one being updated while the other is fixed.
In each iteration, an extra update of $\zeta$ involves solving a weighted least  squares problem, as 
$\bzt^{(m+1)}=\argmin_{\bzt} \frac12 (\bY-\bUp \bzt)^T \Omega^{-(m)} (\bY-\bUp \bzt)
= (\Up^T\Omega^{-(m)}\Up)^{-1} \Up^T\Omega^{-(m)}Y$.
The revised MM algorithm for FLAG-CA is summarized in the appendix \ref{MM_CA}.

\subsubsection{PX-EM Algorithm for FLAG-CA}

The model of PX-EM algorithm for the FLAG-CA method is $Y = \Up \zeta + \delta X \beta + \epsilon$, where $\delta\in\mR^1$ is the expanded parameter.
The complete-data log-likelihood when adjusting for covariates is 
\begin{equation}
\begin{aligned}
\ell(\Gamma) = & \text{logPr} (\bY, \bb|\Gb,\Ge;\tilde X) \\
=& -\frac12 \ln |\Omega| -\frac12 \txvec (Y-\Up\zeta-\delta X\beta)^T \Omega^{-1} \txvec (Y-\Up\zeta-\delta X\beta) \\
=& -\frac n2 \ln |\Ge|
    -\frac12 (\bY -\bUp \bzt -\delta\tilde X\bb)^T
        (\Ge^{-1}\otimes I_{n})
        (\bY -\bUp \bzt -\delta\tilde X\bb) \\
 & -\frac{p-2}2 \ln |\Gb|
    -\frac12 \bb^T
        (\Gb^{-1}\otimes I_{p-2})
        \bb,
\label{EM_ca_ll}
\end{aligned}
\end{equation}
where $\bY=\vY, \bX=I_2\otimes X, \bb=\txvec(\beta)$ are the same transformations as in the previous section, and $\bUp = I_2\otimes \Upsilon \in \mR^{2n \times 2q}, \bzt=\txvec(\zeta)$. Then the posterior distribution of $\bb$ is $\mN(\bb|\mu_{\bb}, \Sigma_{\bb})$, where
$$\Sigma_{\bb}^{-1}= \delta^2 \Ge^{-1} \otimes X^TX +\Gb^{-1} \otimes I_{p-2}, $$
$$\mu_{\bb}= (\delta^2 \Ge^{-1} \otimes X^TX +\Gb^{-1} \otimes I_{p-2})^{-1}
\delta(\Ge^{-1} \otimes X^T) (\bY - \bUp \bzt). $$

In the E-step, the expectation of complete-data log-likelihood in Equation \ref{EM_ca_ll} is taken with respect to $\beta$, given the parameters from last iteration, as

\begin{equation}
\begin{aligned}
\mathcal Q(\Omega|\Omega_{old})
=& -\frac n2 \text{log}|\Ge| -\frac{p-2}2 \text{log}|\Gb| 
  -\frac12 \{ (\bY - \bUp \bzt - \delta \bX \mu_{\bb})^T
    (\Ge^{-1}\otimes I_{n})
    (\bY - \bUp \bzt - \delta \bX \mu_{\bb}) \\
 & +\delta^2 \text{tr}[ (\Ge^{-1} \otimes X^TX) \Sigma_{\bb} ] \} 
  -\frac12 \{ \mu_{\bb}^T (\Gb^{-1}\otimes I_{p-2}) \mu_{\bb}
   + \text{tr}[ (\Gb^{-1}\otimes I_{p-2}) \Sigma_{\bb} ] \} \\
=& -\frac n2 \text{log}|\Ge| -\frac{p-2}2 \text{log}|\Gb| \\
 & -\frac12 \text{tr}\Big[ (\Ge^{-1}\otimes I_{n})
    [ (\bY - \bUp \bzt - \delta \bX \mu_{\bb})(\bY - \bUp \bzt - \delta \bX \mu_{\bb})^T 
    +\delta^2 \tilde X \Sigma_{\bb} \tilde X^T] \Big] \\
 & -\frac12 \text{tr}\Big[ (\Gb^{-1}\otimes I_{p-2})
    ( \mu_{\bb} \mu_{\bb}^T + \Sigma_{\bb} ) \Big] \\
=& -\frac n2 \text{log}|\Ge| -\frac{p-2}2 \text{log}|\Gb| \\
 & -\frac12 \text{tr}\Big[ \Ge^{-1} 
    \begin{pmatrix} \text{tr}[S_{11}] & \text{tr}[S_{12}] \\ \text{tr}[S_{21}] & \text{tr}[S_{22}] \end{pmatrix} \Big] 
   -\frac12 \text{tr}\Big[ \Gb^{-1} 
    \begin{pmatrix} \text{tr}[W_{11}] & \text{tr}[W_{12}] \\ \text{tr}[W_{21}] & \text{tr}[W_{22}] \end{pmatrix} \Big],
\end{aligned}
\end{equation}
where $S=(\bY - \bUp \bzt - \delta \bX \mu_{\bb})(\bY - \bUp \bzt - \delta \bX \mu_{\bb})^T
     +\delta^2 \tilde X \Sigma_{\bb} \tilde X^T = \begin{pmatrix} S_{11} & S_{12} \\ S_{21} & S_{22} \end{pmatrix}$, \\
$W=\mu_{\bb} \mu_{\bb}^T + \Sigma_{\bb}
 = \begin{pmatrix} W_{11} & W_{12} \\ W_{21} & W_{22} \end{pmatrix}$.
 

In the M-step, parameters $\delta, \Gb, \Ge$ are updated similarly, with the only difference being that when adjusting the covariates, $\bY$ is used as a mean effect offset version $(\bY-\bUp\bzt)$ and an extra update for $\zeta$ is added. The revised MM algorithm for FLAG-CA is summarized in the appendix \ref{EM_CA}.


\section{Numerical Examples}

In this section, the proposed methods are evaluated using various simulation settings, and the real data applications.

\subsection{Simulation Studies}

The critical advantage of FLAG is its ability to perform statistical inference on each entry in the precision matrix, which quantifies the uncertainty associated with each edge. To verify the effectiveness of false discovery rate (FDR) control for graph recovery, a simple simulation setting is designed with $p=50, n=300$, and the nonzero entries whose value is 0.15 are randomly generated with the nonzero proportion $\pi$ varies $\{0.1, 0.15, 0.2, 0.3, 0.4, 0.5, 0.6, 0.7\}$.
The results from FLAG are compared with two methods, ANT and GGM estimation with false discovery rate control (GFC, \cite{liu2013gaussian}) which support the statistical inference and FDR control.

\begin{figure}
    \centering
    \includegraphics[width = \textwidth]{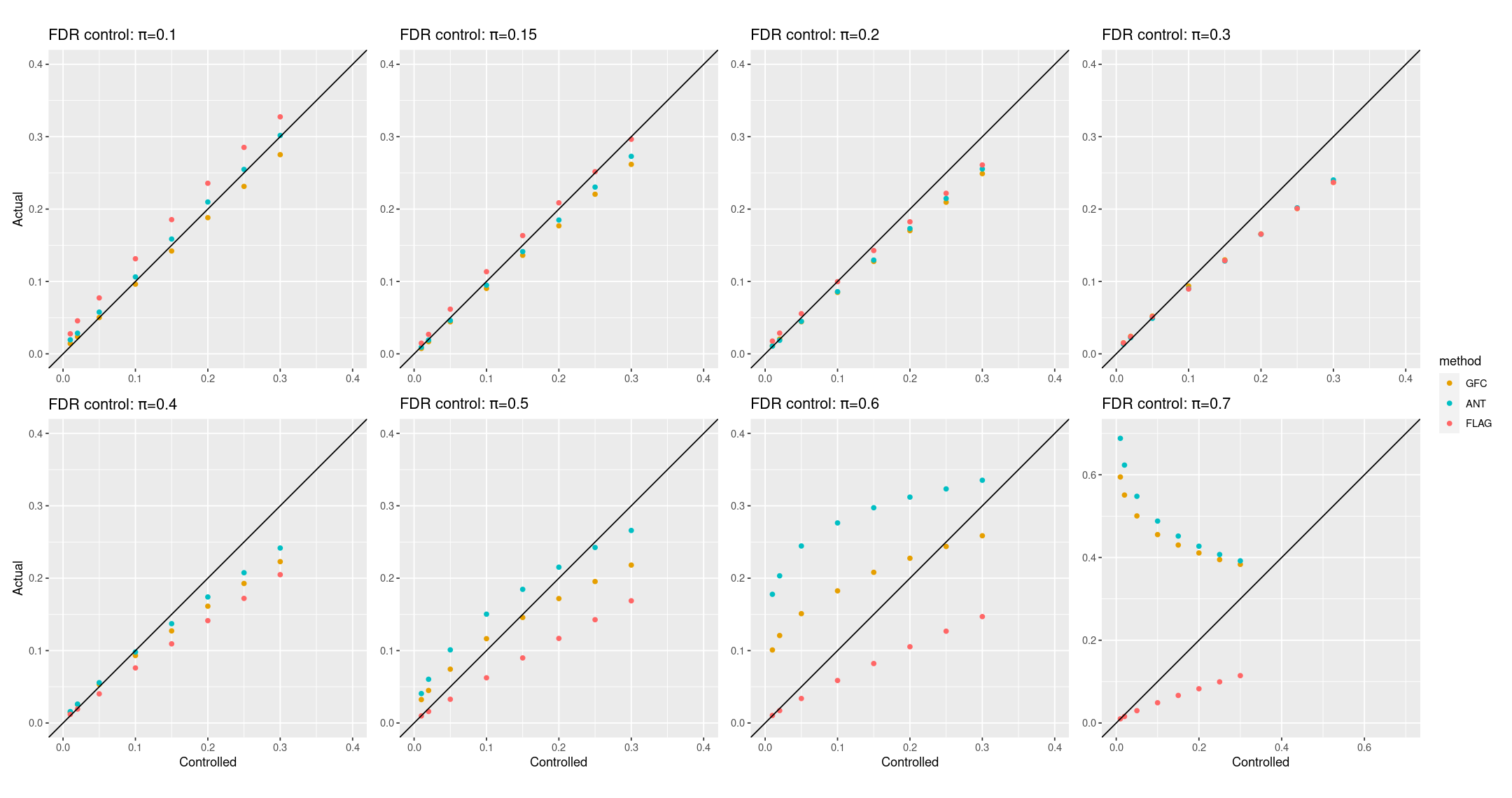}
    \caption{Scatter plots of the estimated partial correlation using different methods, with each data point representing the result from the scaled data in Y versus the result from the centered data in X.}
    \label{fig_simu_1_fdr}
\end{figure}

As shown in the Figure \ref{fig_simu_1_fdr}, the FDR is controlled effectively by FLAG, while the FDR of ANT and GFC are out of control when the nonzero proportion exceeds 0.5.

\subsubsection{Block Magnified Matrix}

To investigate the sensitivity of the methods is to data scaling, unscaled data and scaled data (each column with variance 1) are used with different methods for comparison.
Since the estimated precision $\hat\theta$ from the same method may differ depending on whether the data is scaled or not, the estimated partial correlation $\rij=-\frac{\tij}{\sqrt{\theta_{ii}\theta_{jj}}}$ is used for comparison.

The ground truth is a block magnified matrix as $\Theta = \begin{pmatrix} \alpha_1\Theta_0 &0 &0\\0 & \alpha_2\Theta_0 &0\\0 &0 &\alpha_3\Theta_0 \end{pmatrix}$, where $(\alpha_1,\alpha_2,\alpha_3)=(1,5,25)$. The simulated submatrix $\Theta_0$ has all diagonal elements equal to one, and its off-diagonal elements are non-zero with probability $\pi=0.05$. The non-zero 0ff-diagonal elements are sampled from $\{0.2, 0.4\}$. According to this simulation setting, all the non-zero partial correlations are at the same scale, ranging from $\{0.2, 0.4\}$, which makes it easier for comparison.

\begin{figure}
    \centering
    \includegraphics[width = 0.7\textwidth]{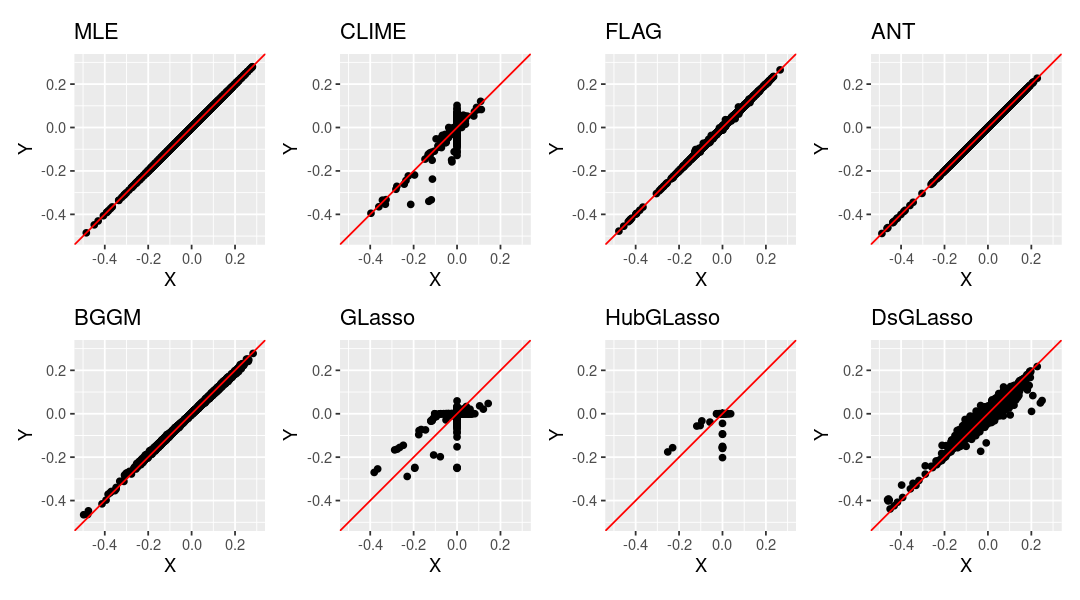}
    \caption{Scatter plots of the estimated partial correlation using different methods, with each data point representing the result from the scaled data in Y versus the result from the centered data in X.}
    \label{fig_simu_1_scatter}
\end{figure}

Figure \ref{fig_simu_1_scatter} shows the results from centered data in X and that from scaled data in Y, with points along the diagonal lines expected. The estimated partial correlation of FLAG is not sensitive to data scaling, compared to CLIME, GLasso, Hub GLasso, and De-sparsified GLasso.
Specifically, the penalty parameter $\lambda$ of the GLasso method, which was tuned by 10-fold cross validation, is 0.063 for the centered data and 0.158 for the scaled data. This indicates different levels of sparsity in matrices when the input data is scaled compared to when it is not. Referring to the subfigure of GLasso, the data points located on the x-axis or y-axis represent the entries that are zero in one setting and nonzero in the other.

\begin{table}[htbp]
  \caption{Relative Frobenius norm error of the estimated partial correlation matrix using different methods, with 100 replications.}
  \label{table_sim_1}
  \centering
  \begin{tabular}{lcc}
    \toprule
    Methods	    & Data centered & Data centered and scaled  \\
    \midrule
    MLE         & 0.859 & 0.859 \\
    CLIME       & 0.162 & 0.166 \\
    FLAG   & 0.178 & 0.178 \\
    ANT     & 0.157 & 0.157 \\
    BGGM        & 0.803 & 0.772 \\
    GLasso      & 0.246 & 0.193 \\
    HubGLasso   & 0.251 & 0.227 \\
    DsGLasso    & 0.628 & 0.574 \\
    \bottomrule
  \end{tabular}
\end{table}

Methods with regularization of the precision matrix are particularly fragile when the entries in the precision matrix are of different scales.
Specifically, when given unscaled data, such kind of methods have false positives in the region with relatively smaller magnitudes of entries, and false negatives in the region with relatively larger magnitudes of entries.
Both the estimation error and recovery performance of our method are not sensitive to data scaling and they are comparable to the outcomes of the well-performing methods in this block-magnified matrix setting.

\paragraph{Hub Structure}

The ground truth for the precision matrix is an adjacency matrix of a weighted graph with a hub structure, where the hub indicates the node that connects with many other nodes with a large degree that exceeds the average \cite{barabasi2013network}.
The hub structure exists widely in real-world applications, such as the structural and functional connectivity hubs in the human brain \cite{van2013network}, a fragile financial instruments that can have a major impact on the financial market by influencing the prices of many related securities, and the source nodes of anomalous activity in the cyber security field \cite{hero2012hub}.

The hub nodes in the ground truth are indexed by $1,...,h$, where the number of hub nodes is smaller than the dimension, i.e., $h<p$.
The precision matrix can be split into blocks as
$\Theta=\begin{pmatrix} \Theta_{aa} &\Theta_{ab} \\ \Theta_{ba} & \Theta_{bb} \end{pmatrix}$, where $a=\{1,...,h\}$ and $b=\{h+1,...,p\}$.
Specifically, $\Theta_{aa}$ encodes the conditional dependence between hub nodes, $\Theta_{ab}$ and $\Theta_{ba}$ correspond to the edges between hub and non-hub nodes, and the dependencies between the non-hub nodes are in block $\Theta_{bb}$.

Based on the conditional Gaussian property as $\Theta_{ba} = -\beta\Theta_{aa}$, where $\Theta_{ba}\in\mR^{(p-h)\times h}, \beta\in\mR^{(p-h)\times h}$, and $\Theta_{aa}\in\mR^{h\times h}$. Once $\Theta_{aa}$ and $\beta$ are generated, the true $\Theta_{ba}$ can be obtained through multiplication.
According to the definition of a hub in a graph, each hub node has many connections with other nodes, and thus $\Theta_{ba}$ is required to have a large proportion of non-zero entries.
To investigate whether the sparsity of the true $\beta$ influences the precision estimation, $h=10$ hubs are separated into five pairs, and the columns in $\beta$ that correspond to the hub nodes with odd indices are fully populated with non-zero elements, while the proportion of non-zero entries in the columns with even indices is varied across $\{0.9, 0.7, 0.5, 0.3, 0.1\}$.
The remaining block matrix $\Theta_{bb}$, which denotes the relationships between non-hub nodes, is a relatively sparse matrix with a non-zero proportion of $\pi=0.3$.
Specifically, the diagonal elements of $\Theta_{bb}$ are set to 50, and the non-zero elements are uniformly generated from $U[3,5]$.
In this simulation, the dimension is set to $p=50$ and the sample size is $n=200$.

\begin{figure}
     \centering
     \begin{subfigure}[b]{0.9\textwidth}
         \centering
         \includegraphics[width=\textwidth]{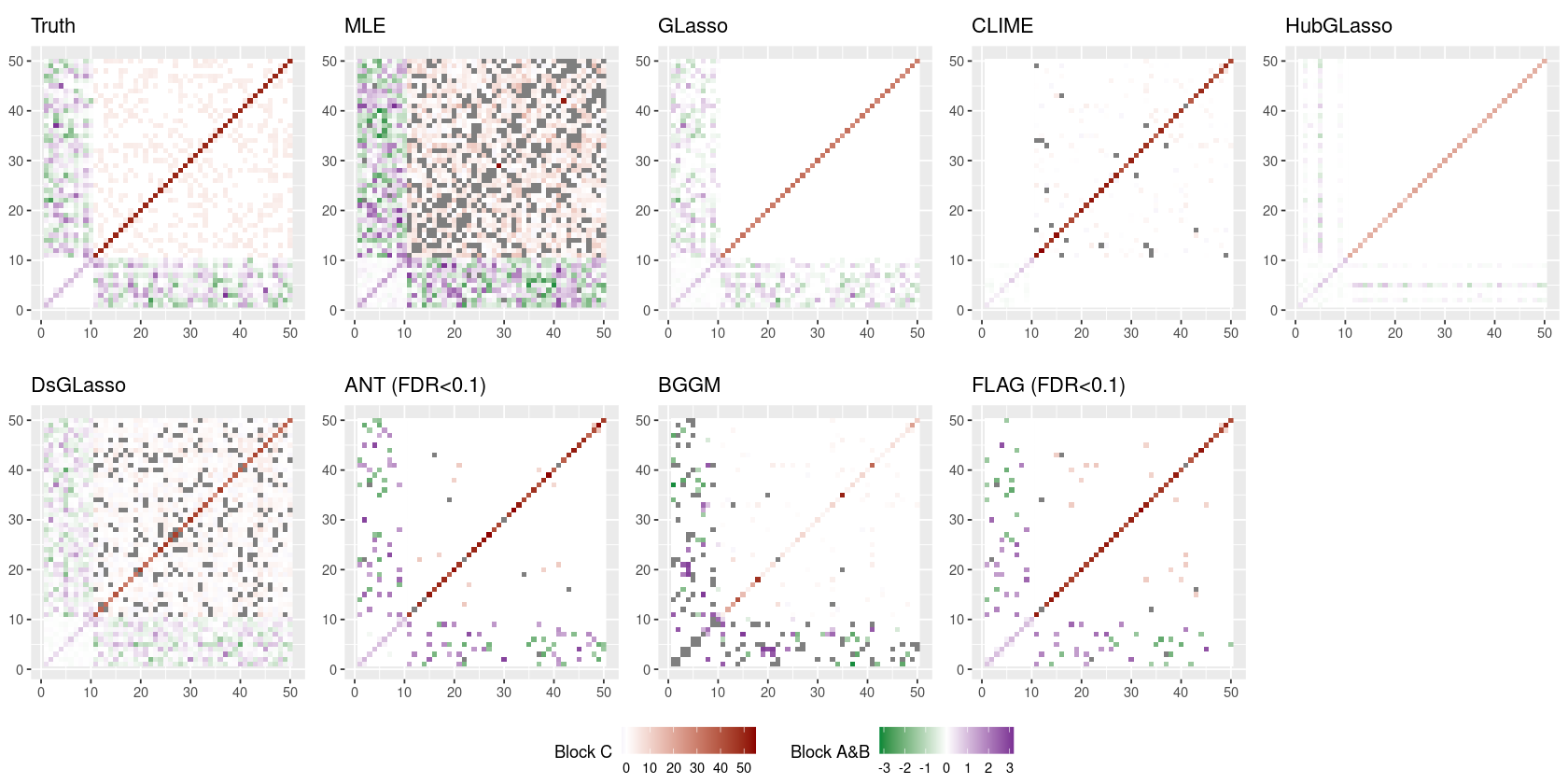}
         \caption{Heatmaps of the estimated precision matrices using different methods.}
         \label{simu_2_prec}
     \end{subfigure}
     \hfill
     \begin{subfigure}[b]{0.92\textwidth}
        \centering
        \includegraphics[width=\textwidth]{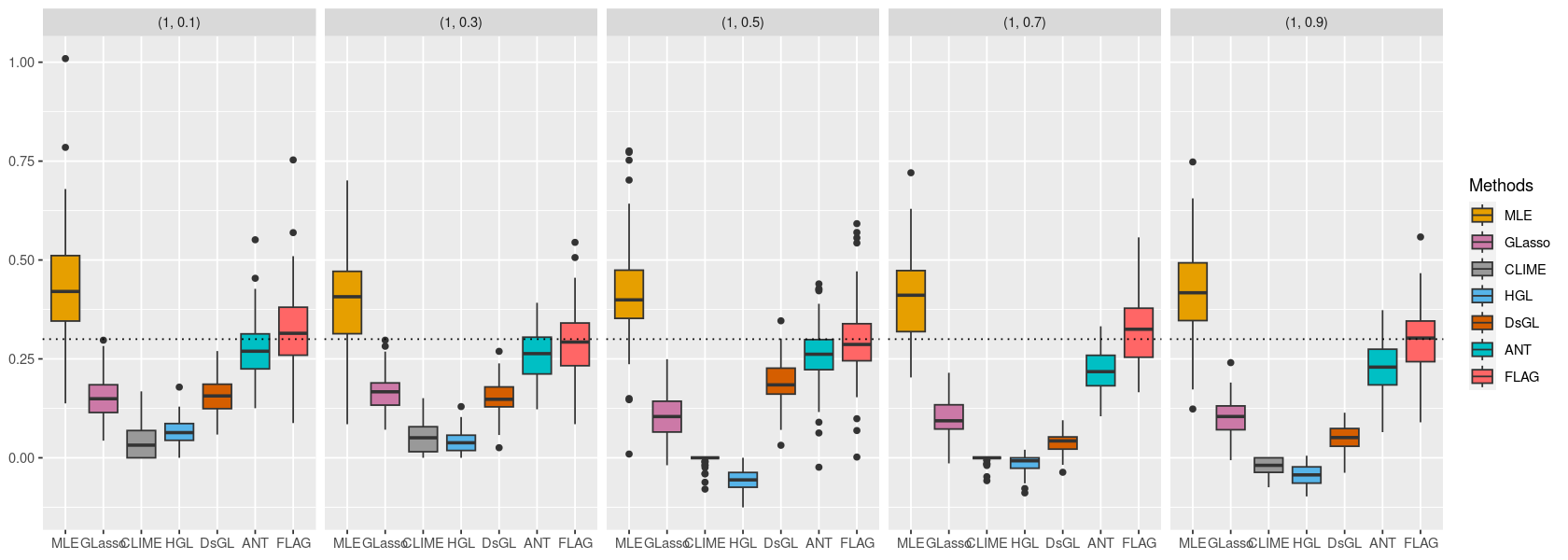}
        \caption{Estimated precision between pair of hubs, when nonzero proportion of underlying regression coefficient varies.}
        \label{simu_2_box}
     \end{subfigure}
     \hfill
    \begin{subfigure}[b]{0.95\textwidth}
        \centering
        \includegraphics[width=\textwidth]{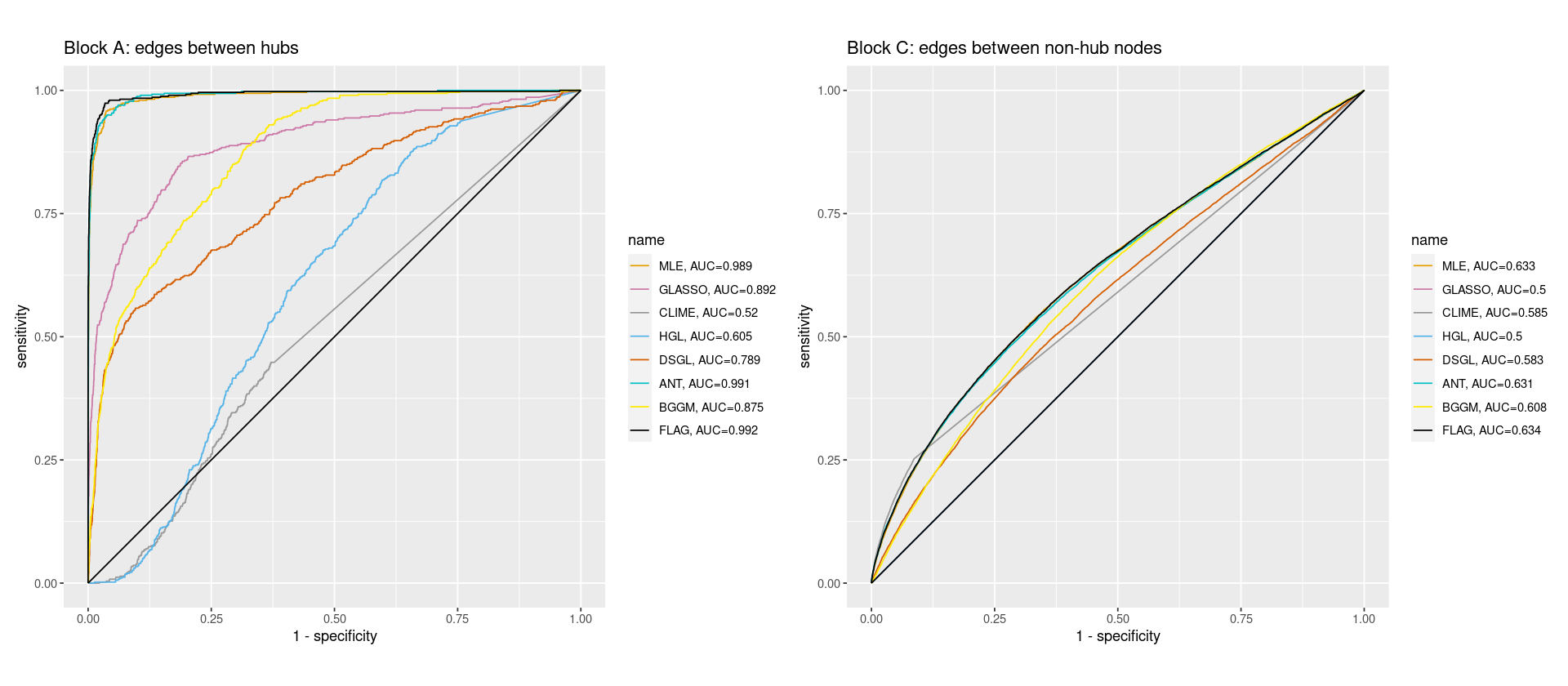}
        \caption{The ROC curves with AUC values for the recovery of edges in the block A and block C.}
        \label{simu2_roc}
     \end{subfigure}
     \caption{The results of precision estimation and graph recovery using different methods.}
\end{figure}

Figure \ref{simu_2_prec} shows the true precision matrix and the estimated precision matrices using different methods. Edges involving the hub nodes that correspond to entries in block matrices A and B are colored in purple for positive values and green for the negative values. Edges between non-hub nodes that correspond to entries in block matrix C are colored in brown. Some entries in the estimated matrices are gray, indicating that the estimated value is far away from the range of the true values.

In the block A, which encodes the conditional dependencies between hub nodes, several methods, including MLE, CLIME, Hub GLasso, ANT, and BGGM, produce false positives. The non-zero entries in block matrix $\Theta_{aa}$ are underestimated by GLasso, CLIME, Hub GLasso, and Desparsified GLasso methods, and overestimated by BGGM method.
In the block B, which captures the edges between hubs and non-hub nodes, results from methods CLIME and Hub GLasso miss the majority of non-zero elements.
In the block C, whose non-zero entries indicate the conditional dependencies between non-hub nodes, several methods, including MLE, GLasso, Hub GLasso, and BGGM, produce inaccurate estimates of the diagonal elements. A large proportion of estimates in the block matrix $\Theta_{bb}$ from MLE and Desparsified GLasso falling far away from the true range.
By contrast, FLAG performs well in both precision matrix estimation and graph recovery, producing estimates that fall within a similar range to the ground truth in all blocks and fewer false positives. More detailed comparisons are provided in the following two parts, based on repeated experiments.

\paragraph{Precision Matrix Estimation}

Figure \ref{simu_2_box} shows the comparisons of the estimated precision between hub nodes as the sparsity of the coefficient $\beta$ varies. 
MLE overestimates the precision values, while the penalized likelihood-based methods, including GLasso, CLIME, Hub GLasso, and Desparsified GLasso, underestimate them.
The underestimation obtained by the ANT method is more obvious as the non-zero proportion increases. To further explain this observation, a detailed comparison between the ANT and FLAG methods is conducted.

\begin{figure}
    \includegraphics[width = 0.8\textwidth]{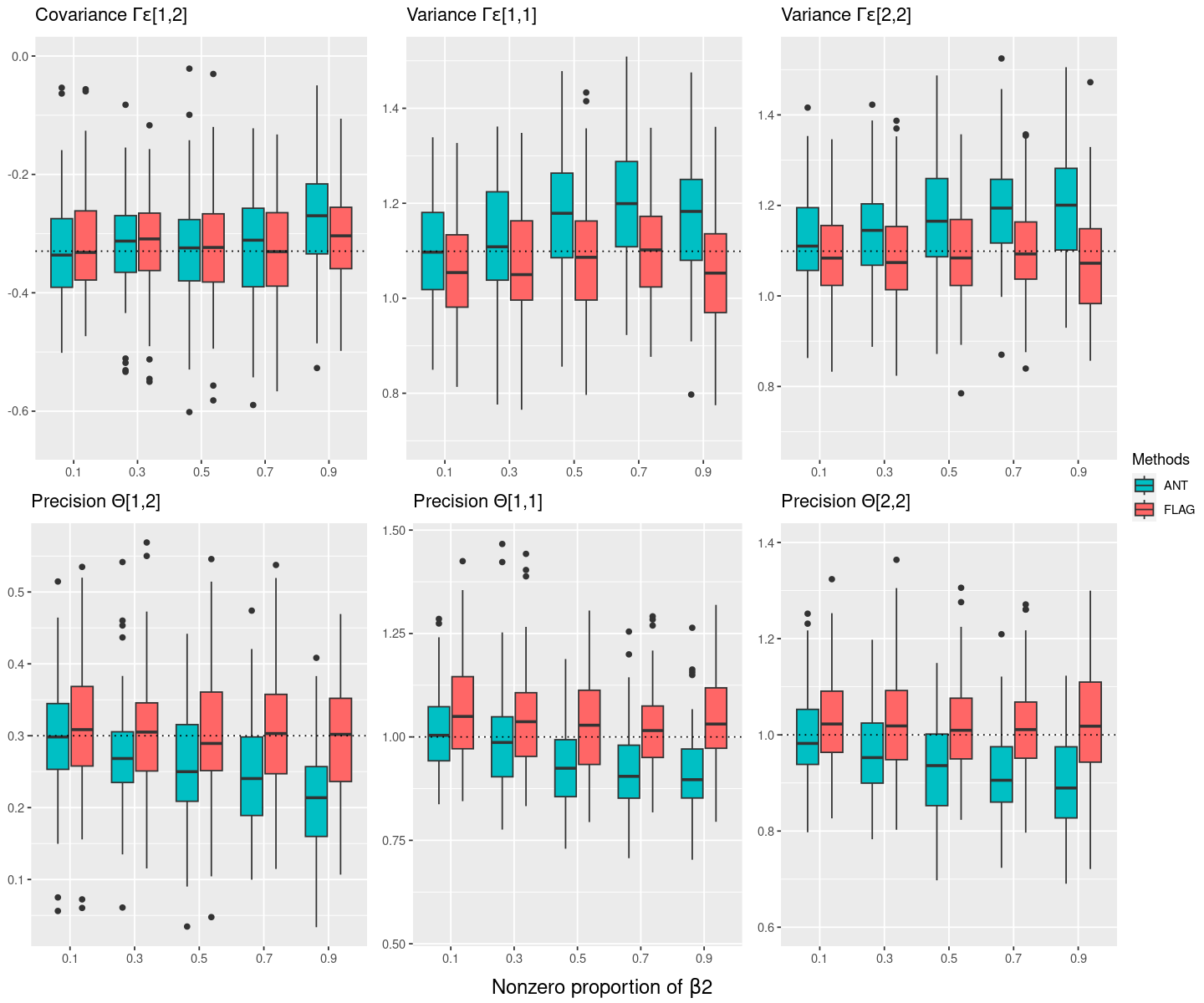}
    \centering
    \caption{A detailed comparison of the estimated covariance of residuals in bivariate regression and the precision of two random variables using the FLAG and ANT methods.}
    \label{fig_simu_2_beta}
\end{figure}

Figure \ref{fig_simu_2_beta} shows a detailed explanation of entries in the precision matrix, with varying sparsity of intrinsic $\beta$, and a comparison between FLAG and ANT.
Based on the sparsity assumption assigned to $\beta$ by ANT, $\beta^{(ANT)}$ has many zero entries, which induces an underestimation of $\var(\Xb)$ and an overestimation of $\var(\e)$.
As a result, the estimated precision by ANT is underestimated, while FLAG can still estimate the precision accurately in this case.

\begin{table}[htbp]
  \caption{Relative Frobenius norm error of estimated precision matrix using different methods, with 100 replications.}
  \label{table_sim_2}
  \centering
  \begin{tabular}{lllll}
    \toprule
    Methods	    & Precision Matrix & Block A & Block B & Block C \\
    \midrule
    MLE         & 0.772 (0.005) & 0.492 (0.007) & 1.129 (0.009) & 0.771 (0.005) \\
    GLasso      & 0.504 (0.007) & 0.320 (0.006) & 0.610 (0.007) & 0.504 (0.007) \\
    CLIME       & 0.286 (0.001) & 0.606 (0.001) & 1 (0) & 0.279 (0.001) \\
    HubGLasso   & 0.664 (3e-4) & 0.549 (0.001) & 0.927 (0.001) & 0.663 (3e-4) \\
    DsGLasso    & 0.412 (0.001) & 0.452 (0.002) & 0.662 (0.002) & 0.411 (0.001) \\
    ANT     & 0.334 (0.001) & 0.181 (0.004) & 0.758 (0.003) & 0.331 (0.001) \\
    BGGM        & 1.067 (0.005) & 59.22 (0.687) & 5.300 (0.074) & 0.819 (0.007) \\
    \textbf{FLAG}   & \textbf{0.329} (0.001) & \textbf{0.160} (0.004) & 0.847 (0.003) & 0.325 (0.001) \\

    \bottomrule
  \end{tabular}
\end{table}
Table \ref{table_sim_2} shows that FLAG has accurate estimation in the whole precision matrix, with a particularly pronuounced advantage in submatrix A, which denotes the conditional dependence among the hub nodes.

\paragraph{Graph Recovery}
As shown in Figure \ref{simu2_roc}, FLAG achieves the best graph recovery in block A and C, with an Area Under the ROC Curve (AUC) of 0.992 in the block determining edges between hub nodes and an AUC of 0.634 in the block of edges between hub nodes and non-hub nodes. It should be noted that all the entries in the block B are non-zero in the ground truth, and therefore no false positive exist.

\begin{figure}
    \includegraphics[width = \textwidth]{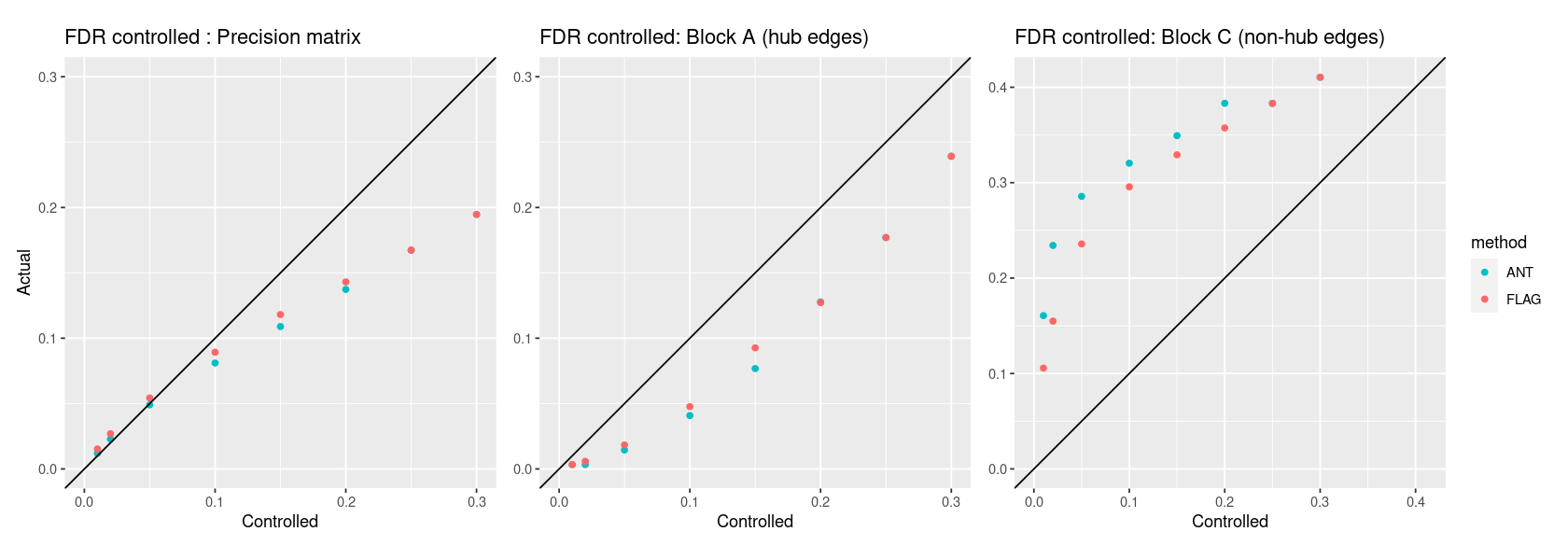}
    \centering
    \caption{The false discovery rate controlled by the ANT and FLAG methods in the entire precision matrix, block matrix A, and block matrix C from left to right.}
    \label{fig_simu_2_fdr}
\end{figure}

The False Discovery Rate (FDR) is well-controlled in the entire precision matrix as shown in the leftmost subplot of Figure \ref{fig_simu_2_fdr}. The actual FDR is relatively conservative in the whole precision matrix due to the dense connections between hubs and non-hub nodes, where the false discovery rate in the block B is zero in this setting. This observation is consistent with the findings of smaller actual FDR than controlled in graphs with hub structures, as reported in \cite{liu2013gaussian}.

In conclusion, FLAG is the only method that performs well in both precision matrix estimation and graph recovery across all blocks, particularly in the edges between hubs, where it outperforms the other methods, without any explicit assumption on the graph structure.
In graphs with a hub structures, hub nodes are crucial components due to their numerous connections and greater influence on other nodes and the graph as a whole. Consequently, the edges of hub nodes are more informative, and FLAG exhibits better performance than other methods in this setting.

\subsubsection{Multiple Graphs}

For each graph, a cluster structure is constructed, and the corresponding precision matrix is a block diagonal matrix.
The dimension for each group is $p=20$, and the sample sizes for the two groups are $n_1=100$, and $n_2=200$. Within each cluster, all nodes are connected to the node with the smallest index to ensure the connectivity, and then the probability of the existence of edges between nodes other than that one is $\pi=0.3$.
The diagonal elements in the precision matrix are set as one, and other non-zero entries are set as 0.2 for easier comparison.


First, the entries of the partial correlation matrix in each group are estimated individually, followed by testing for whether they are equal to zero or not, with the p-values of these tests collected. 


Then, these null cases are tested for whether the partial correlation of the same entry from two groups is equal.
For entries that cannot reject this hypothesis testing, meta-analysis is applied, and the p-values of testing whether the entry after meta-analysis is zero are obtained.
Similarly, entries that are non-zero from both groups in the ground truth are collected and tested following the same routine.
The partial correlation of individual estimation and inference shows a large power as the points deviate away from the diagonal line, with the power in group 2 being larger as it has more samples. When it comes to the result after meta-analysis, the power exceeds the performance of a single group, with the enhancement of power being more obvious for group 1.

\begin{figure}
    \begin{subfigure}[b]{0.45\textwidth}
        \centering
        \includegraphics[width=\textwidth]{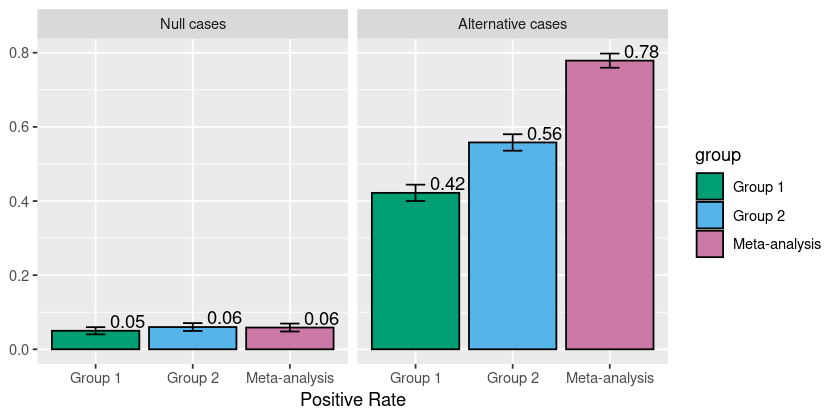}
        \caption{Bar plots of positive rates for the null cases and the alternative cases, comparing the results from FLAG given the data of individual groups or using FLAG-Meta.}
        \label{simu3_bar}
    \end{subfigure}
    \hfill
    \begin{subfigure}[b]{0.52\textwidth}
        \centering
        \includegraphics[width=\textwidth]{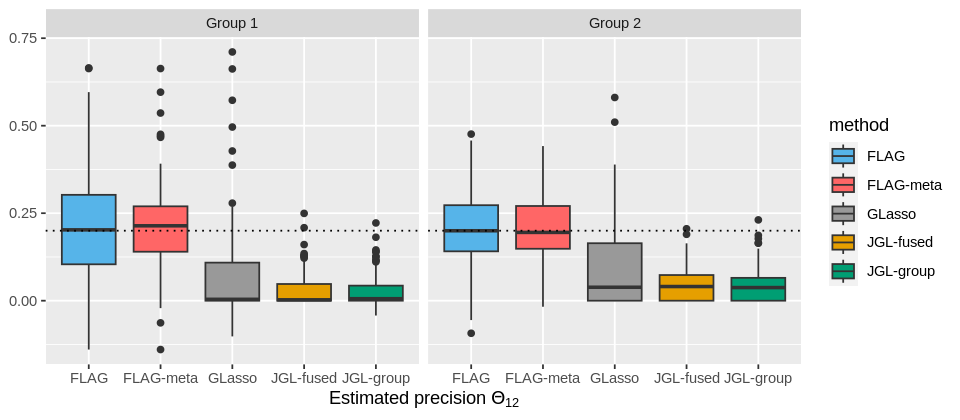}
        \caption{Box plots of the estimates of one nonzero entry in the precision matrix, for Group 1 and Group 2. The estimates are obtained using FLAG-based and GLasso-based methods for comparison.}
    \end{subfigure}
    \hfill
    \begin{subfigure}[b]{\textwidth}
        \centering
        \includegraphics[width=\textwidth]{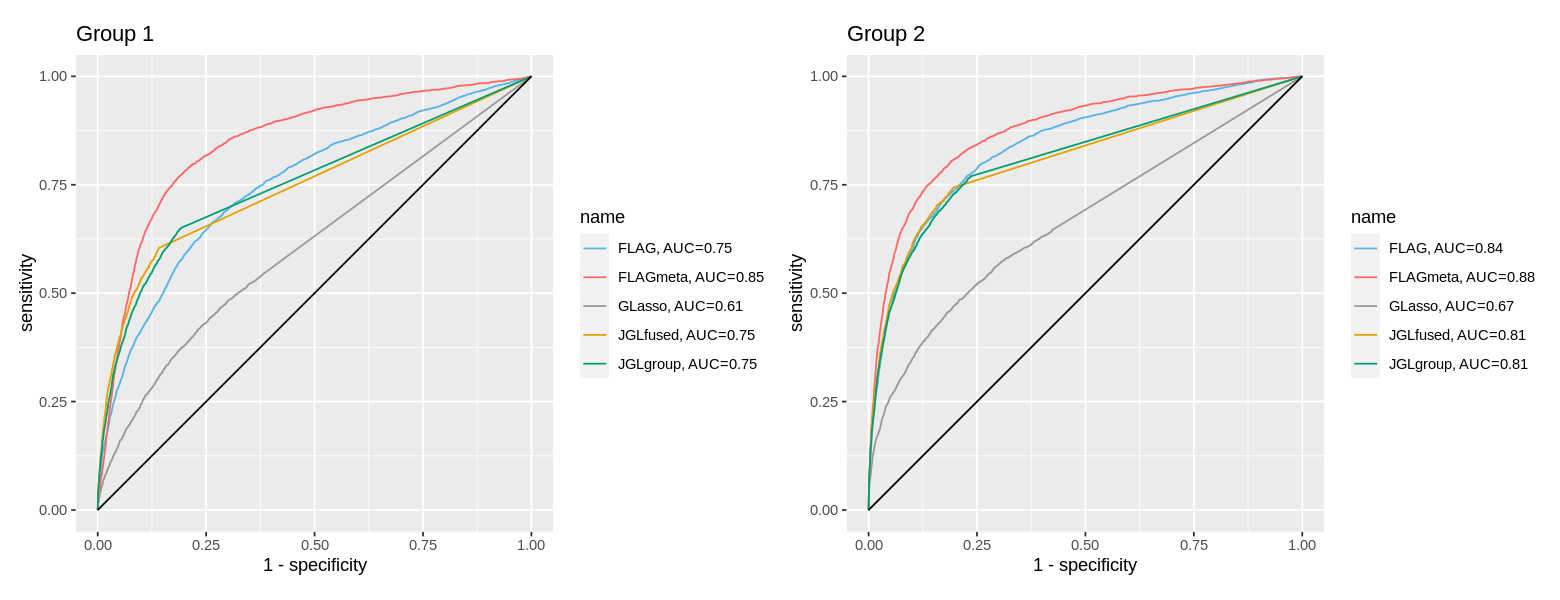}
        \caption{The ROC curves with AUC values of the graph recovery results using FLAG, FLAG-Meta, GLasso, and Joint GLasso (JGL-fused, JGL-group) methods.}
    \end{subfigure}
    \caption{The comparison of statistical inference of precision matrix and graph recovery between FLAG-based and GLasso-based methods. The sample size is 80 for Group 1 and 120 for Group 2.}
\end{figure}

FLAG-Meta has larger power and better graph recovery, smaller standard error for each entry and smaller estimation error of whole precision matrix. The improvement is more obvious in the group 1 with smaller sample size.

\subsection{Real Data Analysis}

\subsubsection{Human Brain Gene Expression Data}

We apply the FLAG method to the spatio-temporal gene expression data of the human brain \cite{kang2011spatio}. Seven high-confidence Autism spectrum disorder (ASD) genes (GRIN2B, DYRK1A, ANK2, TBR1, POGZ, CUL3, SCN2A)\cite{willsey2013coexpression} are selected to analyze the co-expression network among ASD-related genes. The data from period 1 and 2, which correspond to the early stages of brain development, as well as the groups with sample sizes smaller than three are all excluded \cite{lin2017joint}.

Data are integrated as several groups in seven time periods and four brain regions.
Our aim is to discover how the conditional dependence among ASD-related genes changes over time or across region.
The time periods are as follows, 1) Early fetal: [10PCW, 19PCW); 2) Late fetal: [19PCW, 38PCW); 3) Infancy: [0M, 12M); 4) Childhood and adolescence: [1Y, 20Y]; 5) Young adulthood: [20Y, 40Y); 6) Middle adulthood: [40Y, 60Y); 7) Late adulthood: age$\geq$60Y;
The brain regions are 1) Parietal lobe, Occipital lobe, Temporal lobe; 2) Frontal lobe; 3) Striatum, Hippocampus, Amygdala; 4) Thalamus, Cerebellum.

\begin{figure}
    \includegraphics[width = 0.92\textwidth]{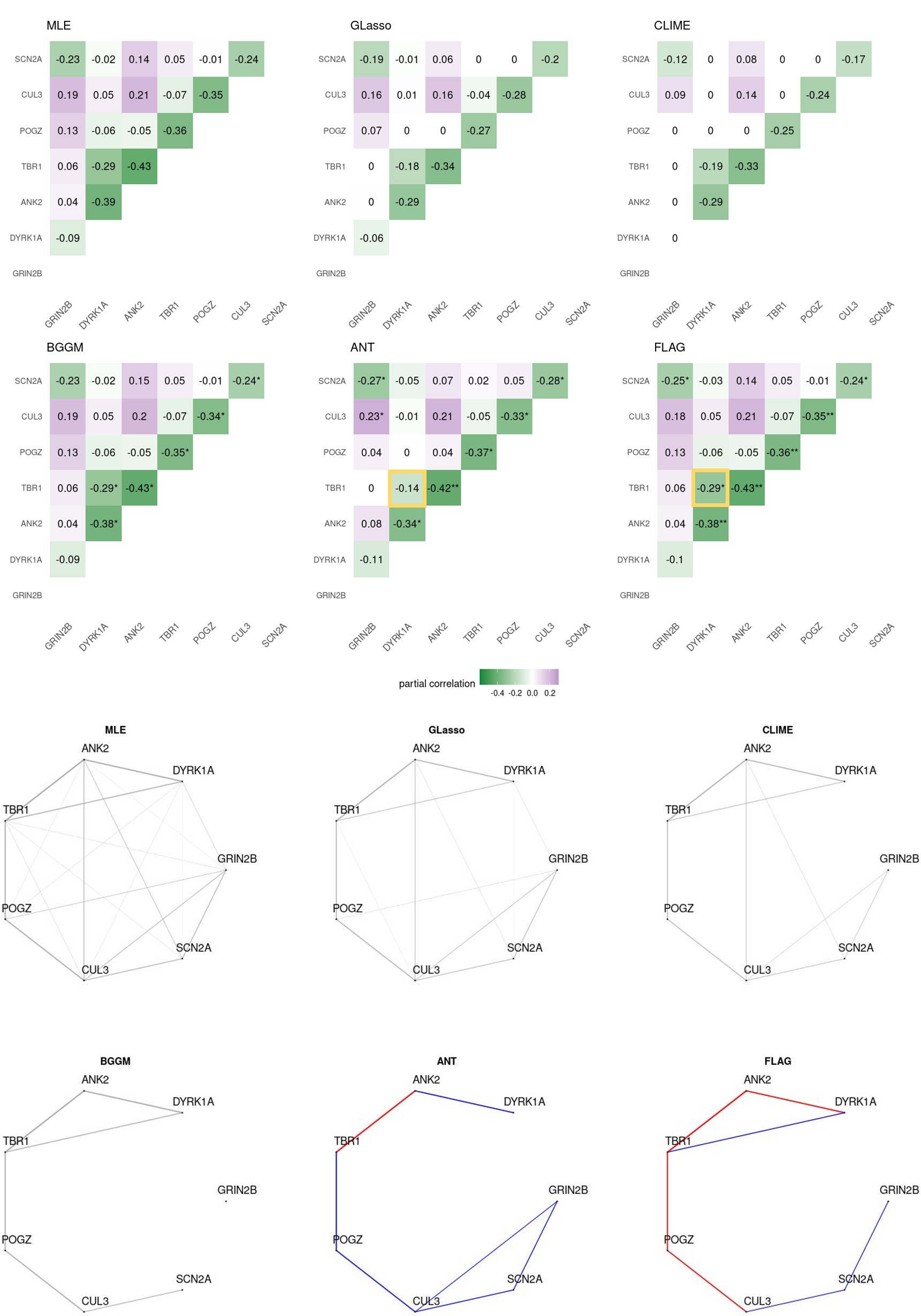}
    \caption{Partial correlation matrices estimated and corresponding graphs recovered by multiple methods using the data from brain region 2, period 13.}
    \label{brain_gene_methods}
    \centering
\end{figure}

To compare the results of different methods in this dataset, we use the group in period 13 and region 2, which has a relatively large sample size of 85, as shown in Figure \ref{brain_gene_methods}.
As the dimension equals seven and the sample size equals 85, the maximum likelihood estimator, i.e., inverse sample covariance as estimated precision matrix, is a good reference for estimation.
The estimation from the CLIME method shows less magnitude than the reference. The magnitude of estimated precision and partial correlation of the gene pair (DYRK1A, TBR1) from the ANT method is about half of the reference's, while the estimation through FLAG method equals the reference's.
The reason for such underestimation from the ANT method is similar to what we observe in the simulation, where the large zero proportion (80\%) in $\beta^{(\text{ANT})}$ induces a smaller $\var(X\beta)$, and a larger $\var(\epsilon)$ (0.386 by ANT and 0.316 by FLAG), resulting in a smaller estimated precision (0.41 by ANT and 1.01 by FLAG) and a smaller magnitude of partial correlation (-0.14 by ANT and -0.29 by FLAG).
In addition, due to the underestimation of precision, the inferred graph from the ANT method omits the edge between DYRK1A and TBR1.
The red lines in the graphs from the FLAG and ANT methods indicate edges of great significance, with a p-value of test smaller than 0.05 after Bonferroni correction, and blue lines indicate the significant edges after controlling the False Discovery Rate (FDR) to be smaller than 0.1.

\begin{figure}
    \centering
    \includegraphics[width = \textwidth]{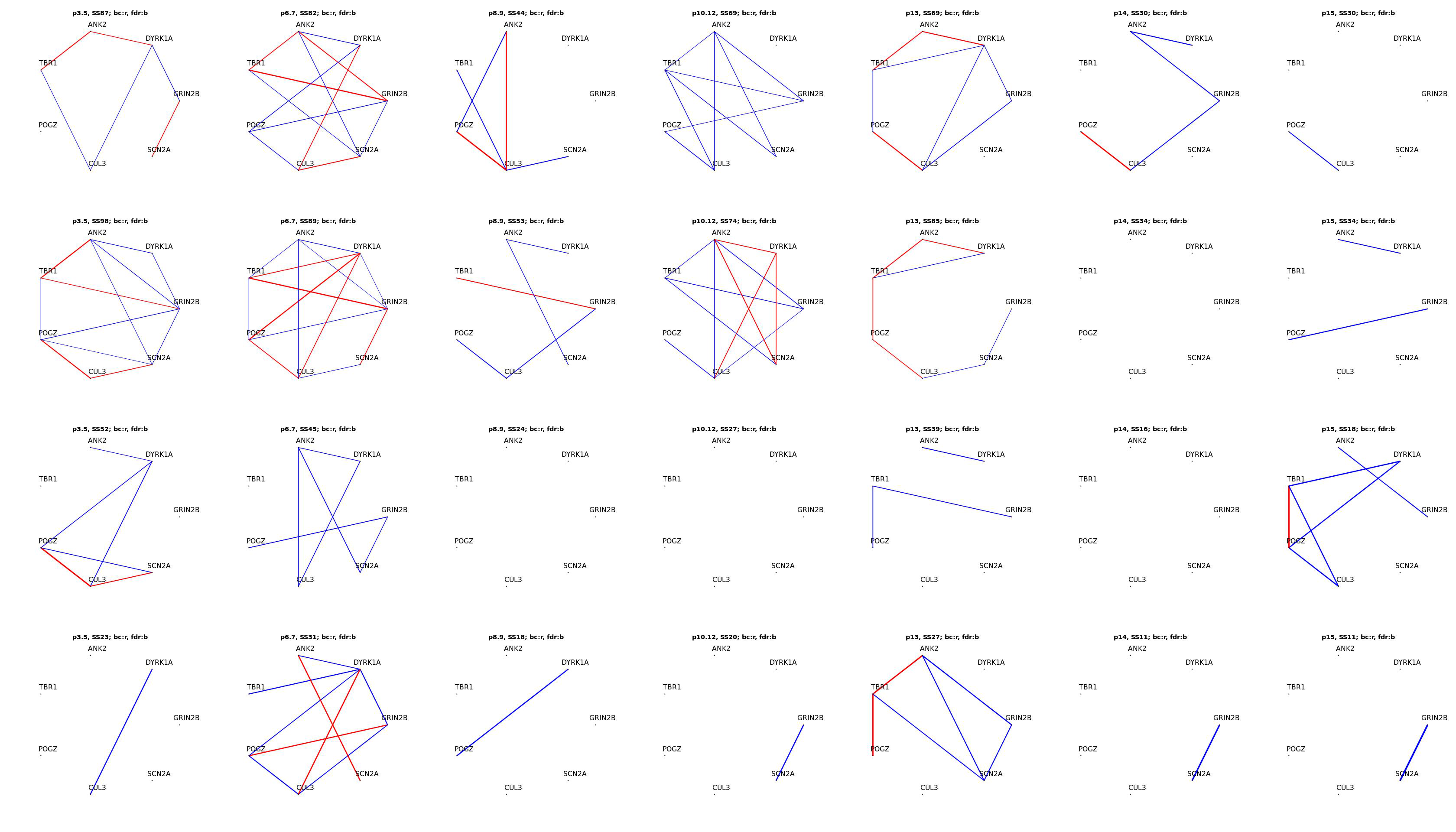}
    \caption{Inferred graphs by FLAG, arranged by region 1,2,3,4 in different rows and time periods in different columns.}
    \label{brain_graph_om}
\end{figure}

Figure \ref{brain_graph_om} shows the temporal varying pattern of conditional dependence between ASD-related genes is shown in each row, and spatial variations are in each column.
The edges inferred by Bonferroni correction are denoted in red, and FDR $\leq$ 0.1 in blue.
The thickness of edges is weighted by its magnitude of partial correlation.
As an example of spatial variation in time period 6-7, the gene pair (DYRK1A, CUL3) has a precision of -1.16, -1.12, -0.96, -2.33, and partial correlation of this pair is 0.37, 0.36, 0.39, 0.57 and in regions 1, 2, 3, 4, respectively.
In period 6-7, the conditional dependence between this pair exists in all regions, and their partial correlation shows consistency in the first three regions, while it is higher in region 4.
Moreover, there are many edges involving the gene DYRK1A, which is evident in the graphs of region 2, where the edge of the pair (DYRK1A, ANK2) exists in almost all the periods except the period 14. This finding is supported by the evidence that DYRK1A plays an important role in the signaling pathway regulating cell proliferation as a protein kinase and may be involved in brain development \cite{di2015chromatin}.

\begin{figure}
    \centering
    \includegraphics[width = 0.7\textwidth]{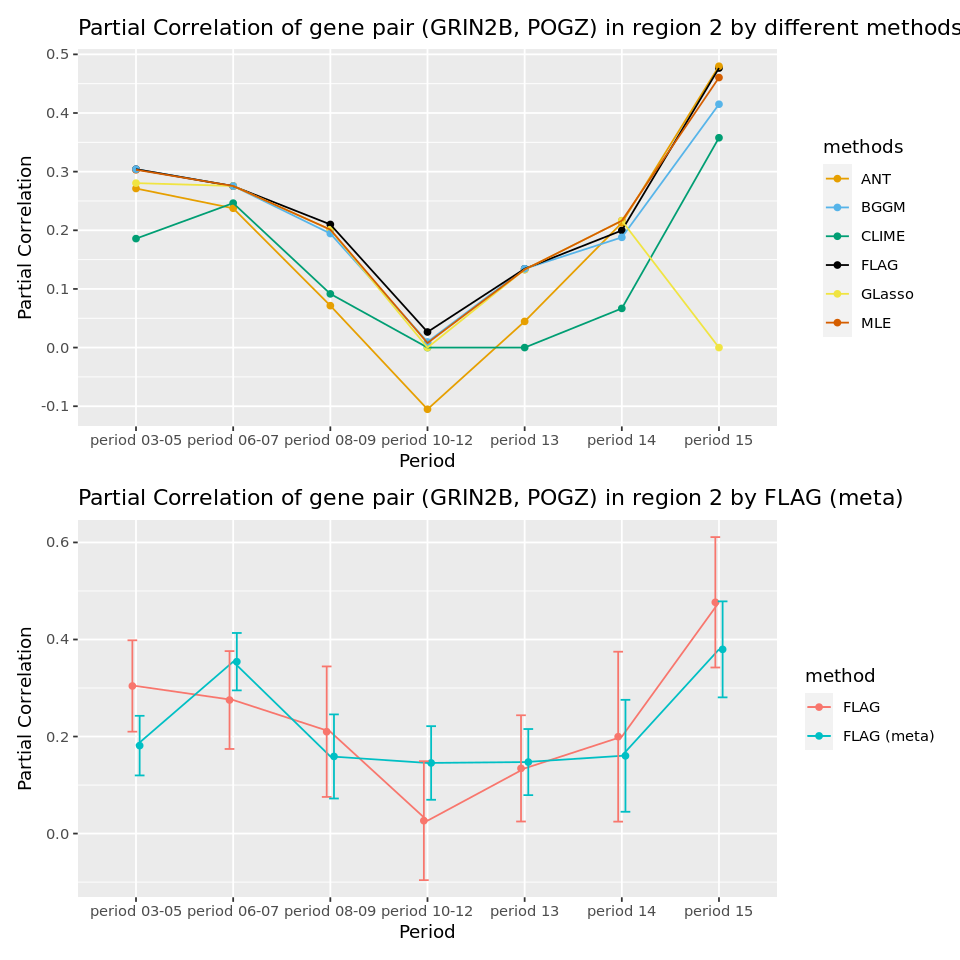}
    \caption{Partial correlation estimated by different methods between gene pair (GRIN2B, POGZ) using the expression data in brain region 2.}
    \label{brain_rho_line}
\end{figure}

As shown in figure \ref{brain_rho_line}, the estimated partial correlation by different methods are compared in the upper subfigure.
Using the maximum likelihood estimation with a relatively large sample size as a reference, the estimation by FLAG is quite similar to the reference's, thus accurate.
However, GLasso method shrinks some precision entries to zero, the estimation from CLIME method has a smaller magnitude, and in period 10-12, ANT method has non-zero estimation against all the other methods.
These cases, such as GLasso method having some false negatives in the results, CLIME method underestimating the magnitude of precision and partial correlation, and some
inaccurately estimated entries from ANT method, are consistent with what we observed in the simulation studies.

\subsubsection{University Webpage Data} \label{Webpage}

The webpage dataset was collected from the computer science departments of four universities by the  World Wide Knowledge Base (Web-$>$KB) project of the Carnegie Mellon University (CMU) Text Learning Group, with pages manually classified into several categories. This raw data has been pre-processed by \cite{cachopo2007improving} with stemming words. The occurrences of terms in 544 student webpages, 374 faculty webpages, 310 course webpages, and 168 project webpages are used in the following analysis.

First, the word count of the $i$-th term in the $j$-th webpage is denoted as $f_{i,j}$, which is used to calculate the following relative frequency of terms in the document (webpage). The Document-Term Matrix (DTM) weighting for terms in $D$ documents is the multiplication between local and global weights, i.e., $x_{i,j}=L_{i,j}G_j$, where the log local weight is $L_{i,j}=\log (f_{i,j}+1)$, and entropy Global weight is $G_j=1+\frac{\Sigma_ip_{i,j}\log p_{i,j}}{\text{D}}$, with $p_{i,j}=\frac{f_{i,j}}{gf_j}$. 100 terms are selected with the largest entropy $-\Sigma_i p_{i,j} \log p_{i,j}$ for the following analysis.

\begin{figure}
    \includegraphics[width = \textwidth]{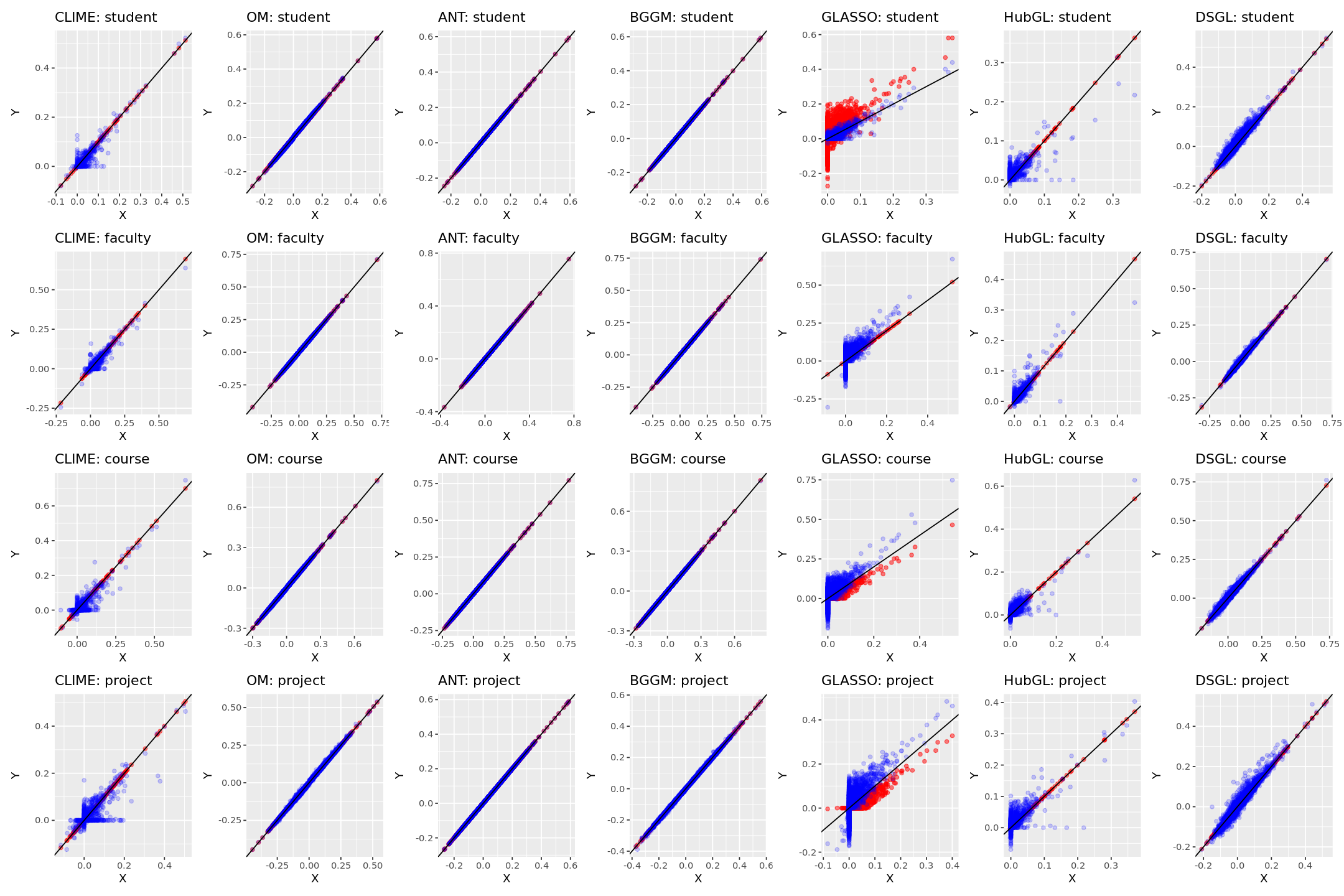}
    \caption{Estimated partial correlations in different categories after applying various standardization methods, with data from each category presented in a separate row.}
    \label{webkb_standardization}
    \centering
\end{figure}

Standardization is to scale data with zero mean and unit variance, but different operations may lead to different outcomes. The document-term matrix weighting is denoted as $X$. Specifically, when the raw count matrix is from all webpages, the weighting matrix is $X^{(all)}$, and when the webpages of a single category are split to be preprocessed, the weighting matrix is $X^{(student)}, X^{(faculty)}, X^{(course)}$, and $X^{(project)}$.
It is obvious that using all webpages or that from each category separately will lead to different weighting due to different term frequencies. Thus, standardizing $X^{(all)}$ and then taking corresponding lines for each category is different from standardizing weights from individual weights separately.
Even when the data is in the same scale, methods with parameters to be tuned, such as CLIME, GLasso, Hub GLasso, and Desparsified GLasso, still have unstable results when the data standardization is different, while FLAG preserves stable results
, as shown in Figure \ref{webkb_standardization}. After comparison, the data standardization is fixed to center and scale data from each category separately in the following analysis.

When taking single category data as input, four inferred graphs by FLAG can be obtained. The common edges in the graphs of all four categories are standard phrases in computer science websites, such as ('comput', 'scienc'), ('home', 'page'), ('high', 'perform'), and ('commun', 'network'). The corresponding precision and partial correlation are far away from zero, and the p-values of tests are much smaller than $1e-4$.
Compared with the results obtained by the ANT method, there are some standard phrases that are omitted by ANT but successfully identified by FLAG. For example, the common phrase 'related area' links the term pair ('relat', 'area'). However, the result from ANT underestimates its precision and fails to identify this edge in the course category data. More precisely, the estimated precision and partial correlation of this pair by ANT are 0.13 and -0.06, respectively, while the estimates are 0.52 and -0.22 by FLAG. This situation is consistent with our finding in the simulation that the underestimation of precision by ANT comes from a large zero proportion (80.6\%, 79.6\%) in the $\beta$, which induces smaller $\var(X\beta)$ and larger $\var(\epsilon)$ ((0.46,0.62) by ANT and (0.36,0.53) by FLAG), and thus leads to smaller estimated precision.

The graphs inferred by FLAG can capture different conditional dependencies in different categories. Taking the term 'data' as an example, the edge ('data', 'educ') in the student category is significant to have a precision of -0.23 and a p-value of corresponding hypothesis test of $5e-4$.
The edge ('data', 'structur') has a p-value of $7e-10$ in the faculty category and $2e-10$ in the course category.
The edge ('data', 'model') in the project category has a p-value of $3e-5$.
The estimated precision and partial correlation have a relatively large standard error due to the small sample size in the project category, which can be alleviated by meta-analysis.

\begin{table}[htbp]
  \begin{subtable}[h]{\textwidth}
    \centering
    \begin{tabular}{lllll}
        \toprule
        Pair of Terms   &$\rho^\text{(student)}$   &$\rho^\text{(faculty)}$   &$\rho^\text{(course)}$    &$\rho^\text{(project)}$\\
        \midrule
        ('data', 'structur')    & 0.16 (0.05)   & \textbf{0.34}(0.05)   & \textbf{0.41}(0.06)   & -0.05 (0.11) \\
        ('data', 'model')       & 0.01 (0.05)   & -0.07 (0.06)  & 0.15 (0.06)   & \textbf{0.40}(0.08) \\
        \bottomrule
    \end{tabular}
    \caption{Estimated partial correlation across categories}
  \end{subtable}

  \begin{subtable}[h]{\textwidth}
    \centering
    \begin{tabular}{|l|*{5}{c|}}\hline
        ('data','model') \textbackslash ('data','structur')
        &student   &faculty   &course    &project \\\hline
        student     & /     & \textbf{9e-3}      & \textbf{5e-4}      & 0.06   \\\hline
        faculty     & 0.29  & /          & 0.35      & \textbf{8e-4} \\\hline
        course      & 0.09  & 0.014     & /          & \textbf{1e-4 }\\\hline
        project     & \textbf{4e-5}  & \textbf{3e-6}      & 0.015     & / \\\hline
    \end{tabular}

    \caption{P-values of testing the difference between partial correlation of the same pair in different categories}
    \label{webkb_table_b}
  \end{subtable}

  \caption{Edge differences of term pairs ('data', 'structure') and ('data', 'model') across categories.}
  \label{table_webkb}
\end{table}

In addition to graph recovery in each category, the inference can be extended to test the differences of partial correlation of the same pair across different categories, as shown in Table \ref{webkb_table_b}, with the null hypothesis $\rho^{(\text{category A})}-\rho^{(\text{category B})}=0$. Specifically, the test of the pair ('data', 'model') between the project category and the result from the faculty category rejects the null hypothesis, indicating that the partial correlation is significant to be different in these two categories.

\begin{figure}
     \centering
     \begin{subfigure}[b]{0.48\textwidth}
         \centering
         \includegraphics[width=\textwidth]{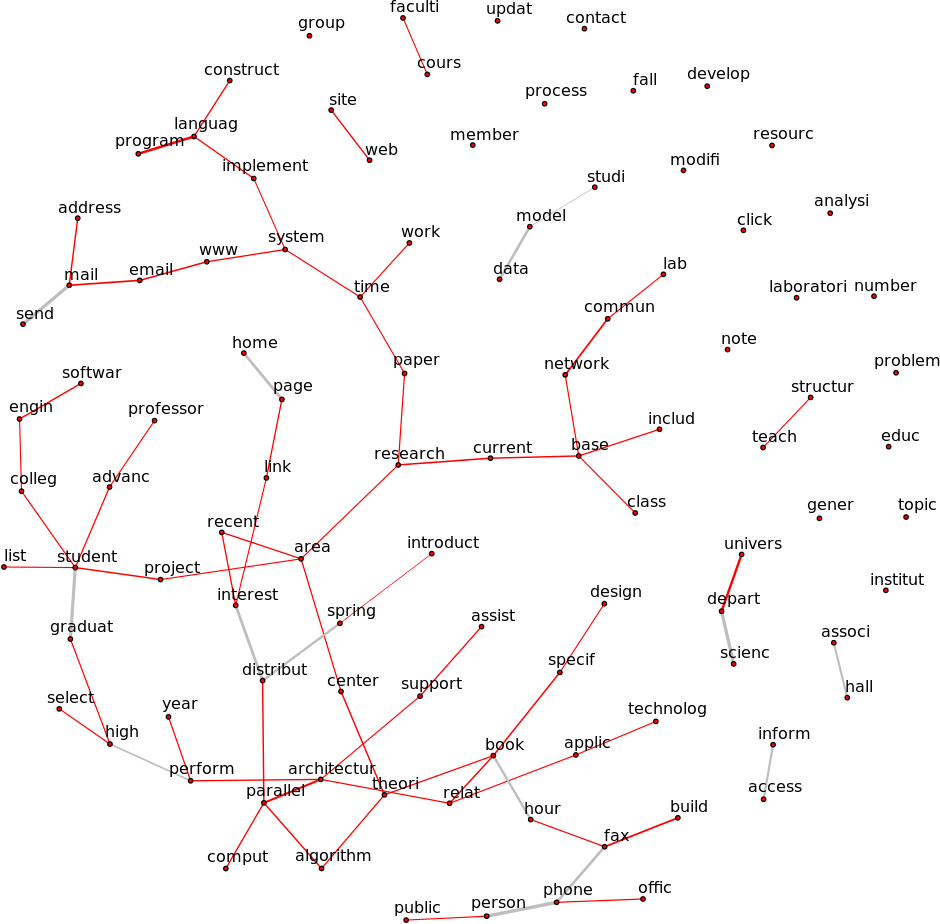}
         \caption{Meta-analysis: project with the result from course category}
         \label{}
     \end{subfigure}
     \hfill
     \begin{subfigure}[b]{0.5\textwidth}
         \centering
         \includegraphics[width=\textwidth]{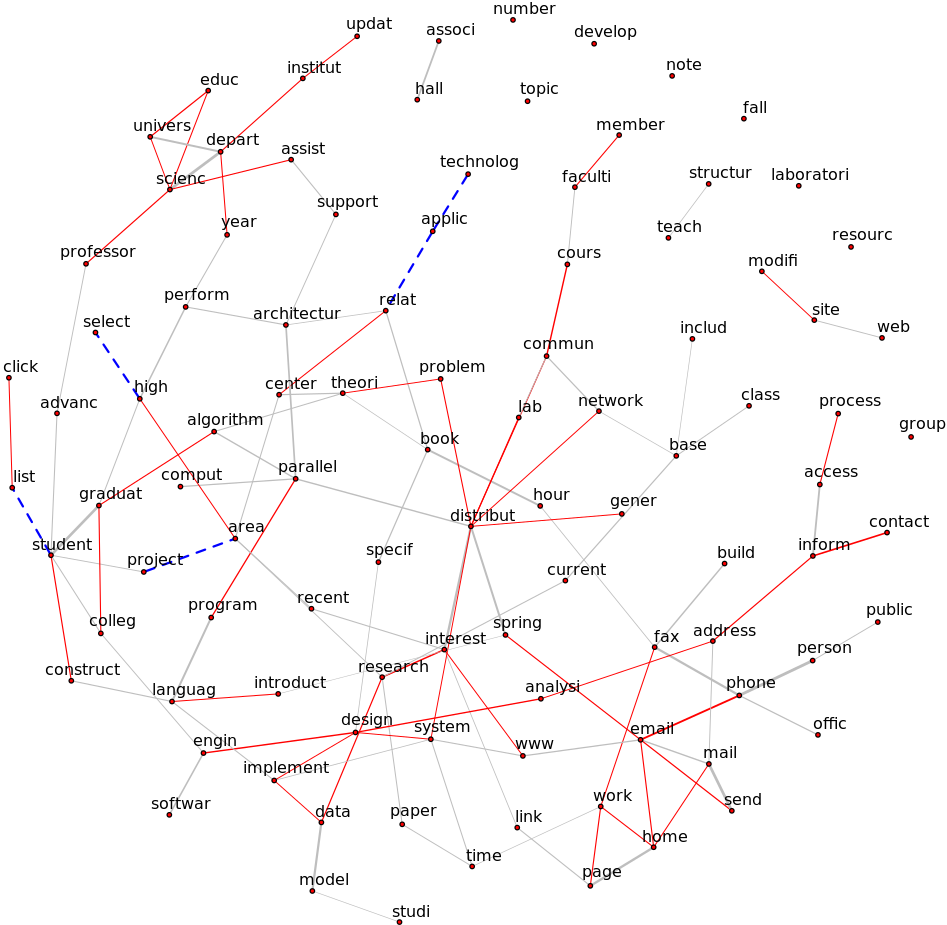}
         \caption{Meta-analysis: (project,course) with the result from faculty category}
         \label{webkb_meta_faculty}
     \end{subfigure}     
     \begin{subfigure}[b]{\textwidth}
         \centering
         \includegraphics[width=0.6\textwidth]{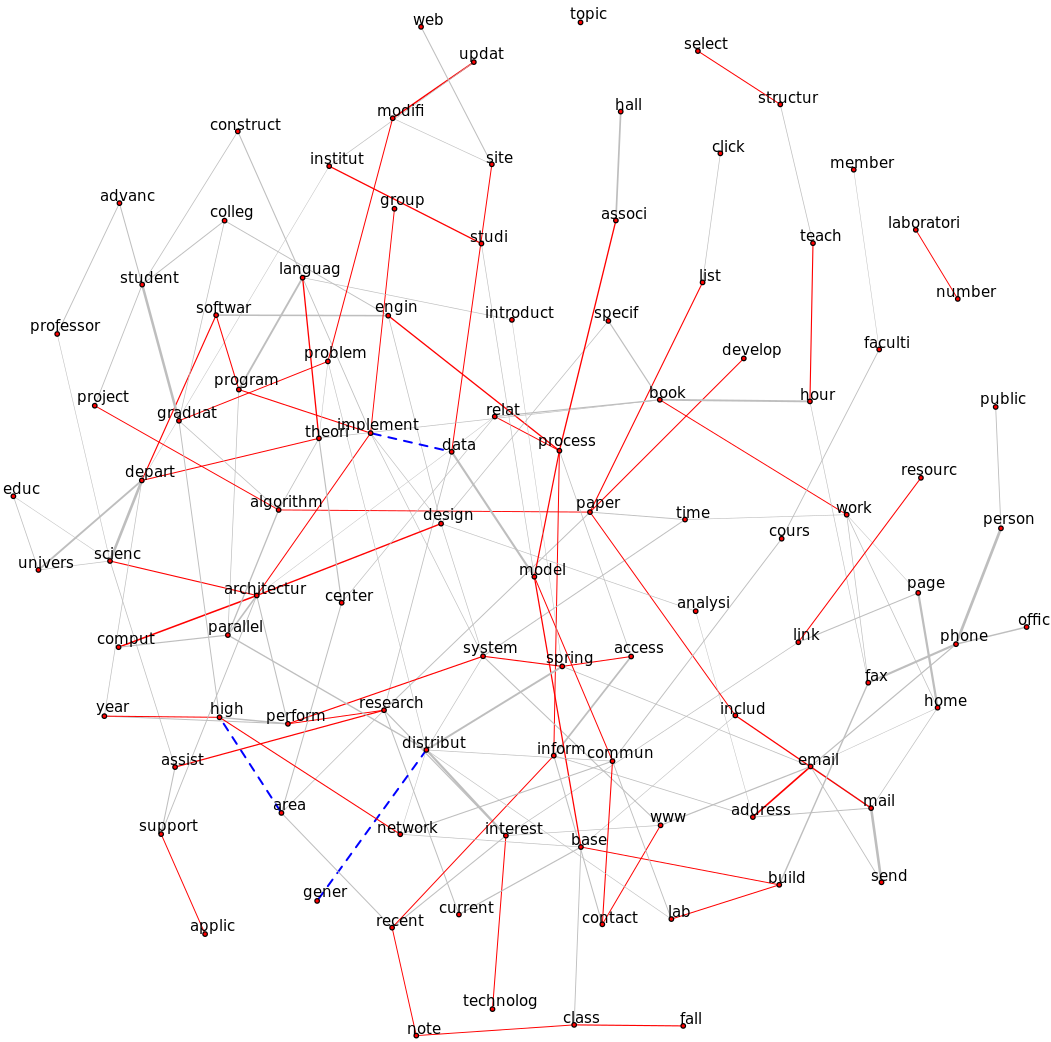}
         \caption{Meta-analysis: (project,course,faculty) with the result from student category}
         \label{webkb_meta_student}
     \end{subfigure}
    \caption{A many-to-one meta-analysis using the FLAG-Meta method, with project category as the pivot for progressive analysis.}
    \label{webkb_meta}
\end{figure}

The results from the project category, due to its small sample size, have relatively large standard error in estimated precision and partial correlation, and thus its inferred graph has few edges due to relative small power. Since all data are from the terms on the webpages of computer science department in universities, it is natural to leverage their common shared properties to enhance the result in the project category. In order to obtain the result after one-to-one meta-analysis and identify how each category contributes to the enhancement of the result, each category is used for meta-analysis in the order of ascending sample size: course, faculty, and student. The whole procedure is shown in Figure \ref{webkb_meta}. In each step, the data from one category is combined with the previous result in grey, and the edges that are only detected after meta-analysis are in red. Blue dotted lines denote the edges that are shown in the previous result but are not significant in the result of meta-analysis.

\begin{table}[htbp]
  \centering
    \begin{tabular}{llll}
    \toprule
    Pair of Terms   &$\rho^\text{(project)}$   &$\rho^\text{(course)}$ &$\rho^\text{(meta)}$\\
    \midrule
    ('engin', 'softwar')        & 0.32 (0.09)   & 0.19 (0.06)   & 0.24 (0.05) \\
    ('language', 'implement')   & 0.30 (0.08)   & 0.17 (0.06)   & 0.21 (0.05) \\
    ('assist', 'support')       & 0.24 (0.09)   & 0.20 (0.06)   & 0.21 (0.05) \\
    \bottomrule
    \end{tabular}
    \caption{Example of edges missed by FLAG given the data from individual groups, but unveiled by FLAG-Meta.}
\end{table}

The first meta-analysis is between the project and course categories. Compared with the graph inferred only based on project data, 61 edges are added. The pairs ('engin', 'softwar') and ('language', 'implement'), whose dependencies are supported by the common phrase 'software engineering' in the computer science field, and concurrence of related words 'implement' and (programming) 'language', are not found by data in a single category but are discovered by meta-analysis between project and course data.
The next step is the meta-analysis between the result of meta-analysis of project and course and the result from the faculty category. As shown in Figure \ref{webkb_meta_faculty}, 46 edges are added in red, while 5 edges are removed in blue dotted lines. The meta-analysis in this stage not only further enlarges the power but also detects some possible false positives like ('high', 'select') and ('area', 'project'). Overall, the meta-analysis provides a result from the project category with smaller standard error and larger power.

For comparison, taking the result shown in Figure \ref{webkb_meta_student} that achieves many-to-one meta-analysis with respect to the project category, the edges of the node 'data' in the single category project data are only with 'model', while the edges in the result of FLAG-Meta are with 'model', 'structur', and 'research'. 

\begin{figure}
    \centering
    \includegraphics[width = 0.7\textwidth]{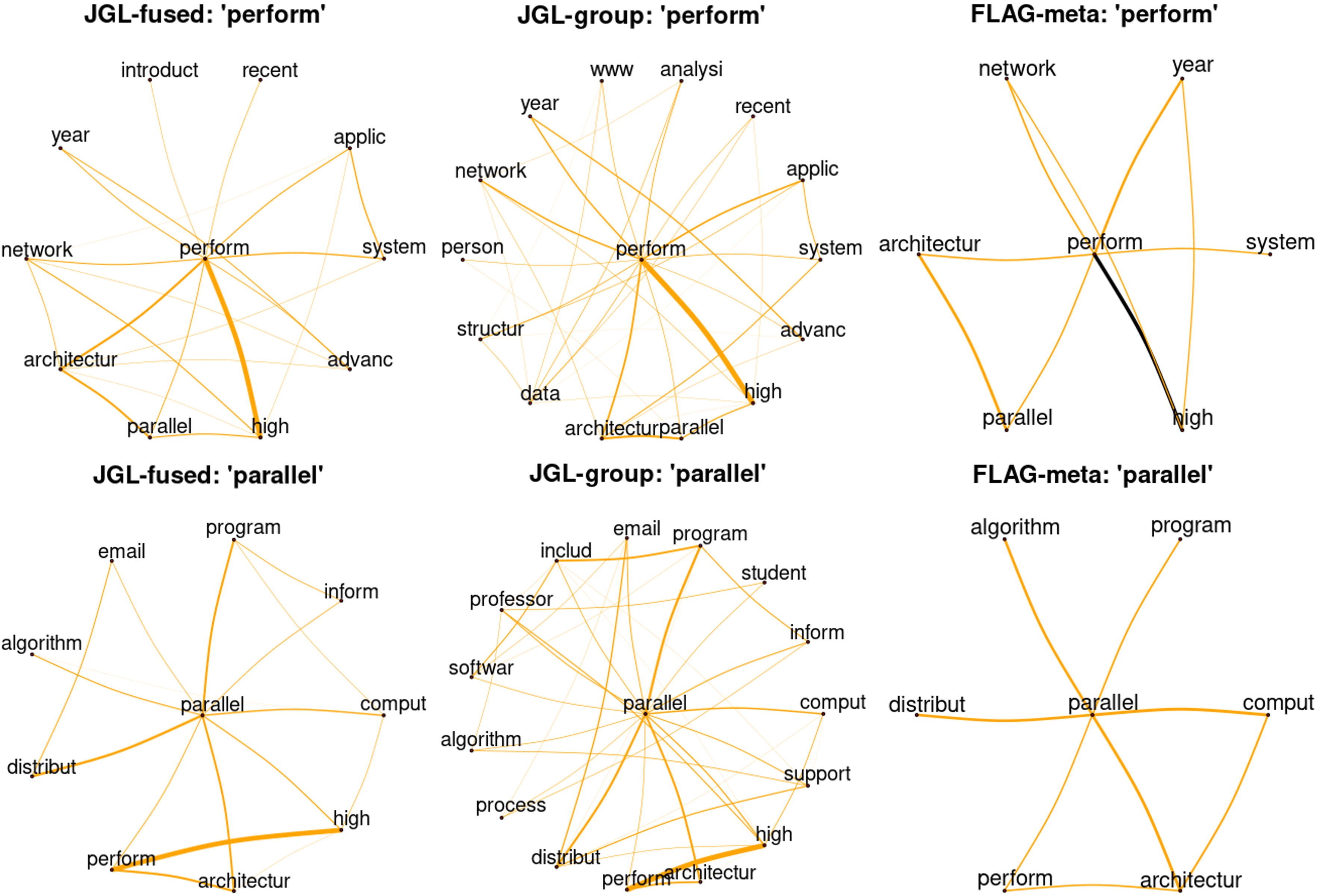}
    \caption{Subgraphs around the terms 'perform' and 'parallel' using the JGL-based methods and FLAG-Meta.}
    \label{webkb_subgraph}
\end{figure}

From Figure \ref{webkb_subgraph}, the joint GLasso fails to recover such reasonable edges of 'data' with 'structur' and 'research', and it involves some false positives like edges between 'data' and 'class', 'develop', 'program'.

\subsubsection{U.S. Stock Prices}

The raw data consists of daily close prices of 99 stocks in the S\&P100 Index from 2018-01-02 to 2022-06-30. The stock with the code 'DOW' in the S\&P 100 Index is excluded due to its start time being on 2019-03-21, with missing data of more than 14 months.

It is preprocessed by taking the logarithmic difference, $Z_{i,j}=\log P_{i,j}-\log P_{i-1,j}$, where $P_{i,j}$ is the close price of the $j$-th stock on the $i$-th day.
The log return is used as the input data in the following analysis, where the outcome is a perceived network \cite{anufriev2015connecting}, and the conditional dependence in the stock network is the return co-movements.

\begin{figure}
    \centering
    \includegraphics[width = 0.75\textwidth]{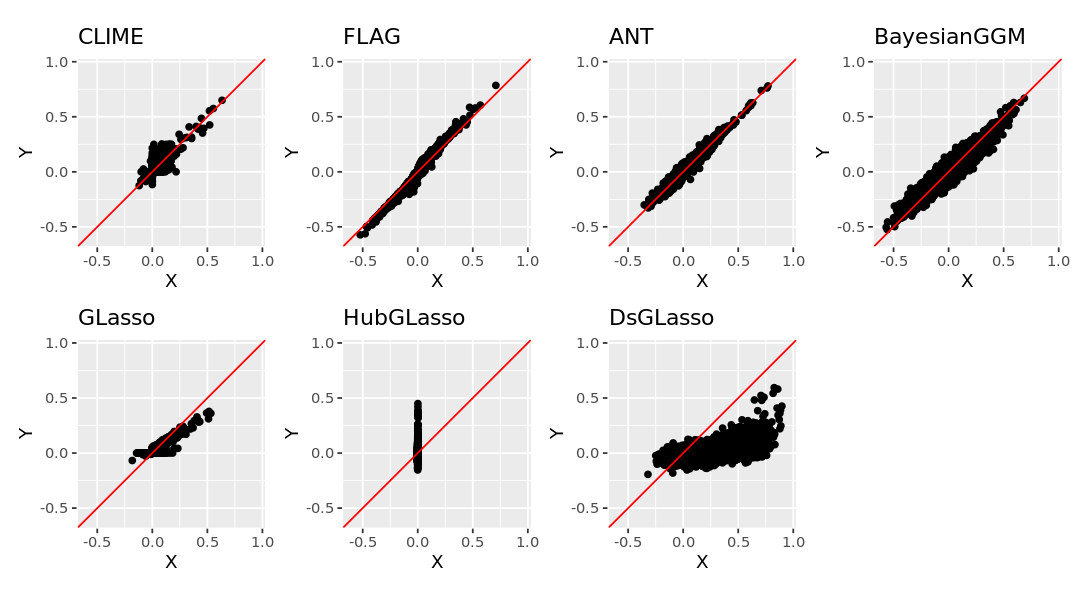}
    \caption{Scatter plots of the estimated partial correlation using different methods, with each data point representing the result from the scaled data in Y versus the raw log-return in X.}
    \label{sp100_scale}
    \centering
\end{figure}

Due to the small variance of the log return, which is around $e^{-4}$, the precision is about $e^4$ to $e^5$. Such a large magnitude increases instability in the estimation of precision, as well as the partial correlation.
From Figure \ref{sp100_scale}, it can be observed that the estimated partial correlation from FLAG is the least sensitive to data scaling, as the scattered points are most tightly clustered around the diagonal line, indicating that the FLAG method provides more consistent results across different scales of input data.
In contrast, the results from regularization-based methods such as CLIME, GLasso, HubGLasso, and DsGLasso heavily rely on the penalty parameter. It is evident that the tuned parameters will vary widely depending on whether the input data is log return or scaled data.
On the one hand, the two types of tuned parameters have no correspondence, and thus the results also vary greatly by each penalty-based method.
On the other hand, given two results from the same method, when the input data is scaled or not, it is difficult to determine which result to use, even though they are expected to reveal the same underlying structure from the data.

Inspired by \cite{bernardini2022new}, we are also interested in whether and how the S\&P100 stock co-movement network shows the impact of Covid-19 pandemic by using a rolling window of one-year length, which shifts one month in each step, as the input data.
Recall the stock market crash in 2020, there were trading curbs on March $9^{th}$, March $12^{nd}$, March $16^{th}$, and March $18^{th}$, which occured 25 years after the previous one in 1997. Such stock market crashes imply increased instability in the market due to the Covid-19 pandemic.

A large complex system transitions from stable to unstable once its connectance increases over a critical level, as suggested by \cite{gardner1970connectance}.
In it common knowledge in the financial market that the correlations between securities,
no matter marginal correlations or partial correlations, increases significantly during market crises, just as the prices of most securities drop together with negative returns. Therefore, it is natural to use the stock network whose edges are weighted by (partial) correlation to evaluate the stability of market.

The stability of a system is quantified using the connectances $C$ and the average interaction strengths $\alpha$, and the system is stable when $\alpha^2nC<1$, where $n$ is the number of variables in the system, and the system is unstable when $\alpha^2nC>1$ , as proposed by May in \cite{may1972will}. The May–Wigner stability theorem has been applied to evaluate the stock network stability in \cite{heiberger2014stock}, with the stability condition $m=\sqrt{nC}\alpha<1$ where $n$ is the number of stocks as the size of the network, and connectances $C$ is the density of connections, and the average interaction strength $\alpha$ equals the average value of strength among nodes, with the weighted degree of a node as its strength for each node.
As for the estimated precision matrix and partial correlation and inferred graphs of Gaussian graphical model, the weight of edges is the magnitude of partial correlation for fair comparison. The stability indicator $m=\sqrt{nC}\alpha$ is calculated by different methods given data in the rolling window along the time.

\begin{figure}
    \centering
    \includegraphics[width = \textwidth]{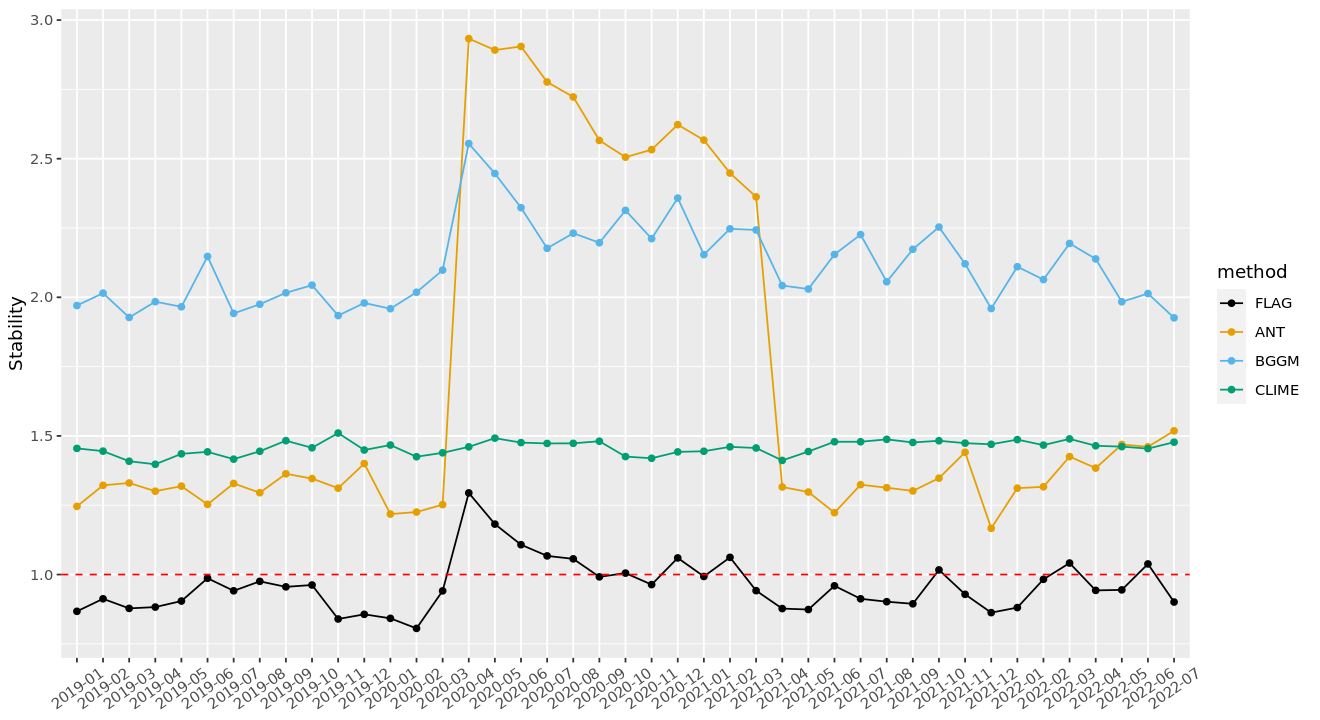}
    \caption{The stability $m$ of the stock network obtained from different methods, using a rolling window of one-year length shifted by one month.}
    \label{sp100_stability}
    \centering
\end{figure}

In Figure \ref{sp100_stability}, the stability of graphs estimated or inferred from different methods are shown in different lines, with each point on the line representing the stability calculated using the most recent one-year market data. For instance, the point at time '2020-04' uses the log-return from [2019-04-01,2020-04-01) as input.
Recall the May–Wigner stability theorem, which states that the system is stable when $m<1$ and unstable when $m>1$.
Among the methods shown, FLAG is the only method whose outcome correctly oscillates around the reference line as one, while the result from GLasso is not shown due to its magnitude of $m$ being too large compared with other methods and the reference value one.
The stability evaluated by the FLAG method increases significantly from February 2020 to April 2020, which highly matches the crashes in the U.S. stock market from 20 February 2020 to 7 April 2020. After this period, FLAG detects that the market stabilizes from March 2021. 
However, the stability calculated by the results from ANT increases dramatically from March 2020 to April 2020 and decreases dramatically from March 2021 to April 2021, indicating that the results are dominated by the data in March 2020 when it is included. This scenario implies the vulnerability of results from the ANT method.
Regarding the point at time '2021-03', the aim is to evaluate the stability of the market in the recent period, but the stability indicator equals 2.36 when input data is the recent 12-month-long, and 1.24 for the recent 11-month-long data as input, making it difficult to determine which result to trust.
The results from BGGM are roughly twice the expected value, although the trend is relatively matches that from the real market. According to the simulation studies, BGGM overestimates the magnitude of the precision matrix, partial correlation, and then such overestimation is propagated to the strength of nodes as the sum of weighted edges, strength of networks, and stability.
The results from the CLIME method are too flat to reflect the dynamic pattern of the market. In conclusion, FLAG can successfully detect the impact of the Covid-19 pandemic on the US stock market with the proper magnitude of stability.

\begin{figure}
    \centering
    \includegraphics[width = \textwidth]{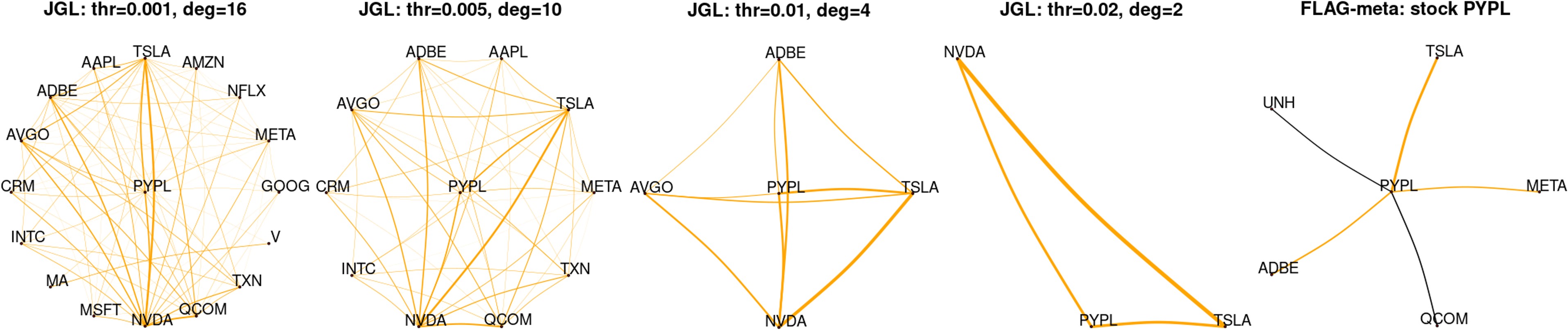}
    \caption{The inferred subgraphs around the stock 'PYPL' in the year 2021, using the JGL and FLAG-Meta methods.}
    \label{PYPL}
    \centering
\end{figure}

The many-to-one meta-analysis is conducted between the results from data in 2021 and that from the other groups (data in 2019 and 2020), compared with the joint group GLasso, with the subgraph of node ‘PYPL’ as an example.
The results from the joint group graphical lasso vary widely depending on the threshold of the estimated precision values, making it difficult to determine the optimal threshold especially in real data.
The results from FLAG-Meta have larger power compared with the results estimated from single-year data.


\section{Discussion}
The Flexible and Accurate method of Gaussian graphical model (FLAG) aims to estimate precision matrix entries accurately and efficiently, and further quantify the uncertainty of each entry, which allows for better leveraging of the common structure across different groups through meta-analysis.
FLAG has no explicit structural assumptions on the precision matrix or the corresponding graphs, making it tuning-free. Its capability of element-wise inference allows for extension to multiple graphs with small computational consumption, making it highly flexible.

Simulation studies in three different settings show that FLAG is not sensitive to data scaling, unlike other methods that require tuning parameters. FLAG is particularly suitable for the data with a hub structure, where it outperforms other methods, especially in the region of edges between hubs, even when the non-zero proportion of underlying coefficients is varied.
FLAG can make inferences to test each edge individually and adjust partial correlation and precision values after cooperating the entries that have common structure across groups to achieve smaller standard error and larger power.
FLAG is accurate, with a small relative error and a large area under the ROC curve in the simulation studies. 

FLAG is capable of unveiling the co-expression relationships between genes in the brain across time and region, identifying the associations between terms in the webpage data from different categories, and revealing the relationships between stocks in the S\&P100 with stability influenced by Covid-19 captured well.


\newpage
\bibliographystyle{plainnat}
\bibliography{ref.bib}

\begin{thebibliography}{53}
\providecommand{\natexlab}[1]{#1}
\providecommand{\url}[1]{\texttt{#1}}
\expandafter\ifx\csname urlstyle\endcsname\relax
  \providecommand{\doi}[1]{doi: #1}\else
  \providecommand{\doi}{doi: \begingroup \urlstyle{rm}\Url}\fi

\bibitem[Anufriev and Panchenko(2015)]{anufriev2015connecting}
Mikhail Anufriev and Valentyn Panchenko.
\newblock Connecting the dots: Econometric methods for uncovering networks with
  an application to the australian financial institutions.
\newblock \emph{Journal of Banking \& Finance}, 61:\penalty0 S241--S255, 2015.

\bibitem[Atay-Kayis and Massam(2005)]{atay2005monte}
Aliye Atay-Kayis and H{\'e}lene Massam.
\newblock A monte carlo method for computing the marginal likelihood in
  nondecomposable gaussian graphical models.
\newblock \emph{Biometrika}, 92\penalty0 (2):\penalty0 317--335, 2005.

\bibitem[Barab{\'a}si(2013)]{barabasi2013network}
Albert-L{\'a}szl{\'o} Barab{\'a}si.
\newblock Network science.
\newblock \emph{Philosophical Transactions of the Royal Society A:
  Mathematical, Physical and Engineering Sciences}, 371\penalty0
  (1987):\penalty0 20120375, 2013.

\bibitem[Benjamini and Hochberg(1995)]{benjamini1995controlling}
Yoav Benjamini and Yosef Hochberg.
\newblock Controlling the false discovery rate: a practical and powerful
  approach to multiple testing.
\newblock \emph{Journal of the Royal statistical society: series B
  (Methodological)}, 57\penalty0 (1):\penalty0 289--300, 1995.

\bibitem[Bernardini et~al.(2022)Bernardini, Paterlini, and
  Taufer]{bernardini2022new}
Davide Bernardini, Sandra Paterlini, and Emanuele Taufer.
\newblock New estimation approaches for graphical models with elastic net
  penalty.
\newblock \emph{Econometrics and Statistics}, 2022.

\bibitem[Bilgrau et~al.(2020)Bilgrau, Peeters, Eriksen, B{\o}gsted, and
  Van~Wieringen]{bilgrau2020targeted}
Anders~Ellern Bilgrau, Carel~FW Peeters, Poul~Svante Eriksen, Martin
  B{\o}gsted, and Wessel~N Van~Wieringen.
\newblock Targeted fused ridge estimation of inverse covariance matrices from
  multiple high-dimensional data classes.
\newblock \emph{The Journal of Machine Learning Research}, 21\penalty0
  (1):\penalty0 946--997, 2020.

\bibitem[Cachopo et~al.(2007)]{cachopo2007improving}
Ana Margarida de Jesus~Cardoso Cachopo et~al.
\newblock Improving methods for single-label text categorization.
\newblock \emph{Instituto Superior T{\'e}cnico, Portugal}, 2007.

\bibitem[Cai et~al.(2013)Cai, Li, Liu, and Xie]{cai2013covariate}
T~Tony Cai, Hongzhe Li, Weidong Liu, and Jichun Xie.
\newblock Covariate-adjusted precision matrix estimation with an application in
  genetical genomics.
\newblock \emph{Biometrika}, 100\penalty0 (1):\penalty0 139--156, 2013.

\bibitem[Cai et~al.(2011)Cai, Liu, and Luo]{cai2011constrained}
Tony Cai, Weidong Liu, and Xi~Luo.
\newblock A constrained $\ell_1$ minimization approach to sparse precision
  matrix estimation.
\newblock \emph{Journal of the American Statistical Association}, 106\penalty0
  (494):\penalty0 594--607, 2011.

\bibitem[Chen et~al.(2016)Chen, Ren, Zhao, and Zhou]{chen2016asymptotically}
Mengjie Chen, Zhao Ren, Hongyu Zhao, and Harrison Zhou.
\newblock Asymptotically normal and efficient estimation of covariate-adjusted
  gaussian graphical model.
\newblock \emph{Journal of the American Statistical Association}, 111\penalty0
  (513):\penalty0 394--406, 2016.

\bibitem[Danaher et~al.(2014)Danaher, Wang, and Witten]{danaher2014joint}
Patrick Danaher, Pei Wang, and Daniela~M Witten.
\newblock The joint graphical lasso for inverse covariance estimation across
  multiple classes.
\newblock \emph{Journal of the Royal Statistical Society: Series B (Statistical
  Methodology)}, 76\penalty0 (2):\penalty0 373--397, 2014.

\bibitem[Di~Vona et~al.(2015)Di~Vona, Bezdan, Islam, Salichs, Lopez-Bigas,
  Ossowski, and de~la Luna]{di2015chromatin}
Chiara Di~Vona, Daniela Bezdan, Abul~BMMK Islam, Eulalia Salichs, Nuria
  Lopez-Bigas, Stephan Ossowski, and Susana de~la Luna.
\newblock Chromatin-wide profiling of dyrk1a reveals a role as a gene-specific
  rna polymerase ii ctd kinase.
\newblock \emph{Molecular cell}, 57\penalty0 (3):\penalty0 506--520, 2015.

\bibitem[Dobra et~al.(2011)Dobra, Lenkoski, and Rodriguez]{dobra2011bayesian}
Adrian Dobra, Alex Lenkoski, and Abel Rodriguez.
\newblock Bayesian inference for general gaussian graphical models with
  application to multivariate lattice data.
\newblock \emph{Journal of the American Statistical Association}, 106\penalty0
  (496):\penalty0 1418--1433, 2011.

\bibitem[Engelke and Hitz(2020)]{engelke2020graphical}
Sebastian Engelke and Adrien~S Hitz.
\newblock Graphical models for extremes.
\newblock \emph{Journal of the Royal Statistical Society Series B: Statistical
  Methodology}, 82\penalty0 (4):\penalty0 871--932, 2020.

\bibitem[Epskamp et~al.(2018)Epskamp, Waldorp, M{\~o}ttus, and
  Borsboom]{epskamp2018gaussian}
Sacha Epskamp, Lourens~J Waldorp, Ren{\'e} M{\~o}ttus, and Denny Borsboom.
\newblock The gaussian graphical model in cross-sectional and time-series data.
\newblock \emph{Multivariate behavioral research}, 53\penalty0 (4):\penalty0
  453--480, 2018.

\bibitem[Fan et~al.(2009)Fan, Feng, and Wu]{fan2009network}
Jianqing Fan, Yang Feng, and Yichao Wu.
\newblock Network exploration via the adaptive lasso and scad penalties.
\newblock \emph{The annals of applied statistics}, pages 521--541, 2009.

\bibitem[Feng and Ning(2019)]{feng2019high}
Huijie Feng and Yang Ning.
\newblock High-dimensional mixed graphical model with ordinal data: Parameter
  estimation and statistical inference.
\newblock In \emph{The 22nd international conference on artificial intelligence
  and statistics}, pages 654--663. PMLR, 2019.

\bibitem[Foulley and Van~Dyk(2000)]{foulley2000px}
Jean-Louis Foulley and David~A Van~Dyk.
\newblock The px-em algorithm for fast stable fitting of henderson's mixed
  model.
\newblock \emph{Genetics Selection Evolution}, 32:\penalty0 1--21, 2000.

\bibitem[Friedman et~al.(2008)Friedman, Hastie, and
  Tibshirani]{friedman2008sparse}
Jerome Friedman, Trevor Hastie, and Robert Tibshirani.
\newblock Sparse inverse covariance estimation with the graphical lasso.
\newblock \emph{Biostatistics}, 9\penalty0 (3):\penalty0 432--441, 2008.

\bibitem[Gardner and Ashby(1970)]{gardner1970connectance}
Mark~R Gardner and W~Ross Ashby.
\newblock Connectance of large dynamic (cybernetic) systems: critical values
  for stability.
\newblock \emph{Nature}, 228\penalty0 (5273):\penalty0 784--784, 1970.

\bibitem[Guo et~al.(2011)Guo, Levina, Michailidis, and Zhu]{guo2011joint}
Jian Guo, Elizaveta Levina, George Michailidis, and Ji~Zhu.
\newblock Joint estimation of multiple graphical models.
\newblock \emph{Biometrika}, 98\penalty0 (1):\penalty0 1--15, 2011.

\bibitem[Hao et~al.(2018)Hao, Sun, Liu, and Cheng]{hao2018simultaneous}
Botao Hao, Will~Wei Sun, Yufeng Liu, and Guang Cheng.
\newblock Simultaneous clustering and estimation of heterogeneous graphical
  models.
\newblock \emph{Journal of Machine Learning Research}, 2018.

\bibitem[Hastie et~al.(2009)Hastie, Tibshirani, Friedman, and
  Friedman]{hastie2009elements}
Trevor Hastie, Robert Tibshirani, Jerome~H Friedman, and Jerome~H Friedman.
\newblock \emph{The elements of statistical learning: data mining, inference,
  and prediction}, volume~2.
\newblock Springer, 2009.

\bibitem[Heiberger(2014)]{heiberger2014stock}
Raphael~H Heiberger.
\newblock Stock network stability in times of crisis.
\newblock \emph{Physica A: Statistical Mechanics and its Applications},
  393:\penalty0 376--381, 2014.

\bibitem[Hero and Rajaratnam(2012)]{hero2012hub}
Alfred Hero and Bala Rajaratnam.
\newblock Hub discovery in partial correlation graphs.
\newblock \emph{IEEE Transactions on Information Theory}, 58\penalty0
  (9):\penalty0 6064--6078, 2012.

\bibitem[Hunter and Lange(2000)]{hunter2000quantile}
David~R Hunter and Kenneth Lange.
\newblock Quantile regression via an mm algorithm.
\newblock \emph{Journal of Computational and Graphical Statistics}, 9\penalty0
  (1):\penalty0 60--77, 2000.

\bibitem[Jankova and Van De~Geer(2015)]{jankova2015confidence}
Jana Jankova and Sara Van De~Geer.
\newblock Confidence intervals for high-dimensional inverse covariance
  estimation.
\newblock \emph{Electronic Journal of Statistics}, 9\penalty0 (1):\penalty0
  1205--1229, 2015.

\bibitem[Kang et~al.(2011)Kang, Kawasawa, Cheng, Zhu, Xu, Li, Sousa, Pletikos,
  Meyer, Sedmak, et~al.]{kang2011spatio}
Hyo~Jung Kang, Yuka~Imamura Kawasawa, Feng Cheng, Ying Zhu, Xuming Xu, Mingfeng
  Li, Andr{\'e}~MM Sousa, Mihovil Pletikos, Kyle~A Meyer, Goran Sedmak, et~al.
\newblock Spatio-temporal transcriptome of the human brain.
\newblock \emph{Nature}, 478\penalty0 (7370):\penalty0 483--489, 2011.

\bibitem[Lee and Liu(2015)]{lee2015joint}
Wonyul Lee and Yufeng Liu.
\newblock Joint estimation of multiple precision matrices with common
  structures.
\newblock \emph{The Journal of Machine Learning Research}, 16\penalty0
  (1):\penalty0 1035--1062, 2015.

\bibitem[Lin et~al.(2017)Lin, Wang, Yang, and Zhao]{lin2017joint}
Zhixiang Lin, Tao Wang, Can Yang, and Hongyu Zhao.
\newblock On joint estimation of gaussian graphical models for spatial and
  temporal data.
\newblock \emph{Biometrics}, 73\penalty0 (3):\penalty0 769--779, 2017.

\bibitem[Liu et~al.(1998)Liu, Rubin, and Wu]{liu1998parameter}
Chuanhai Liu, Donald~B Rubin, and Ying~Nian Wu.
\newblock Parameter expansion to accelerate em: the px-em algorithm.
\newblock \emph{Biometrika}, 85\penalty0 (4):\penalty0 755--770, 1998.

\bibitem[Liu(2013)]{liu2013gaussian}
Weidong Liu.
\newblock Gaussian graphical model estimation with false discovery rate
  control.
\newblock \emph{THE ANNALS of STATISTICS}, pages 2948--2978, 2013.

\bibitem[Ma and Michailidis(2016)]{ma2016joint}
Jing Ma and George Michailidis.
\newblock Joint structural estimation of multiple graphical models.
\newblock \emph{The Journal of Machine Learning Research}, 17\penalty0
  (1):\penalty0 5777--5824, 2016.

\bibitem[May(1972)]{may1972will}
Robert~M May.
\newblock Will a large complex system be stable?
\newblock \emph{Nature}, 238\penalty0 (5364):\penalty0 413--414, 1972.

\bibitem[Meinshausen and B{\"u}hlmann(2006)]{meinshausen2006high}
Nicolai Meinshausen and Peter B{\"u}hlmann.
\newblock High-dimensional graphs and variable selection with the lasso.
\newblock \emph{The annals of statistics}, 34\penalty0 (3):\penalty0
  1436--1462, 2006.

\bibitem[Mohan et~al.(2012)Mohan, Chung, Han, Witten, Lee, and
  Fazel]{mohan2012structured}
Karthik Mohan, Mike Chung, Seungyeop Han, Daniela Witten, Su-In Lee, and Maryam
  Fazel.
\newblock Structured learning of gaussian graphical models.
\newblock \emph{Advances in neural information processing systems}, 25, 2012.

\bibitem[Mohan et~al.(2014)Mohan, London, Fazel, Witten, and
  Lee]{mohan2014node}
Karthik Mohan, Palma London, Maryam Fazel, Daniela Witten, and Su-In Lee.
\newblock Node-based learning of multiple gaussian graphical models.
\newblock \emph{The Journal of Machine Learning Research}, 15\penalty0
  (1):\penalty0 445--488, 2014.

\bibitem[Price et~al.(2015)Price, Geyer, and Rothman]{price2015ridge}
Bradley~S Price, Charles~J Geyer, and Adam~J Rothman.
\newblock Ridge fusion in statistical learning.
\newblock \emph{Journal of Computational and Graphical Statistics}, 24\penalty0
  (2):\penalty0 439--454, 2015.

\bibitem[Price et~al.(2021)Price, Molstad, and Sherwood]{price2021estimating}
Bradley~S Price, Aaron~J Molstad, and Ben Sherwood.
\newblock Estimating multiple precision matrices with cluster fusion
  regularization.
\newblock \emph{Journal of Computational and Graphical Statistics}, 30\penalty0
  (4):\penalty0 823--834, 2021.

\bibitem[Ren et~al.(2015)Ren, Sun, Zhang, and Zhou]{ren2015asymptotic}
Zhao Ren, Tingni Sun, Cun-Hui Zhang, and Harrison~H Zhou.
\newblock Asymptotic normality and optimalities in estimation of large gaussian
  graphical models.
\newblock \emph{The Annals of Statistics}, 43\penalty0 (3):\penalty0 991--1026,
  2015.

\bibitem[Rothman et~al.(2008)Rothman, Bickel, Levina, and
  Zhu]{rothman2008sparse}
Adam~J Rothman, Peter~J Bickel, Elizaveta Levina, and Ji~Zhu.
\newblock Sparse permutation invariant covariance estimation.
\newblock \emph{Electronic Journal of Statistics}, 2:\penalty0 494--515, 2008.

\bibitem[Rue and Held(2005)]{rue2005gaussian}
Havard Rue and Leonhard Held.
\newblock \emph{Gaussian Markov random fields: theory and applications}.
\newblock Chapman and Hall/CRC, 2005.

\bibitem[Saegusa and Shojaie(2016)]{saegusa2016joint}
Takumi Saegusa and Ali Shojaie.
\newblock Joint estimation of precision matrices in heterogeneous populations.
\newblock \emph{Electronic journal of statistics}, 10\penalty0 (1):\penalty0
  1341, 2016.

\bibitem[Shan and Kim(2018)]{shan2018joint}
Liang Shan and Inyoung Kim.
\newblock Joint estimation of multiple gaussian graphical models across
  unbalanced classes.
\newblock \emph{Computational Statistics \& Data Analysis}, 121:\penalty0
  89--103, 2018.

\bibitem[Tan et~al.(2014)Tan, London, Mohan, Lee, Fazel, and
  Witten]{tan2014learning}
Kean~Ming Tan, Palma London, Karthik Mohan, Su-In Lee, Maryam Fazel, and
  Daniela Witten.
\newblock Learning graphical models with hubs.
\newblock \emph{Journal of machine learning research: JMLR}, 15:\penalty0 3297,
  2014.

\bibitem[Van~den Heuvel and Sporns(2013)]{van2013network}
Martijn~P Van~den Heuvel and Olaf Sporns.
\newblock Network hubs in the human brain.
\newblock \emph{Trends in cognitive sciences}, 17\penalty0 (12):\penalty0
  683--696, 2013.

\bibitem[Wasserman(2004)]{wasserman2004all}
Larry Wasserman.
\newblock \emph{All of statistics: a concise course in statistical inference},
  volume~26.
\newblock Springer, 2004.

\bibitem[Williams(2021)]{williams2021bayesian}
Donald~R Williams.
\newblock Bayesian estimation for gaussian graphical models: Structure
  learning, predictability, and network comparisons.
\newblock \emph{Multivariate Behavioral Research}, 56\penalty0 (2):\penalty0
  336--352, 2021.

\bibitem[Willsey et~al.(2013)Willsey, Sanders, Li, Dong, Tebbenkamp, Muhle,
  Reilly, Lin, Fertuzinhos, Miller, et~al.]{willsey2013coexpression}
A~Jeremy Willsey, Stephan~J Sanders, Mingfeng Li, Shan Dong, Andrew~T
  Tebbenkamp, Rebecca~A Muhle, Steven~K Reilly, Leon Lin, Sofia Fertuzinhos,
  Jeremy~A Miller, et~al.
\newblock Coexpression networks implicate human midfetal deep cortical
  projection neurons in the pathogenesis of autism.
\newblock \emph{Cell}, 155\penalty0 (5):\penalty0 997--1007, 2013.

\bibitem[Yi et~al.(2022)Yi, Zhang, Lin, and Ma]{yi2022information}
Huangdi Yi, Qingzhao Zhang, Cunjie Lin, and Shuangge Ma.
\newblock Information-incorporated gaussian graphical model for gene expression
  data.
\newblock \emph{Biometrics}, 78\penalty0 (2):\penalty0 512--523, 2022.

\bibitem[Yin and Li(2011)]{yin2011sparse}
Jianxin Yin and Hongzhe Li.
\newblock A sparse conditional gaussian graphical model for analysis of
  genetical genomics data.
\newblock \emph{The annals of applied statistics}, 5\penalty0 (4):\penalty0
  2630, 2011.

\bibitem[Zhao and Duan(2019)]{zhao2019cancer}
Haitao Zhao and Zhong-Hui Duan.
\newblock Cancer genetic network inference using gaussian graphical models.
\newblock \emph{Bioinformatics and biology insights}, 13:\penalty0
  1177932219839402, 2019.

\bibitem[Zhou et~al.(2019)Zhou, Hu, Zhou, and Lange]{zhou2019mm}
Hua Zhou, Liuyi Hu, Jin Zhou, and Kenneth Lange.
\newblock \text{MM} algorithms for variance components models.
\newblock \emph{Journal of Computational and Graphical Statistics}, 28\penalty0
  (2):\penalty0 350--361, 2019.

\end{thebibliography}

\newpage
\begin{center}
{\large\bf SUPPLEMENTAL MATERIALS}
\end{center}

\subsection*{A Algorithms}

\subsubsection*{A.1 PX-EM Algorithm} \label{supp_pxem}

In the M-step of PX-EM algorithm,
$$\frac{\partial \mQ(\gamma|\gamma_{old})} {\partial \Ge}
=\frac{\partial \big ( 
    -\frac n2 \text{log}|\Ge|
    -\frac12 \text{tr}\Big[ \Ge^{-1} 
    \begin{pmatrix} \text{tr}[S_{11}] & \text{tr}[S_{12}] \\ \text{tr}[S_{21}] & \text{tr}[S_{22}] \end{pmatrix} \Big]
    \big )} {\partial \Ge} =0, $$

\begin{equation}
\begin{aligned}
&\partial \big ( 
    -\frac n2 \text{log}|\Ge|
    -\frac12 \text{tr}\Big[ \Ge^{-1} 
    \begin{pmatrix} \text{tr}[S_{11}] & \text{tr}[S_{12}] \\ \text{tr}[S_{21}] & \text{tr}[S_{22}] \end{pmatrix} \Big]
    \big ) \\
=& -\frac n2 \text{tr}[ \Ge^{-1} \partial(\Ge) ]
    -\frac12 \text{tr}\Big[ - \Ge^{-1}
    \begin{pmatrix} \text{tr}[S_{11}] & \text{tr}[S_{12}] \\ \text{tr}[S_{21}] & \text{tr}[S_{22}] \end{pmatrix}
    \Ge^{-1} \partial(\Ge)
    \Big] \\
=& \text{tr}\Big[ (-\frac n2 \Ge^{-1}
                  +\frac12 \Ge^{-1}
    \begin{pmatrix} \text{tr}[S_{11}] & \text{tr}[S_{12}] \\ \text{tr}[S_{21}] & \text{tr}[S_{22}] \end{pmatrix}
    \Ge^{-1} )
    \partial(\Ge) \Big], \\
\end{aligned}
\end{equation}

When $(-\frac n2 \Ge^{-1}
                  +\frac12 \Ge^{-1}
    \begin{pmatrix} \text{tr}[S_{11}] & \text{tr}[S_{12}] \\ \text{tr}[S_{21}] & \text{tr}[S_{22}] \end{pmatrix}
    \Ge^{-1} ) = 0$,
we have $\Ge = \frac1n \begin{pmatrix} \text{tr}[S_{11}] & \text{tr}[S_{12}] \\ \text{tr}[S_{21}] & \text{tr}[S_{22}] \end{pmatrix}$.
Similarly, we have $\Gb = \frac1{p-2} \begin{pmatrix} \text{tr}[W_{11}] & \text{tr}[W_{12}] \\ \text{tr}[W_{21}] & \text{tr}[W_{22}] \end{pmatrix}$.
The expanded parameter is updated by
$\delta=\frac {\bY^T(\Ge^{-1}\otimes X) \mu_{\bb}}
                {\text{tr}[(\Ge^{-1} \otimes X^TX)
                (\mu_{\bb}\mu_{\bb}^T + \Sigma_{\bb})]}$.

Several terms that are used for updating the parameters can also be calculated efficiently as follow,
$$\log |\Sigma_{\bb}|
= -\log|\Sigma_{\bb}^{-1}|
= -\log\det \begin{pmatrix} A & B \\ C & H \end{pmatrix}
= -\sum_{i=1}^{p-2} \log (a_ih_i-c_ib_i),
$$

\begin{equation}
\begin{aligned}
\tr[\bSig_{11}]
= \sum_{i=1}^{p-2} \frac{h_i}{a_ih_i-c_ib_i},
\tr[\bSig_{12}]
= - \sum_{i=1}^{p-2} \frac{b_i}{a_ih_i-c_ib_i}, \\
\tr[\bSig_{21}]
= - \sum_{i=1}^{p-2} \frac{c_i}{a_ih_i-c_ib_i},
\tr[\bSig_{22}]
= \sum_{i=1}^{p-2} \frac{a_i}{a_ih_i-c_ib_i},
\end{aligned}
\end{equation}

\begin{equation}
\begin{aligned}
\tr[X^TX \bSig_{11}]
= \sum_{i=1}^{p-2} \frac{q_i h_i}{a_ih_i-c_ib_i},
\tr [X^TX [\bSig_{12}]
= - \sum_{i=1}^{p-2} \frac{q_i b_i}{a_ih_i-c_ib_i}, \\
\tr[X^TX \bSig_{21}]
= - \sum_{i=1}^{p-2} \frac{q_i c_i}{a_ih_i-c_ib_i},
\tr[X^TX \bSig_{22}]
= \sum_{i=1}^{p-2} \frac{q_i a_i}{a_ih_i-c_ib_i},
\end{aligned}
\end{equation}

\begin{equation}
\begin{aligned}
& \tr[ (\Ge^{-1} \otimes X^TX) \Sigma_{\bb} ]
= \tr \Bigg[
    \begin{pmatrix} (\Ge^{-1})_{11} \diag(q) &
    (\Ge^{-1})_{12} \diag(q) \\
    (\Ge^{-1})_{21} \diag(q) &
    (\Ge^{-1})_{22} \diag(q) \end{pmatrix} \\
&    \begin{pmatrix} \diag( d \oslash (a\odot d - c\odot b) ) &
    \diag( -b \oslash (a\odot d - c\odot b) ) \\
    \diag( -c \oslash (a\odot d - c\odot b) ) &
    \diag( a \oslash (a\odot d - c\odot b) ) \end{pmatrix}
\Bigg] \\
=& \sum_{i=1}^{p-2} \frac{q_i
            [ (\Ge^{-1})_{11} h_i
              - (\Ge^{-1})_{12} c_i
              - (\Ge^{-1})_{21} b_i
              + (\Ge^{-1})_{22} a_i
            ]}{a_ih_i-c_ib_i}.
\end{aligned}
\end{equation}

\subsubsection*{A.2 Low-rank Update} \label{supp_low_rank}

\begin{equation}
\begin{aligned}
M_\beta
&= \Phi^{(m)} \textbf{diag} \Big\{ \textbf{tr}[(\lambda_l D+I_n)^{-1} D] 
+ \textbf{tr}\Big[(\frac1{\lambda_l }I_2 - (U^TZ_{\{ij\}})^T (\lambda_l D+I_n)^{-1} U^TZ_{\{ij\}} )^{-1} \\
& \Big( (U^TZ_{\{ij\}})^T (\lambda_l D+I_n)^{-1} D (\lambda_l D+I_n)^{-1} U^TZ_{\{ij\}} \Big) \Big] \\
&- \textbf{tr}  \Big( (U^TZ_{\{ij\}})^T (\lambda_l D+I_n)^{-1} U^TZ_{\{ij\}} \Big) \\
&- \textbf{tr}\Big[ \Big( (U^TZ_{\{ij\}})^T (\lambda_l D+I_n)^{-1} U^TZ_{\{ij\}} \Big)
(\frac1{\lambda_l }I_2 - (U^TZ_{\{ij\}})^T (\lambda_l D+I_n)^{-1} U^TZ_{\{ij\}} )^{-1} \\
& \Big( (U^TZ_{\{ij\}})^T (\lambda_l D+I_n)^{-1} U^TZ_{\{ij\}}  \Big)  \Big] ,l=1,2 \Big\} \Phi^{(m)T}, \\
M_\epsilon
&= \Phi^{(m)} \textbf{diag} \Big\{
\textbf{tr}[(\lambda_l D+I_n)^{-1}] 
+ \textbf{tr}\Big[(\frac1{\lambda_l }I_2 - (U^TZ_{\{ij\}})^T (\lambda_l D+I_n)^{-1} U^TZ_{\{ij\}} )^{-1} \\
& \Big( (U^TZ_{\{ij\}})^T (\lambda_l D+I_n)^{-1} (\lambda_l D+I_n)^{-1} U^TZ_{\{ij\}} \Big) \Big] ,l=1,2
    \Big\} \Phi^{(m)T}.
\end{aligned}
\end{equation}

The expressions between bigger brackets can be calculated efficiently in $\mathcal{O}(n)$ time complexity.

To further simplify the matrix $N_\beta$, we can vectorize it as
\begin{equation*}
\begin{aligned}
   \txvec(N_\beta)
=& (\Phi^{-T}\Lambda\Phi^{-1}) \otimes ( E^T U^T ) \Omega^{-1} \txvec Y \\
=& (\Phi^{-T}\Lambda\Phi^{-1}) \otimes ( E^T U^T ) (\Phi\otimes U) \\
&  [\Lambda\otimes D+I_2\otimes I_n-\Lambda\otimes(U^TZ_{\{ij\}}(U^TZ_{\{ij\}})^T)]^{-1}(\Phi\otimes U)^T
    \txvec Y \\
=& (\Phi^{-T}\Lambda) \otimes E^T 
   [\Lambda\otimes D+I_2\otimes I_n-\Lambda\otimes(U^TZ_{\{ij\}}(U^TZ_{\{ij\}})^T)]^{-1}
   \txvec(U^TY\Phi).
\end{aligned}
\end{equation*}
Denote $\txvec(G)=[\Lambda\otimes D+I_2\otimes I_n-\Lambda\otimes(U^TZ_{\{ij\}}(U^TZ_{\{ij\}})^T)]^{-1}
   \txvec(U^TY\Phi)$, $G\in\mathbb R^{n\times2}$, then we have
\begin{equation}
\begin{aligned}
\txvec(N_\beta)
= (\Phi^{-T}\Lambda) \otimes E^T \txvec(G)
= \txvec(E^T G \Lambda \Phi^{-1})
\end{aligned}
\end{equation}

Similarly, the vectorization of $N_\epsilon$ is
\begin{equation}
\begin{aligned}
   \txvec(N_\epsilon)
=& (\Phi^{-T}\Phi^{-1}) \otimes (U^{-1}) \Omega^{-1} \txvec Y \\
=& (\Phi^{-T}) \otimes ( I_n ) 
    [\Lambda\otimes D+I_2\otimes I_n-\Lambda\otimes(U^TZ_{\{ij\}}(U^TZ_{\{ij\}})^T)]^{-1} 
    \txvec(U^TY\Phi) \\
=& (\Phi^{-T}) \otimes I_n \txvec(G) \\
=& \txvec(G \Phi^{-1})
\end{aligned}
\end{equation}

\subsubsection*{A.3 FLAG-CA Methods}

\begin{algorithm}
\caption{MM Algorithm with Eigen-decomposition for FLAG-CA}
\hspace*{\algorithmicindent} \textbf{Input}: X, Y. \\
\hspace*{\algorithmicindent} \textbf{Output}: $\hat\Gb, \hat\Ge.$
\begin{algorithmic}[1]
\STATE Eigen decomposition: $U^TXX^TU=D=\diag(d).$
\STATE Transform data: $\tilde Y\leftarrow U^TY.$
\STATE Initialization: $\Gb^{(0)}=\Ge^{(0)}=\frac12\cov(Y)$.
\REPEAT
\STATE Simultaneous congruence decomposition: $(\Lambda^{(m)}, \Phi^{(m)}) \leftarrow (\Gb^{(m)},\Ge^{(m)}); $
\STATE $ \Omega^{(m)}=(\Phi^{-(m)}\otimes U^{-1})^T (\Lambda^{(m)}\otimes D +I_2\otimes I_n) (\Phi^{-(m)}\otimes U^{-1}); $
\STATE $ \zeta \leftarrow (\Up^T\Omega^{-(m)}\Up)^{-1} \Up^T\Omega^{-(m)}Y; $
\STATE $ \text{Cholesky }L_\beta^{(m)}L_\beta^{(m)T}\leftarrow\Phi^{(m)}\diag( \tr(D(\lambda_l^{(m)}D+I_n)^{-1}), l=1,2 )\Phi^{(m)T}; $
\STATE $ \text{Cholesky }L_\epsilon^{(m)}L_\epsilon^{(m)T}\leftarrow\Phi^{(m)}\diag( \tr((\lambda_l^{(m)}D+I_n)^{-1}), l=1,2 )\Phi^{(m)T}; $
\STATE $ N_\beta^{(m)}\leftarrow D^{\frac12} [(\tilde Y \Phi^\mth) \oslash (d\lambda^{(m)T}+\mathbf{1}_n\mathbf{1}_2^T)] \Lambda^{(m)}\Phi^{-(m)}; $
\STATE $ N_\epsilon^{(m)}\leftarrow [(\tilde Y \Phi^\mth) \oslash (d\lambda^{(m)T}+\mathbf{1}_n\mathbf{1}_2^T)] \Phi^{-(m)}; $
\STATE $ \Gb^{(m+1)} \leftarrow L_\beta^{-(m)T} (L_\beta^{(m)T} N_\beta^{(m)T} N_\beta^{(m)} L_\beta^{(m)})^{\frac12} L_\beta^{-(m)}; $
\STATE $ \Ge^{(m+1)} \leftarrow L_\epsilon^{-(m)T} (L_\epsilon^{(m)T} N_\epsilon^{(m)T} N_\epsilon^{(m)} L_\epsilon^{(m)})^{\frac12} L_\epsilon^{-(m)}. $
\UNTIL{the log-likelihood $\ell(\Gamma)$ stop increasing or maximum iteration reached}
\label{MM_CA}
\end{algorithmic}
\end{algorithm}

\begin{algorithm}
\caption{PX-EM algorithm with eigen-decomposition for FLAG-CA}
\begin{algorithmic}[1]
\STATE Initialization: $\Gb=\Ge=\frac{\cov(Y)}{2}$,
\STATE Eigen-decomposition $X^TX=VQV^T$. \\
\REPEAT
\STATE E-step: set $\delta^{(m)}=1$,
\begin{equation*}
\begin{aligned}
\Sigma_{\bb}=
& \begin{pmatrix} V^T & 0 \\ 0 & V^T \end{pmatrix} \\
&\hspace{-0.3cm} \begin{pmatrix} \diag( \delta^2(\Ge^{-1})_{11}q +                                         (\Gb^{-1})_{11}\mathbbm{1}_{p-2} ) &
        \diag( \delta^2(\Ge^{-1})_{12}q +                                       (\Gb^{-1})_{12}\mathbbm{1}_{p-2} ) \\
        \diag( \delta^2(\Ge^{-1})_{21}q +                                       (\Gb^{-1})_{21}\mathbbm{1}_{p-2} ) &
        \diag( \delta^2(\Ge^{-1})_{22}q +                                       (\Gb^{-1})_{22}\mathbbm{1}_{p-2} )
    \end{pmatrix} ^{-1} \\
& \begin{pmatrix} V & 0 \\ 0 & V \end{pmatrix},
\end{aligned}
\end{equation*}
$$\mu_{\bb}= \Sigma_{\bb}
\delta(\Ge^{-1} \otimes X^T)\tilde Y, $$
$$ELBO^{(m)}=Q(\Omega^{(m)})+\frac12\text{log}|\Sigma_{\bb}|.$$
\STATE M-step: Update the model parameters by
$$\delta^{(t+1)} \leftarrow \frac {\tilde Y^T(\Ge^{-1}\otimes X) \mu_{\bb}}
        { \mu_{\bb}^T (\Ge^{-1} \otimes X^TX) \mu_{\bb}
        + \tr[ (\Ge^{-1} \otimes X^TX) \Sigma_{\bb} ] }, $$
$$ \zeta^{(t+1)} \leftarrow (\Up^T \Up)^{-1} \Up^T (Y-\delta X \mu_{\bb}), $$
$$\Ge^{(t+1)} \leftarrow \frac1n \begin{pmatrix} \tr[S_{11}] & \tr[S_{12}] \\ \tr[S_{21}] & \tr[S_{22}] \end{pmatrix},$$
$$\Gb^{(t+1)} \leftarrow \frac1{p-2} \begin{pmatrix} \tr[W_{11}] & \tr[W_{12}] \\ \tr[W_{21}] & \tr[W_{22}] \end{pmatrix}.$$
\STATE Reduction-step: Rescale
$\Gb^{(t+1)} \leftarrow (\delta^{(t+1)})^2\Gb^{(t+1)}$
and reset $\delta^{(t+1)}=1.$
\UNTIL{the incomplete-data log-likelihood $ELBO^{(m)}$ stop increasing}
\label{EM_CA}
\end{algorithmic}
\end{algorithm}

\subsection*{B Simulation Studies}

\subsubsection*{B.1 Hub Structure}

\subsubsection*{B.2 Multiple Graphs}
\begin{figure}
    \centering
    \includegraphics[width=0.7\textwidth]{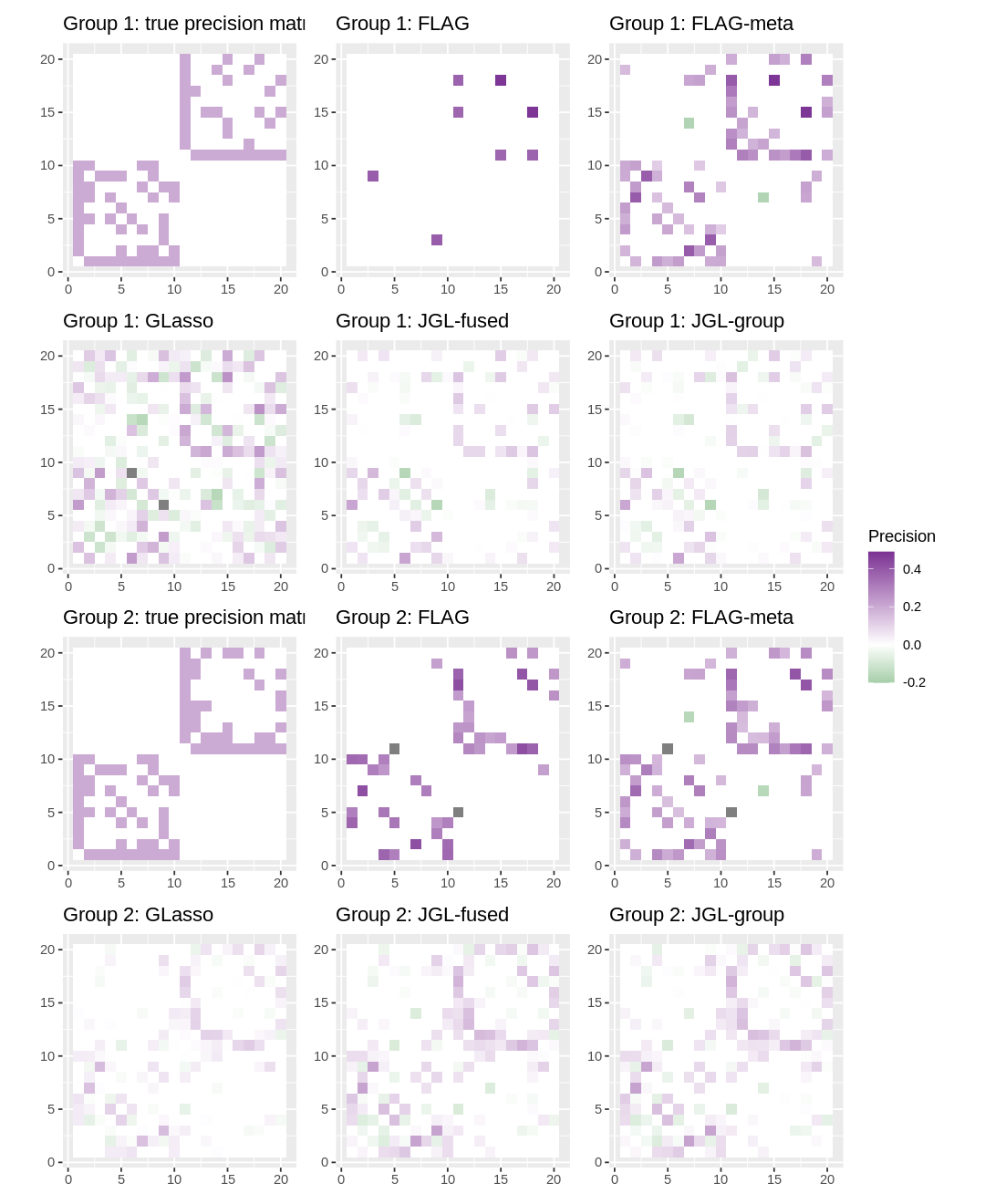}
    \caption{Comparison of estimated precision matrices using FLAG-based and joint GLasso-based methods for group 1 and 2, with the results for different groups shown in separate rows.}
\end{figure}

\subsection*{C Real Data Analysis}

\subsubsection*{C.1 Human Brain Gene Expression Data}

\subsubsection*{C.2 University Webpage Data}

\subsubsection*{C.3 U.S. Stock Prices}

\end{document}